# US HL-LHC Accelerator Upgrade Project

## MQXFA Series Magnet Production Specification

Paolo Ferracin[2], Giorgio Ambrosio[1], Giorgio Apollinari[1], James Blowers[1], Ruben Carcagno[1], Dan Cheng[2], Katherine Ray[2], Soren Prestemon[2], Michael Solis[2], Giorgio Vallone[2]

[1]Fermi National Accelerator Laboratory, Batavia, IL 60510 USA
[2]Lawrence Berkeley National Laboratory, Berkeley, CA 94720 USA.

## TABLE OF CONTENTS

## 1. Scope

The purpose of this document is to define the specifications for the structure fabrication and assembly of MQXFA series magnets [1] to be used by the US High-Luminosity LHC Accelerator Upgrade Project (AUP). Magnets fabricated according to these specifications are expected to allow MQXFA magnets to meet the MQXFA Functional Requirements Specification [2]. These specifications for the fabrication of MQXFA series magnets are based on the R&D performed by the US LHC Accelerator Research Program (LARP) and the development performed by AUP [3], in collaboration with CERN, through fabrication and test of pre-series coils and magnets.

## 2. Magnet Design

The design of the MQXFA quadrupole magnet, described in details in [1], relies on an aluminum shell pre-stressed at room temperature with bladders and interference keys (i.e. bladder and key technology), which has been demonstrated in the previous successful series of LARP magnets, in particular the HQ quadrupole magnet. The cross-section of the structure of MQXF is a direct scale-up from the HQ models. As shown Fig. 2.1, MQXF features an aperture of 150 mm and provides a nominal field gradient of 132.2 T/m over a magnetic length of 4.2 m (MQXFA).

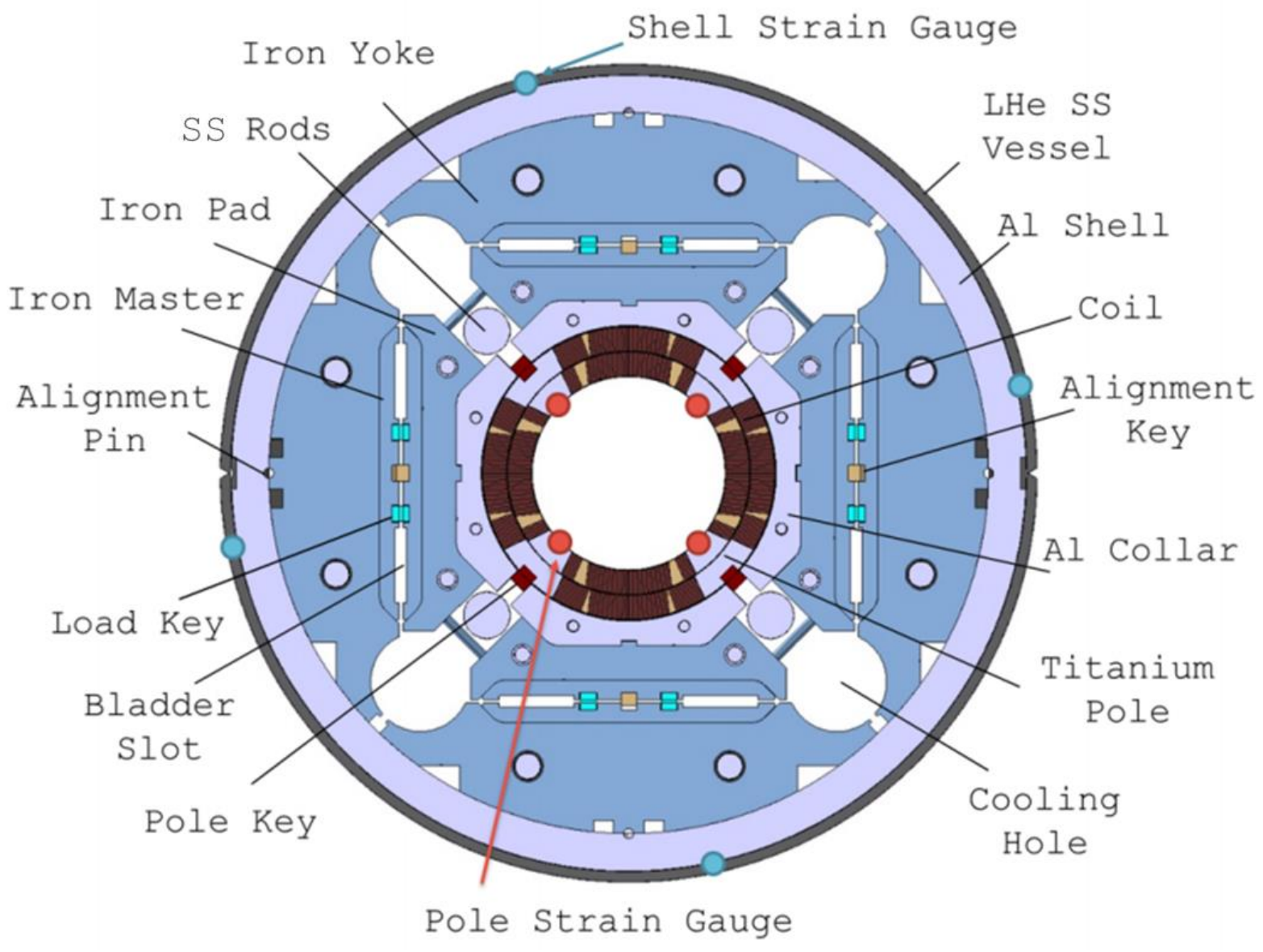

Figure 2.1: Cross section of the MQXF.

The design of the structure comprises the iron shell-yoke sub-assembly, composed by four iron yokes surrounded by a 29 mm thick aluminum shell, and the coil-pack sub-assembly, which consists of four aluminum collars and iron pads bolted around the coils with G11 pole alignment keys. The iron yoke, iron pads and aluminum collars are made of about 50 mm thick laminations assembled with tie rods. In the end coil regions, the iron pad laminations are replaced by stainless steel pad laminations to reduce the conductor peak field in the ends. Between each pad and yoke, the master package on each quadrant contains two interference keys to balance the azimuthal tension in the outer shell with the azimuthal compression in the inner coils.

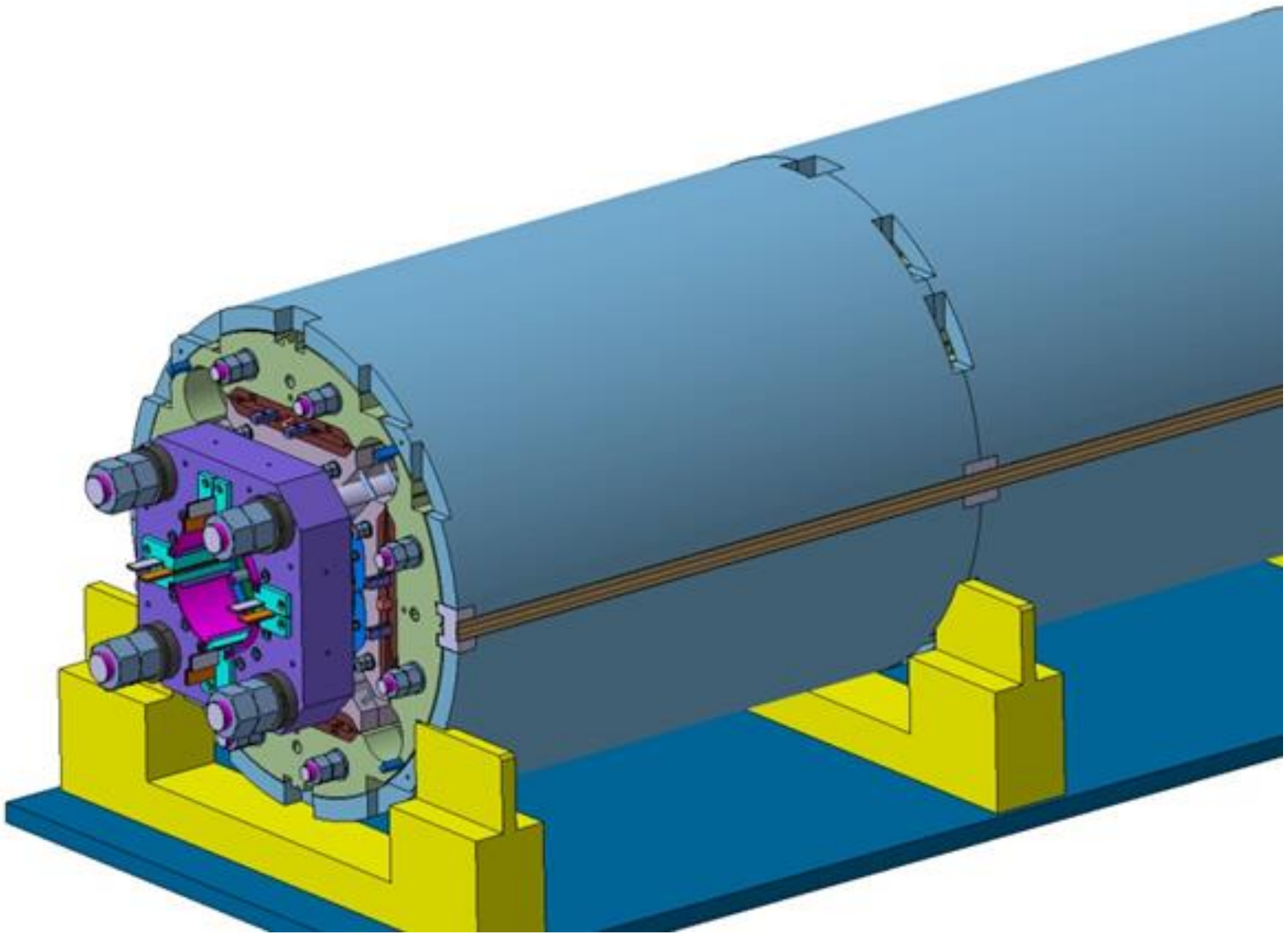

Figure 2.2: 3D view of MQXFA end region.

Maintaining contact between the coils and poles pieces at all stages is achieved by the azimuthal pre-load applied on the shell. The pre-load is obtained by a system of water-pressurized bladders and keys which pre-compress the coil-pack and pre-tension the aluminum shell at room temperature. During the pre-load operation, the pressurized bladders open up the master package and allow inserting the load keys with shims of the designed thickness, thus creating an interference between the coil-pack and the shell-yoke sub-assembly when the bladders are deflated and removed. The final pre-load is achieved during the cool-down phase, when the tensioned aluminum shell increases its stress because of its high thermal contraction.

In operational conditions the MQXFA magnet will experience a total electro-magnetic force in the axial direction (Z direction, parallel to the magnet's bore) of 1.17 MN at 16.47 kA (nominal current plus margin). Axial pre-stress is therefore designed to withstand the total axial forces generated by the coil ends. Four tensioned steel rods within the pads' gaps are connected to end-plates to provide the axial pre-stress (see Fig. 2.2). Similar to the azimuthal pre-stress, the room temperature axial pre-stress is tuned to counteract after cool-down the axial Lorentz force.

To summarize, the main features of the MQXF structural design are:

- Shell-based support structure relying on "bladder and key" technology to perform azimuthal pre-loads, which allows reversible assembly process and tunable preload;
- G11 alignment keys inserted into the pole pieces provide coil alignment by assuring azimuthal contact between coil and collars after cool-down;
- The aluminum shell provides additional pre-load to the coil during cool-down;
- Axial pre-load is provided by four SS rods and two end-plates.

## 3. Structure Components Specifications

All the structure components are procured by LBNL. This section describes the major components, including shell, yoke, pads and collars. A list of all the major components, with part number and quantity is provided in Table 3.1

Table 3.1. List of major components, and quantities per magnet assembly

| Component | Part Number | Type | Qty | Notes |
|---|---|---|---|---|
| **Shells** | | | | |
| MQXFA Short Magnet Shell | SU-1010-1072 | Part | 2 | |
| MQXFA Long Magnet Shell | SU-1010-1073 | Part | 6 | |
| **Yokes** | | | | |
| MQXFA RE YOKE HALF-STACK ASSY | SU-1010-0097 | Assy | 4 | |
| MQXFA-2 YOKE HALF-STACK ASSEMBLY – LE | 27L179 | Assy | 4 | |
| Yoke pre-stack Assembly, 588 | 27L410 | Assy | 24 | |
| Yoke pre-stack assembly, 255 | 27L412 | Assy | 8 | |
| Yoke, type 1 | 27L175 | Part | 4 | |
| Yoke, type 2 | 27L176 | Part | 328 | In Pre-assy |
| Yoke, type 3 | 27L177 | Part | 24 | |
| Yoke, type 4 | 27L178 | Part | 4 | |
| Yoke, type 5 | SU-1010-0196 | Part | 4 | |
| Yoke, type 6 | SU-1010-0197 | Part | 4 | |
| **Master Keys** | | | | |
| Master Key half, bladder cutout | 27L213 | Part | 8 | |
| Master Key half, no bladder cutout | 27L214 | Part | 8 | |
| **Load Pads** | | | | |
| MQXFA Load Pad Thru Prestack, 1050 | 27L417 | Assy | | |
| MQXFA Load Pad Thru Prestack, 1100 | 27L418 | Assy | | |
| MQXFA Load Pad Thru Prestack, 1088 | 27L419 | Assy | | |
| Load Pad Lamination, Thru, type 1 (304CO) | 27L253 | Part | 2 | |
| Load Pad Lamination, Thru, type 2 (304CO) | 27L254 | Part | 24 | In Pre-assy |
| Load Pad Lamination, Thru, type 3 (304CO) | 27L255 | Part | 2 | |
| Load Pad Lamination, Thru, type 4 (ARMCO) | 27L229 | Part | 162 | In Pre-assy |
| Load Pad Lamination, Thru, type 5 (ARMCO) | 27L230 | Part | 6 | |
| Load Pad Lamination, Thru, type 6 (ARMCO) | 27L223 | Part | 2 | In Pre-assy |
| MQXFA Load Pad Threaded Prestack, 1050 | 27L552 | Assy | | |
| MQXFA Load Pad Threaded Prestack, 1100 | 27L553 | Assy | | |
| MQXFA Load Pad Threaded Prestack, 1088 | 27L554 | Assy | | |
| Load Pad Laminat.., Threaded, type 1 (304CO) | 27L250 | Part | 2 | |
| Load Pad Laminat., Threaded, type 2 (304CO) | 27L251 | Part | 24 | In Pre-assy |
| Load Pad Laminat., Threaded, type 3 (304CO) | 27L252 | Part | 2 | |
| Load Pad, Threaded, type 4 (ARMCO) | 27L227 | Part | 162 | In Pre-assy |
| Load Pad, Threaded, type 5 (ARMCO) | 27L228 | Part | 6 | |
| Load Pad, Threaded, type 6 (ARMCO) | 27L224 | Part | 2 | In Pre-assy |
| **Collars** | | | | |
| Collar Lamination Type 1 | SU-1010-1058 | Part | 4 | |
| Collar Lamination Type 2 | SU-1010-1059 | Part | 344 | |
| Collar Lamination Type 3 | SU-1010-1060 | Part | 4 | |
| Collar Lamination Type 4 | SU-1010-1061 | Part | 12 | |
| **Axial Loading** | | | | |
| Axial Endplate, Lead End | 27K557 | Part | 1 | |
| Axial Endplate, Return End | 27K564 | Part | 1 | |
| Axial Rods | 27L236 | Part | 4 | |

## 3.1. Magnet Shells

The MQXFA magnet aluminum shell are divide into eight segments: six long sections in the central section, and two half-length sections on each end (see Fig. 3.1 and 3.2).

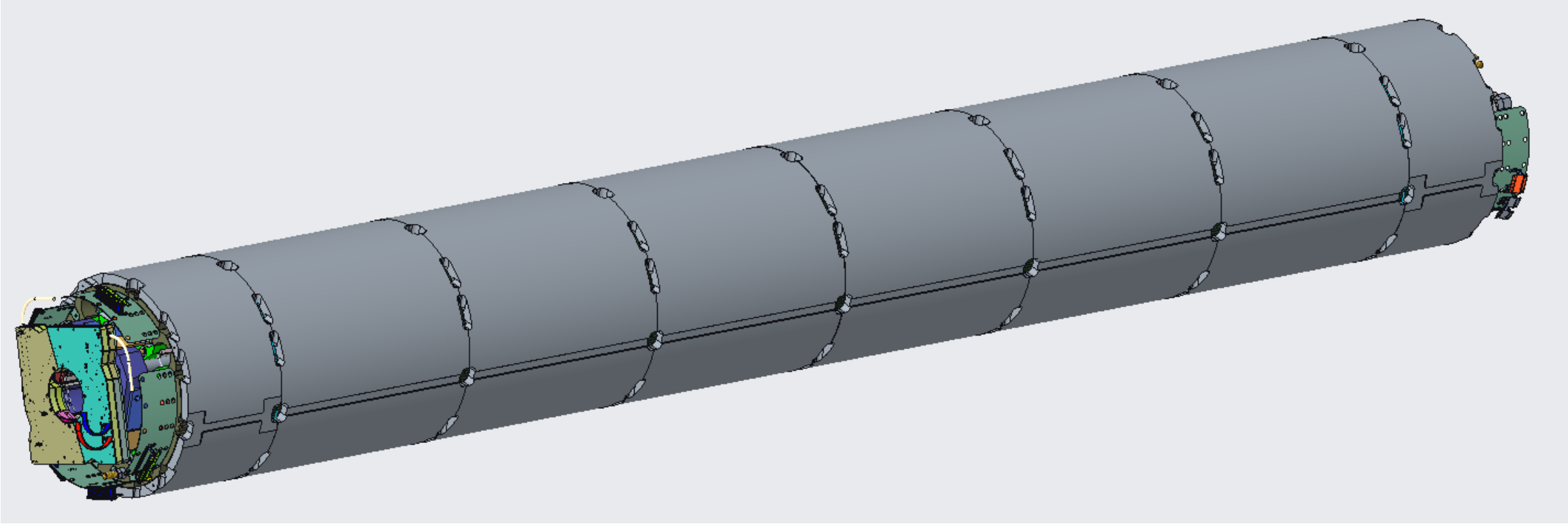

Fig. 3.1: MQXFA magnet with 8 shell segments.

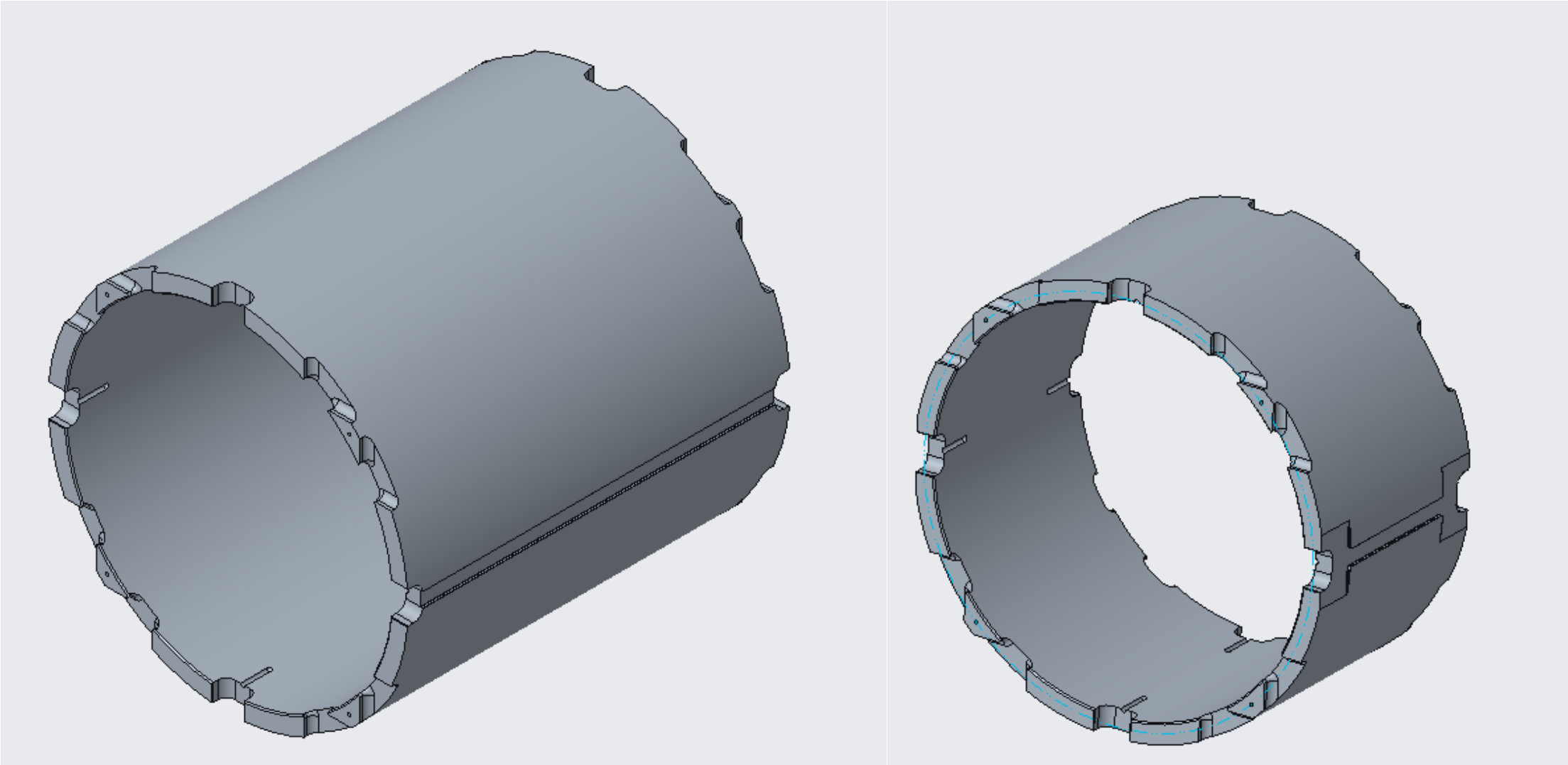

Fig. 3.2: Long shell segment and short shell segments in the MQXFA magnet.

Table 3.2 MQXFA Shells

| Name | Drawing ID | Qty. per Magnet |
|---|---|---|
| Long Shell | SU-1010-1072 | 2 |
| Short Shell | SU-1010-1073 | 6 |

### 3.1.1. Magnet Shells Features

#### 3.1.1.1. Inner and Outer Diameter

The MQXFA magnet shells, both long and short, share the same 556 mm +0.1/-0.0 inner diameter and 614 mm +0.1/-0.0 outer diameter. Due to the nature of the AL7075-T6 material and the thin-walled geometry, the dimensional tolerances called out in the drawings only apply to the restrained condition of the shell (see Fig. 3.3). The ID and OD for the long shells are scanned in a ring pattern at 5 positions spaced down the bore: 108 mm, 217 mm, 325 mm, 434 mm and 542 mm from the face of the bore. The ID and OD for the short shells are scanned in a ring pattern at 5 positions spaced down the bore: 56 mm, 108 mm, 162 mm, 217 mm and 217 mm from the face of the bore.

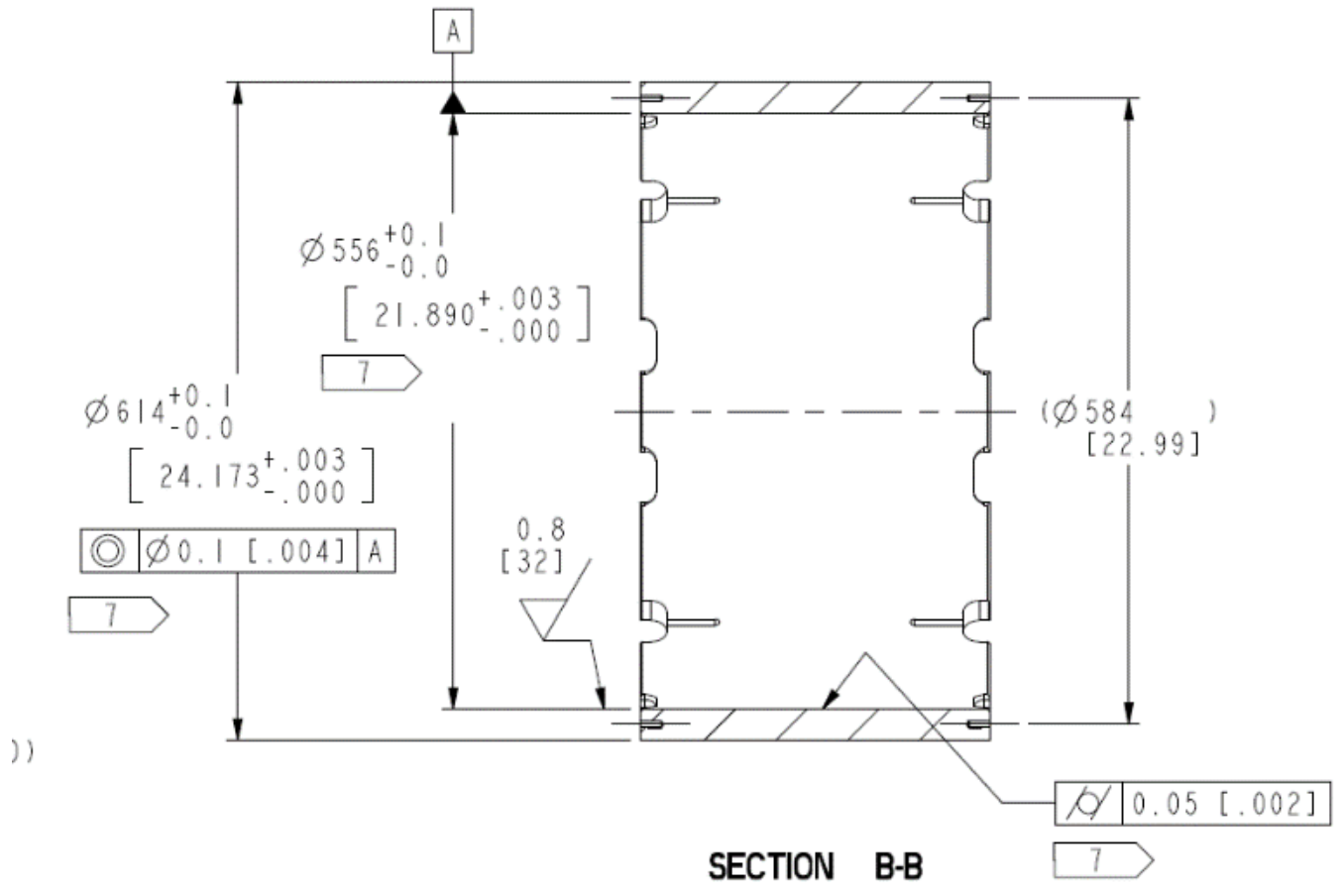

Fig. 3.3: Shell Inner and Outer Diameter. The "7" flag in the dimension a note stating the tolerance requirement applies in the restrained condition.

For the shell ID and OD, the following specifications are set:

- ***The shell ID at 5 evenly spaced rings shall have an average diameter of 556 mm +0.1/-0.0***
- ***The shell OD at 5 evenly spaced rings shall have an average diameter of 614 mm +0.1/-0.0***

#### 3.1.1.2. End Cut-outs

The end cuts out serve multiple purposes. The V-shaped cut out allow for the cradle tooling to directly support the yokes during assembly and integration. The 54 mm U-shaped cut out is ultimately used for the tack block when welding the cold mass stainless steel shell over the magnet. During magnet assembly, these cuts are also used to align shells during the stacking process. The pin slot at the base of the U-shaped cut out allows the shells clock their position in relation to the yoke stacks in the shell-yoke assembly. Only one pin slot is required for use, which is the top slot show in Fig. 3.4.

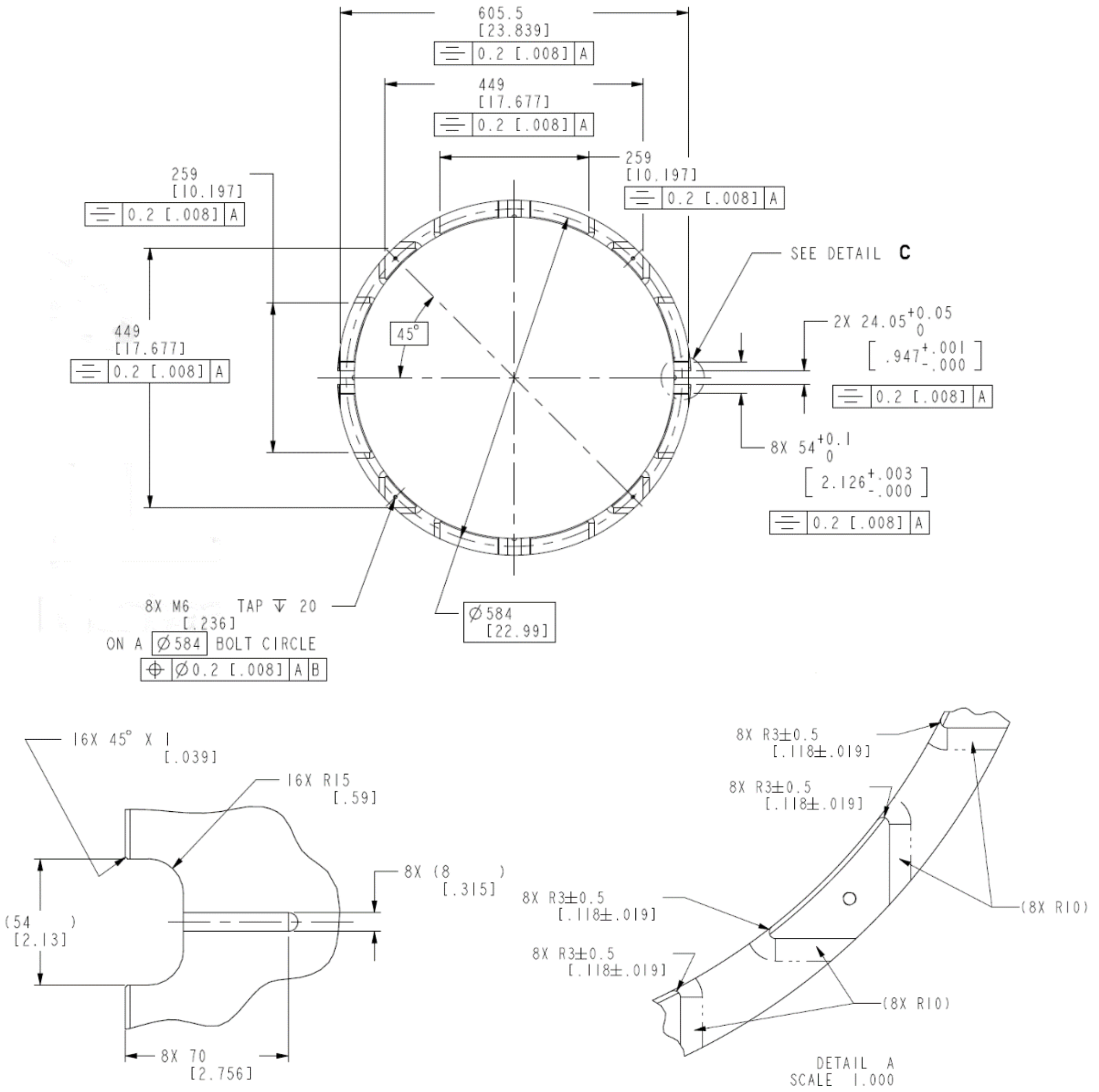

Fig. 3.4: End cutouts Geometry.

#### 3.1.1.3. Welding Slot

The welding slot accommodates the backing strip for when the stainless steel shell of the cold mass is welded around the magnet (see Fig. 3.5). The tolerance of 24.05 mm +.05/-0 must be held tightly in order to maintain a close fit with the backing strip. While long and short shells have identical longitudinal cross section, both ends of the short shells “flare” out in order to reduce the stress concentrations in the end shells (MQXFA Final Design Report, US-HiLumi-Doc-948).

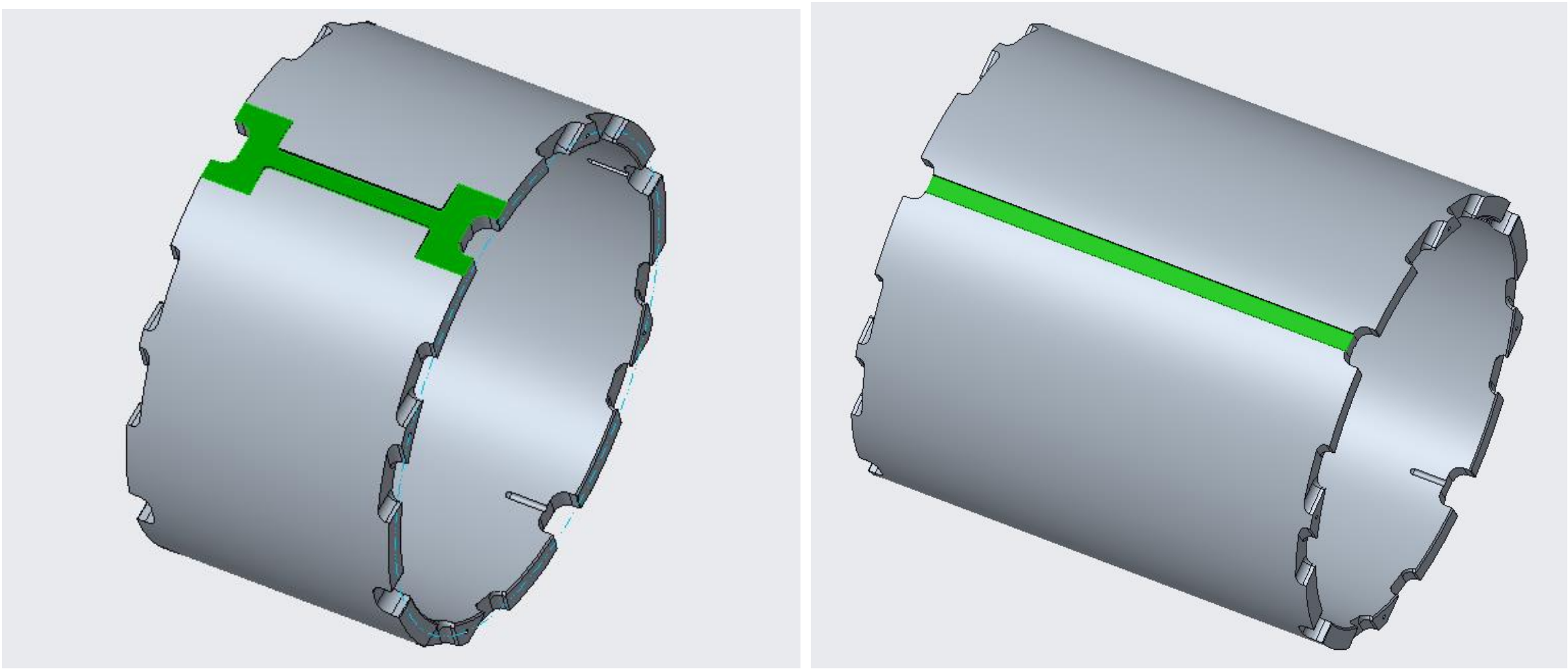

Fig 3.5: Short shell welding slot (left). Long shell welding slot (right).

#### 3.1.2. QC Plan

The quality control of specifications for the magnet shells are documented in the QC Plan for MQXFA Structures Parts Document [4].

### 3.2. Yokes

The iron yoke of MQXFA is composed by laminations shown in Fig. 3.6.

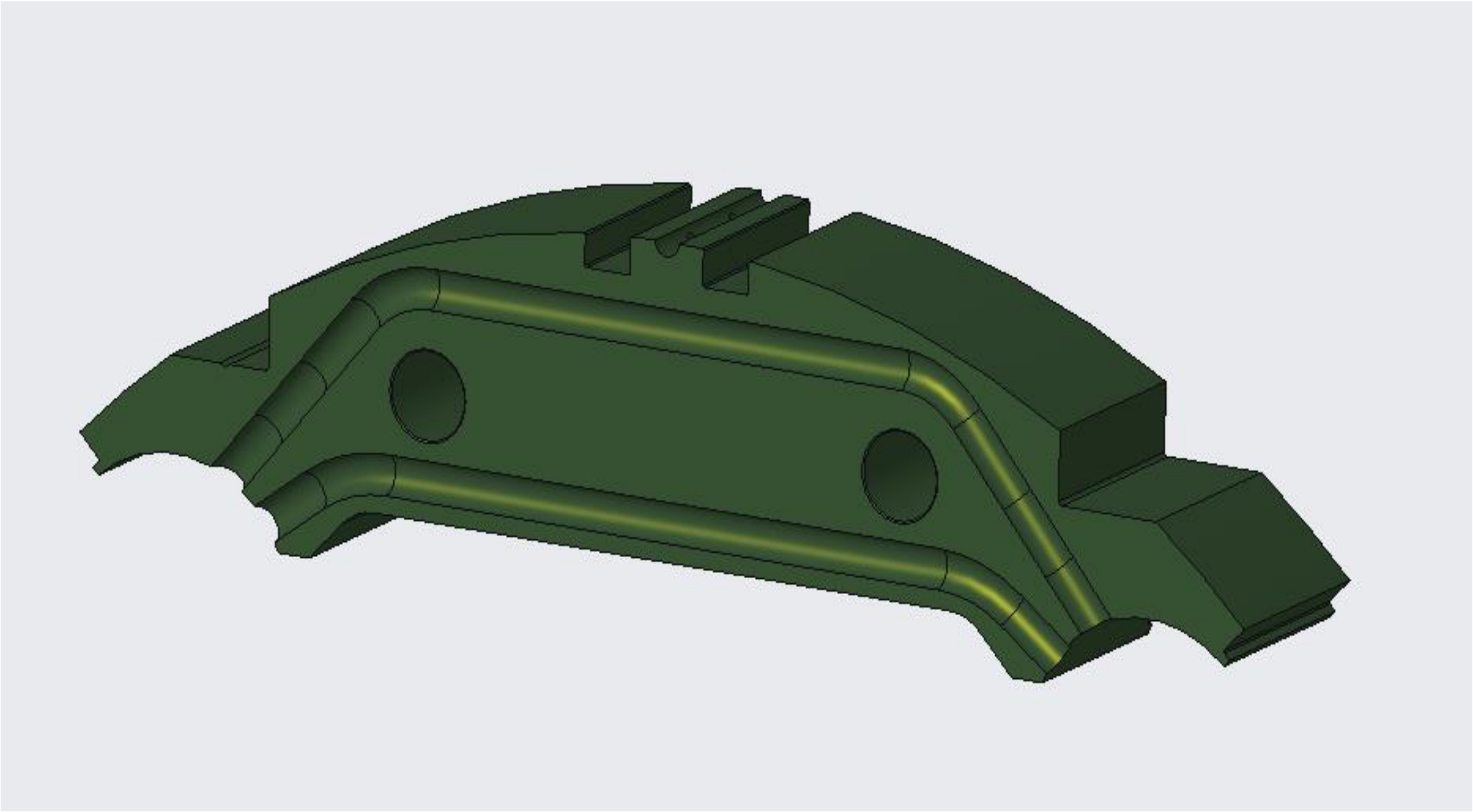

Fig 3.6: Yoke lamination (Type-3, with cooling grooves).

### 3.2.1. Yoke Features

The details of the yoke design are shown in Fig. 3.7 and described in the following sections.

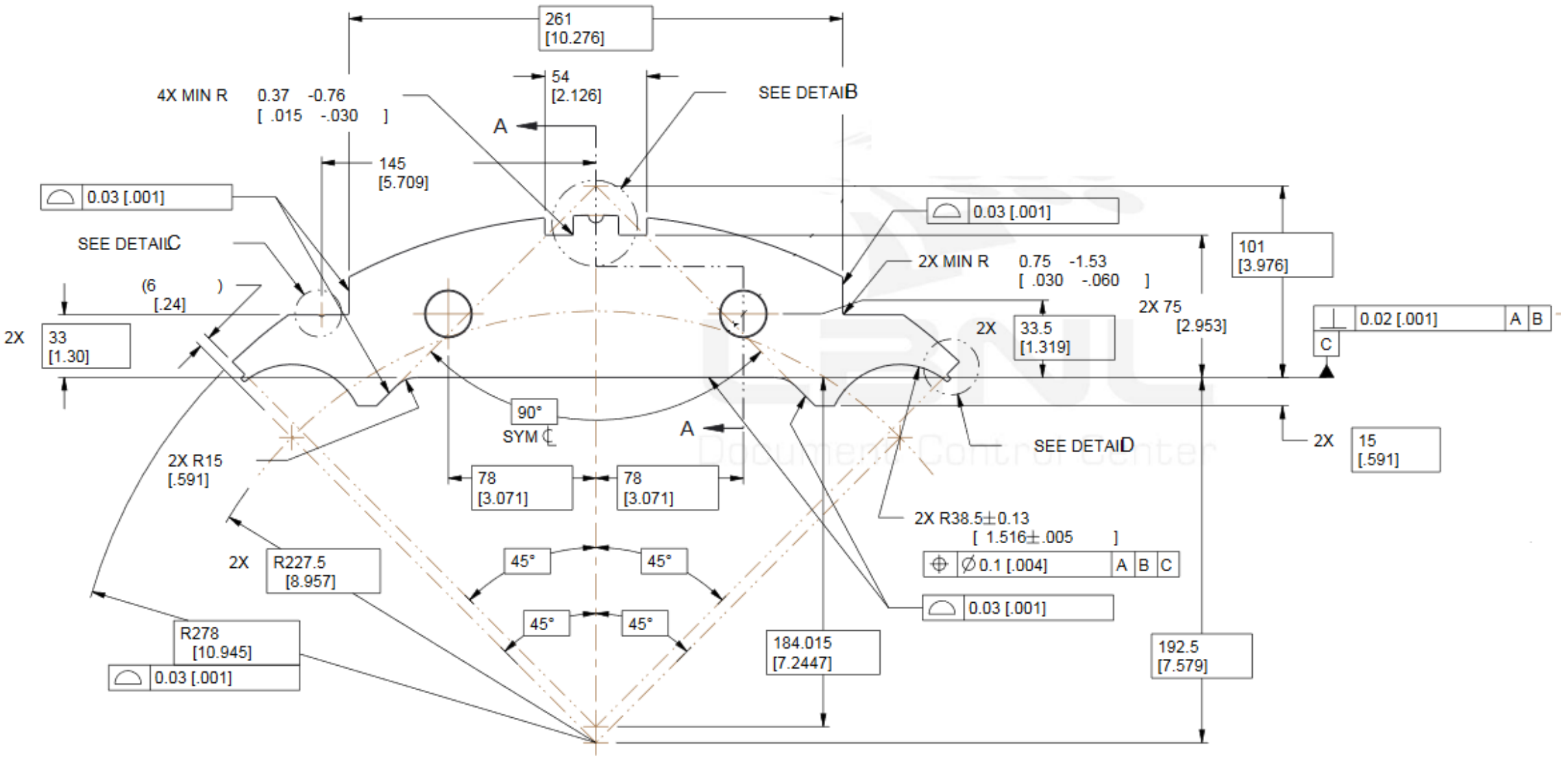

Fig. 3.7: Common yoke fabrication features.

#### 3.2.1.1. Radial Profile

All yokes share the same 278 mm radial profile (see Fig. 3.8). The profile mates with the inner radius of the magnet shells. When the MQXFA magnet is preloaded and energized it will transfer a large amount of azimuthal forces from the yokes to the magnet shells. The precision of the yokes outer profile is crucial in distributing forces uniformly, thus reducing potential stress points.

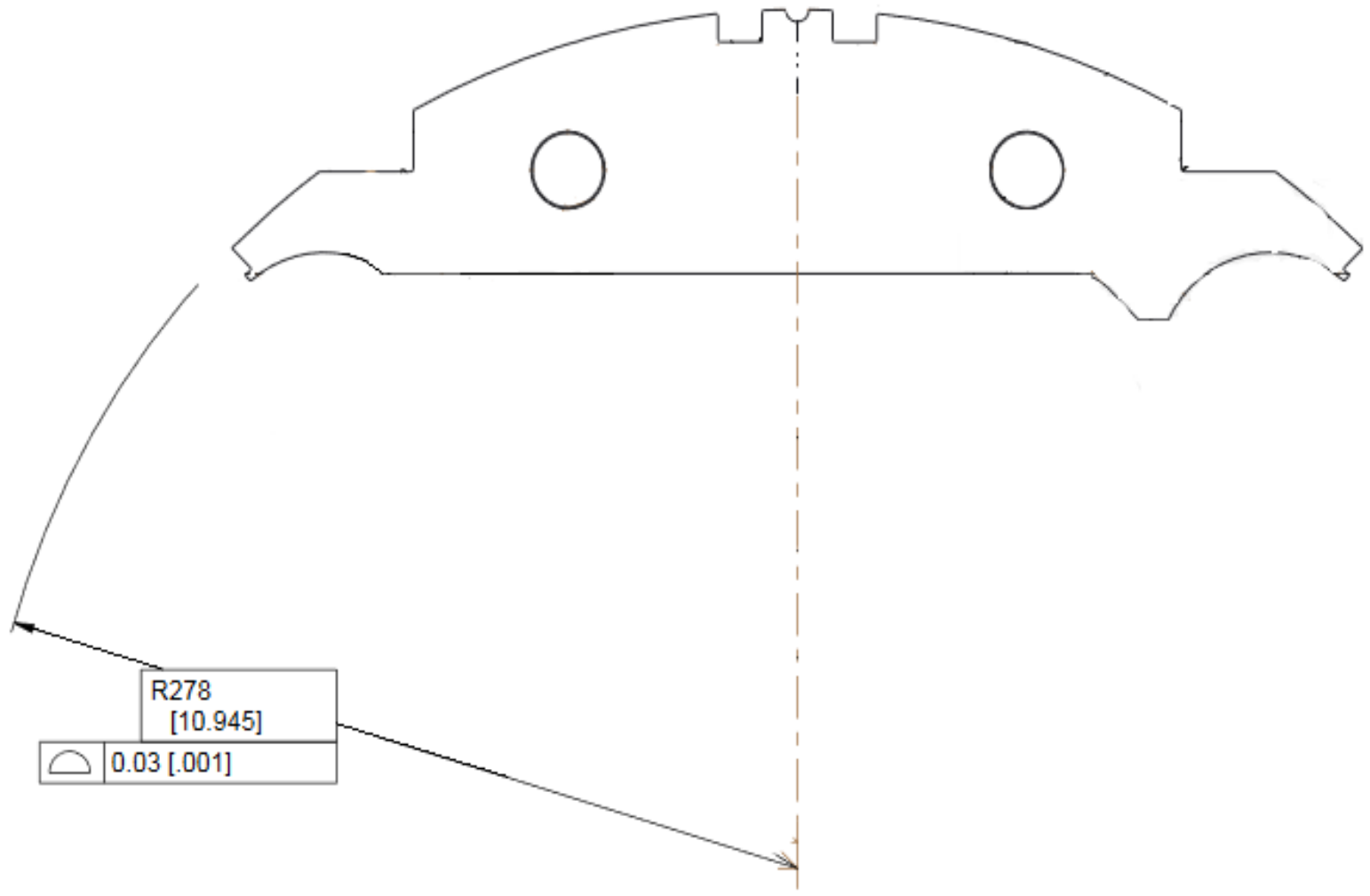

Fig. 3.8: Yoke outer 278 mm radius with precision profile.

<u>For the yoke outer radius, the following specifications are set:</u>

- ***The yoke radial profile shall maintain a 0.03 [.001"] tolerance.***

#### 3.2.1.2. Master Key Interface Profile

The master key interface carries a profile tolerance of 0.03 mm (see Fig.3.9). The precision of the profile is critical in order to transfer uniformly the load from the shell to coil pack.

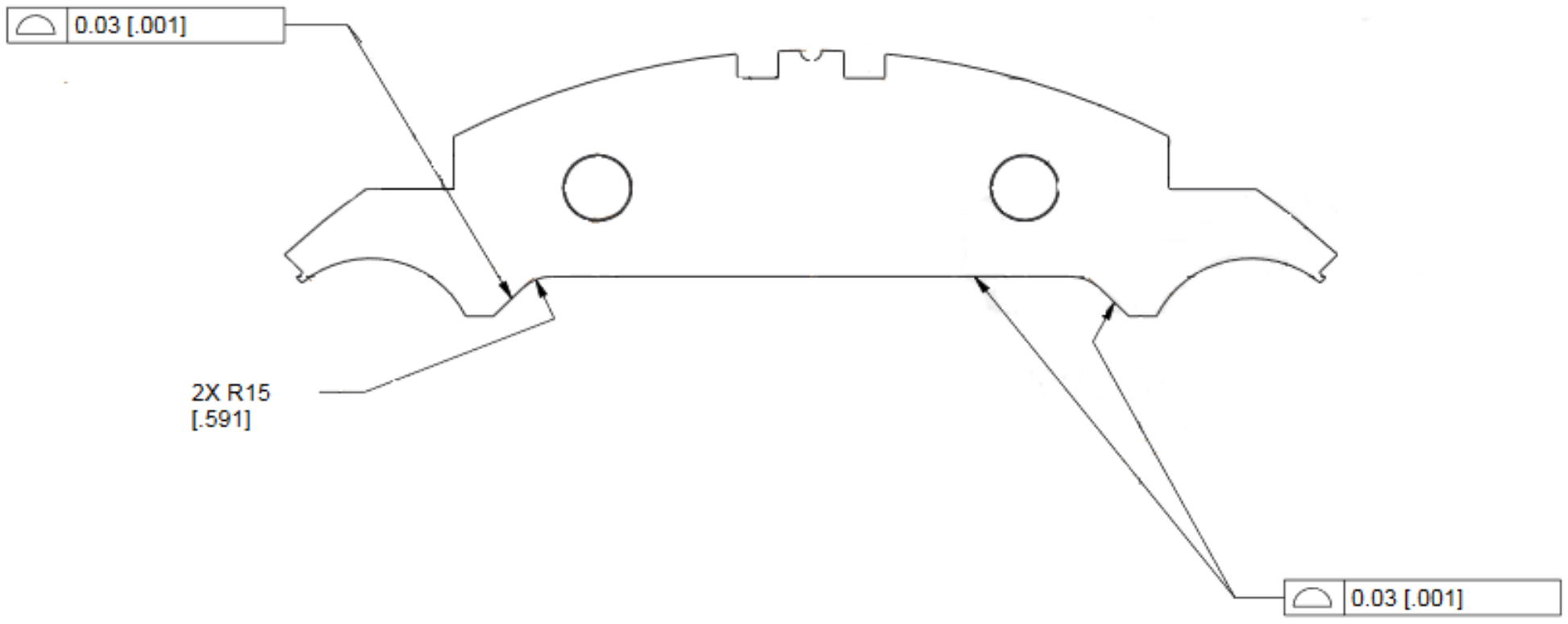

Fig. 3.9: Yoke Master Key interface with precision profile.

For the yoke master key interface, the following specifications are set:

- ***The yoke master key interface profile shall maintain a 0.03 [.001"] tolerance.***

#### 3.2.1.3. Cooling holes

The yoke 38.5 mm radius cooling holes provide crucial assembly and magnet cooling functionality (see Fig. 3.10). The key functionalities of the cooling holes is to provide space for heat exchangers, bus bars, and free LHe path through the length of the MQXFA magnet structure while cooling the structure. The cooling holes are not exposed to high azimuthal load bearing forces, although they must maintain a 77 mm diameter opening after preloaded assembly.

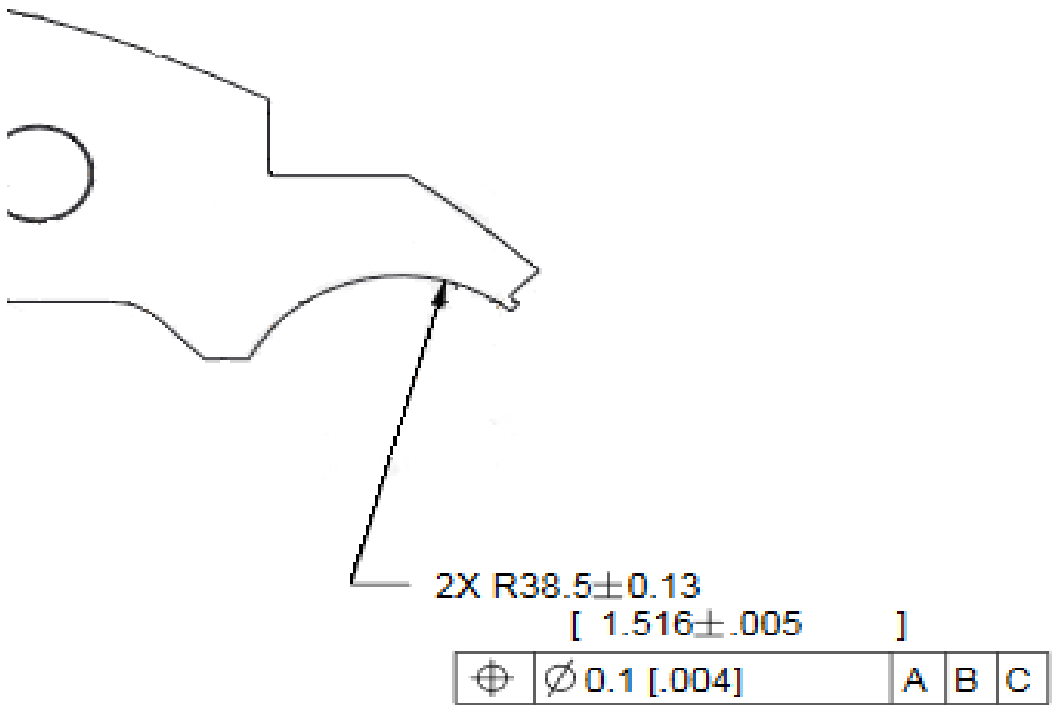

Fig. 3.10 Yoke 38.5 mm cooling hole opening.

During magnet pre-load, the bladder and key operation pushes further apart the yoke quadrants, thus increasing the cooling hole opening by 0.200 mm on the radius (based on FE model).

#### 3.2.1.4. Tie Rod/Bushing Holes

The two 24.1 mm holes symmetric to the center line are for securing stacks in position and precise alignment by compression (see Fig. 3.11). The close-fitting bushings lock the lamination stacks into one complete unit when exposed to forces perpendicular to the direction of the holes such as stack lifting operations.

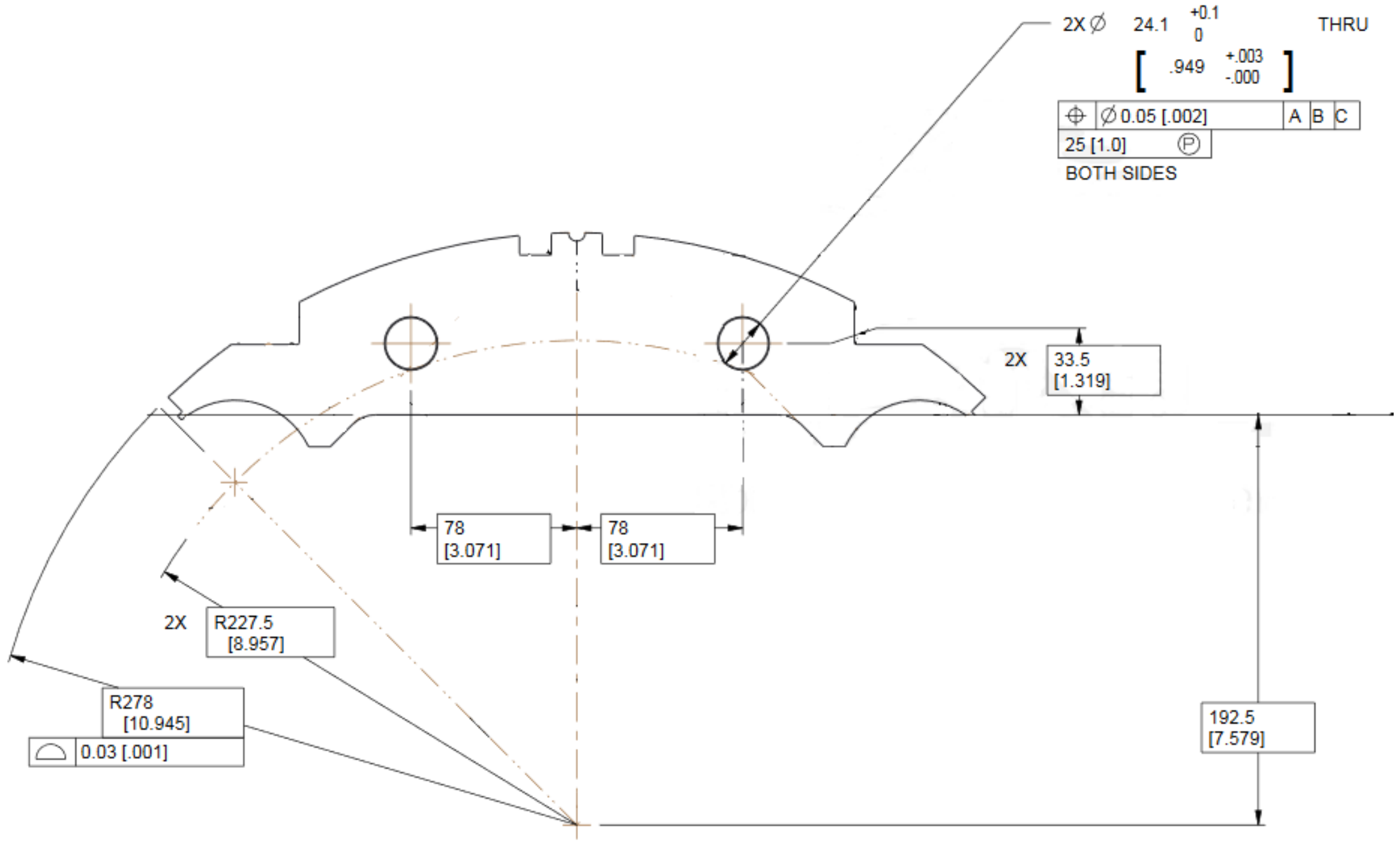

Fig. 3.11: Yoke 24.1mm tie rod/bushing holes.

### 3.2.1.5. Yoke Gap Key Groove

The 3 mm x 9.5 mm yoke gap grooves are load bearing precision features of the yokes (see Fig. 3.12). The groove is designed to hold the yoke gap key in place between adjacent yoke stacks when the assembling the shell-yoke subassembly. Bladder pressure is used to lightly preload the yokes within the magnet shells using precision gap key/shim sets to maintain a uniform opening prior to the insertion of the coil pack subassembly. These gap key/shim sets are removed after the magnet has been fully preloaded.

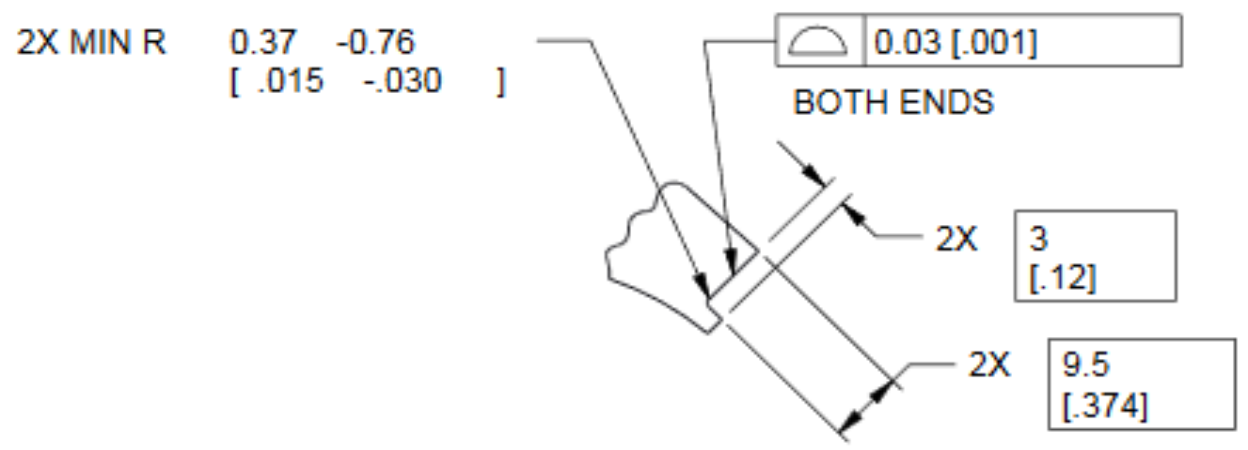

3.12: Yoke 3mm x 9.5mm gap key grooves.

### 3.2.1.6. Cooling Groove

The 5.56 mm cooling groove (see Fig. 3.13) is a feature found on certain yoke laminations (Types 1, 3, 4, 5 and 6). The cooling grooves in these laminations, when fully assembled, are designed to allow a total of 150 cm$^2$ of free helium path between the 77 mm cooling holes along the length of the magnet, as specified in the MQXFA Functional Requirements Specifications document [2].

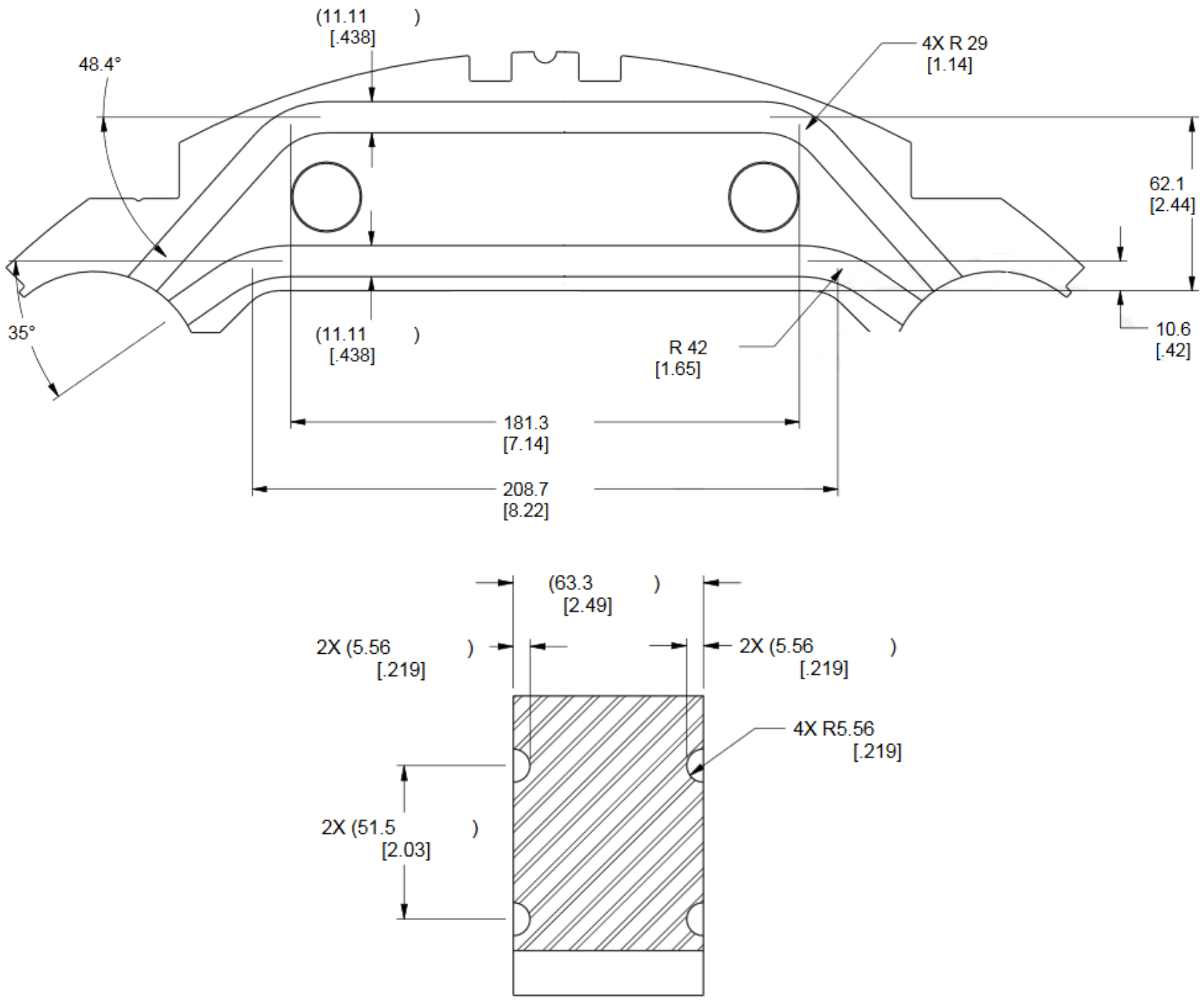

Fig. 3.13: Yoke cooling groove geometry.

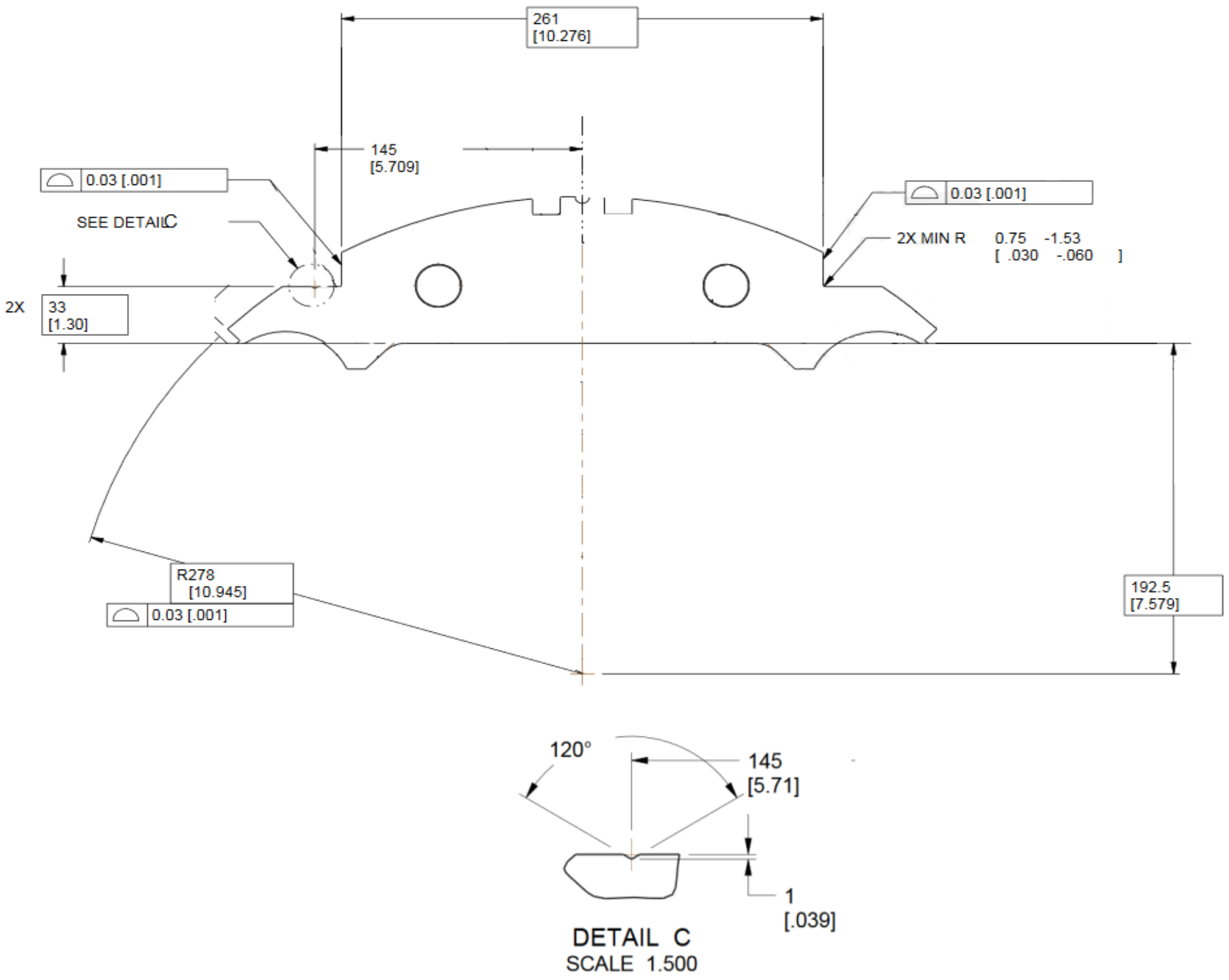

Fig 3.14: Yoke load bearing shoulders.

#### 3.2.1.1. Load Bearing Shoulder

The yokes load bearing shoulder (see Fig. 3.14) are notches in the 278 mm radial profile. The shoulder aligns with the end cutout geometry of the magnet shells. The shell end cutouts allow for cradle tooling to pass through the shell and rest directly on the yoke shoulders. The precision surface is required to ensure straightness when the weight of the 15,000 LB magnet is being supported on cradles.

### 3.2.2. Yoke Subassemblies

#### 3.2.2.1. Pre-Stacks

The pre-stacks (see Fig. 3.15) are an intermediary step in the yoke stacking process. The yokes are integrated into the magnet assembly in separate half-length stacks. These half-length assemblies are comprised of 46 yoke laminations stacked in series. It was found during the MQXFA prototype magnet efforts that assembling the individual yokes required a significant effort to measure and eliminate large variances in lengths of half stacks due to tolerance stack ups. Assembling in pre-stacks allows the supplier to mitigate the tolerance stacking by checking and correcting stack lengths at the pre-stack assembly level.

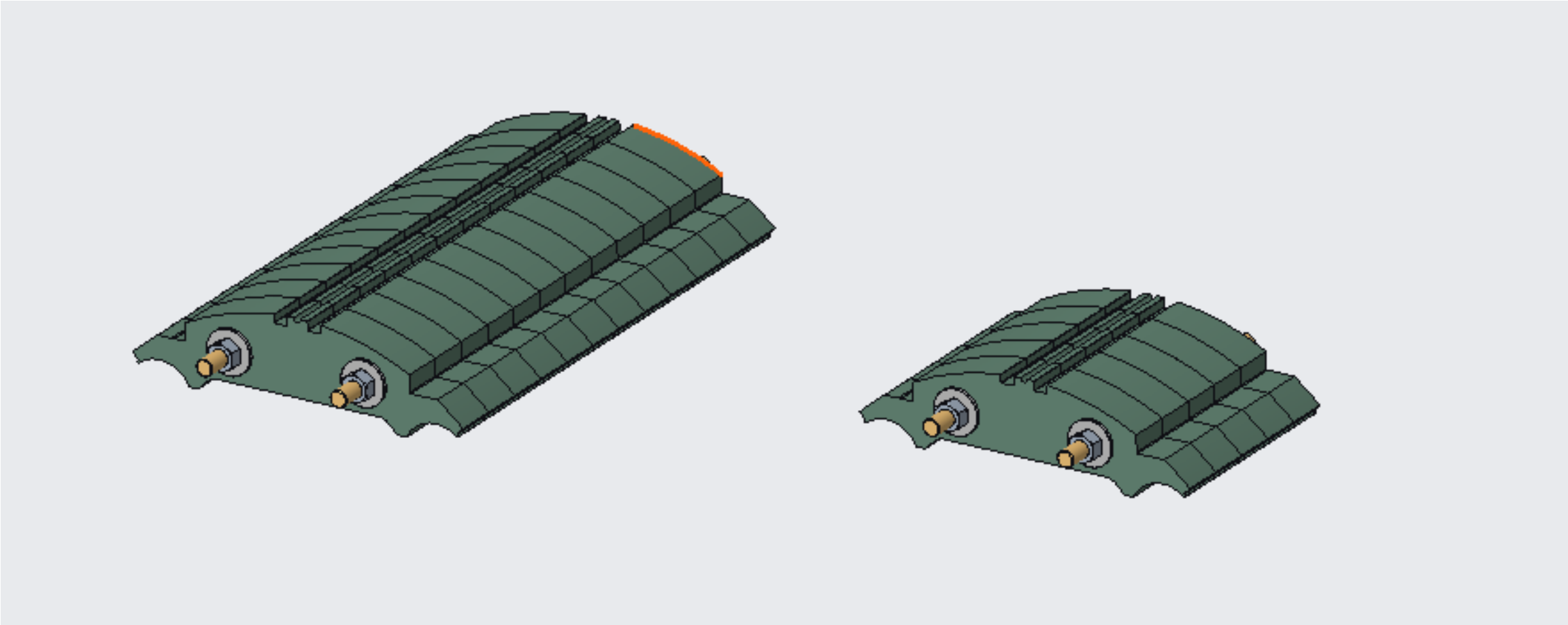

3.15: Yoke 588 mm and 245 mm pre-stacks.

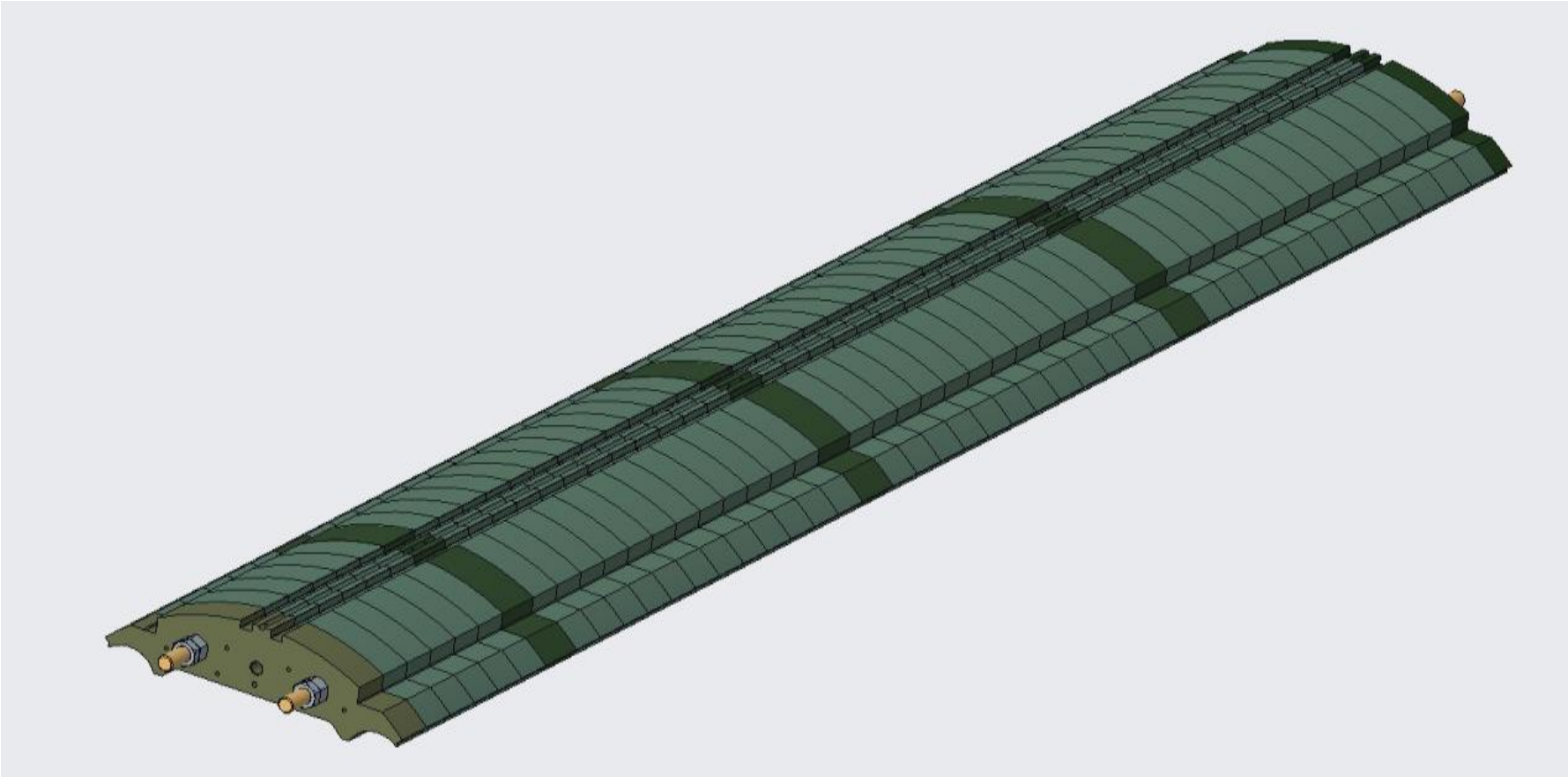

3.16: Example of a yoke half-stack.

For the yoke pre-stacks, the following specifications are set:

- ***588 mm Pre-stacks width shall be 588 mm +/- 0.13 with a Parallelism of 0.03 [.001"]***
- ***245 mm Pre-stacks width shall be 245 mm +/- 0.13 with a Parallelism of 0.03 [.001"]***

#### 3.2.2.2. Half-Stacks

Half-length yoke subassemblies are assembled using four pre-stack assemblies and five intermediate yoke laminations (see Fig 3.16). No additional machining or adjustments are to be performed prior to assembling, since the pre-stacks and laminations will have been machined/assembled within their respective tolerances.

For the yoke pre-stacks, the following specifications are set:

- ***Yoke half-stacks width shall be 2281.5 mm +/- 0.25 with a Parallelism of 0.05***

### 3.2.3. QC Plan

The quality control of specifications for the yokes are documented in the QC Plan for MQXFA Structures Parts Document [4].

## 3.3. Master Keys

The master keys are half-length (2.3 m long) flat components made of ARMCO Grade 4 Pure Iron (see Fig. 3.17). The master keys come in two varieties, with or without cutouts for the bladder junction blocks. The master keys are a critical component in coil pack integration assembly process.

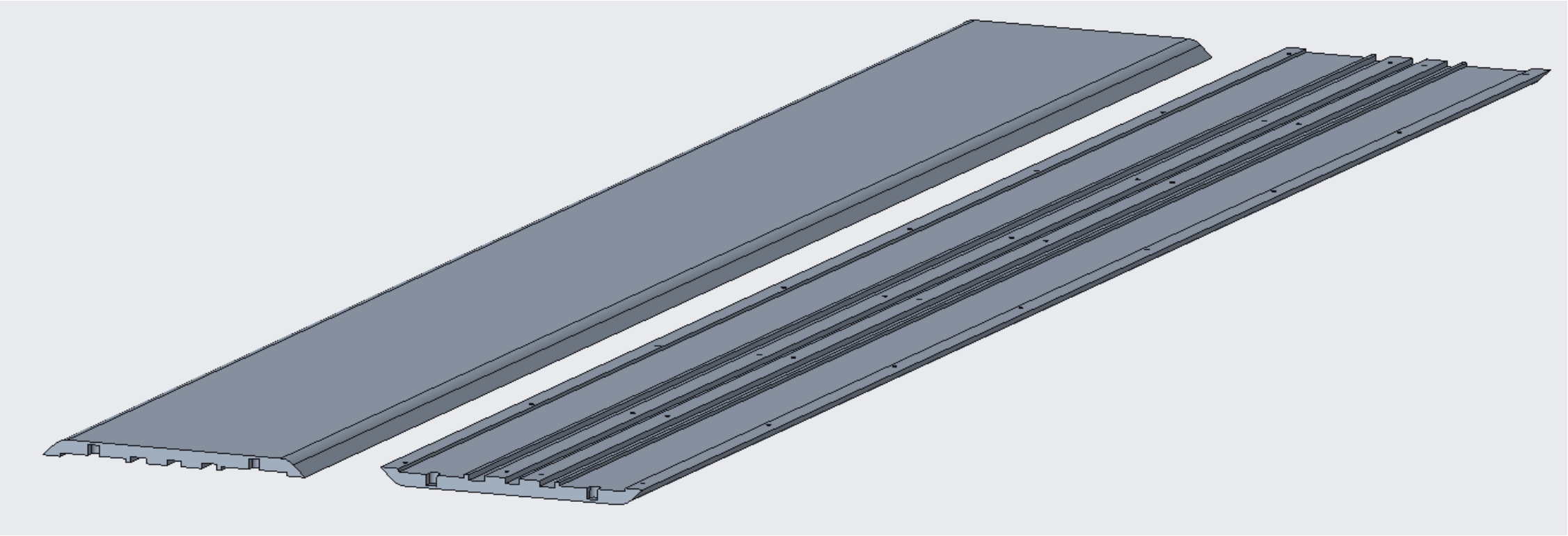

Fig 3.17: Master Key with bladder cutouts, top & bottom.

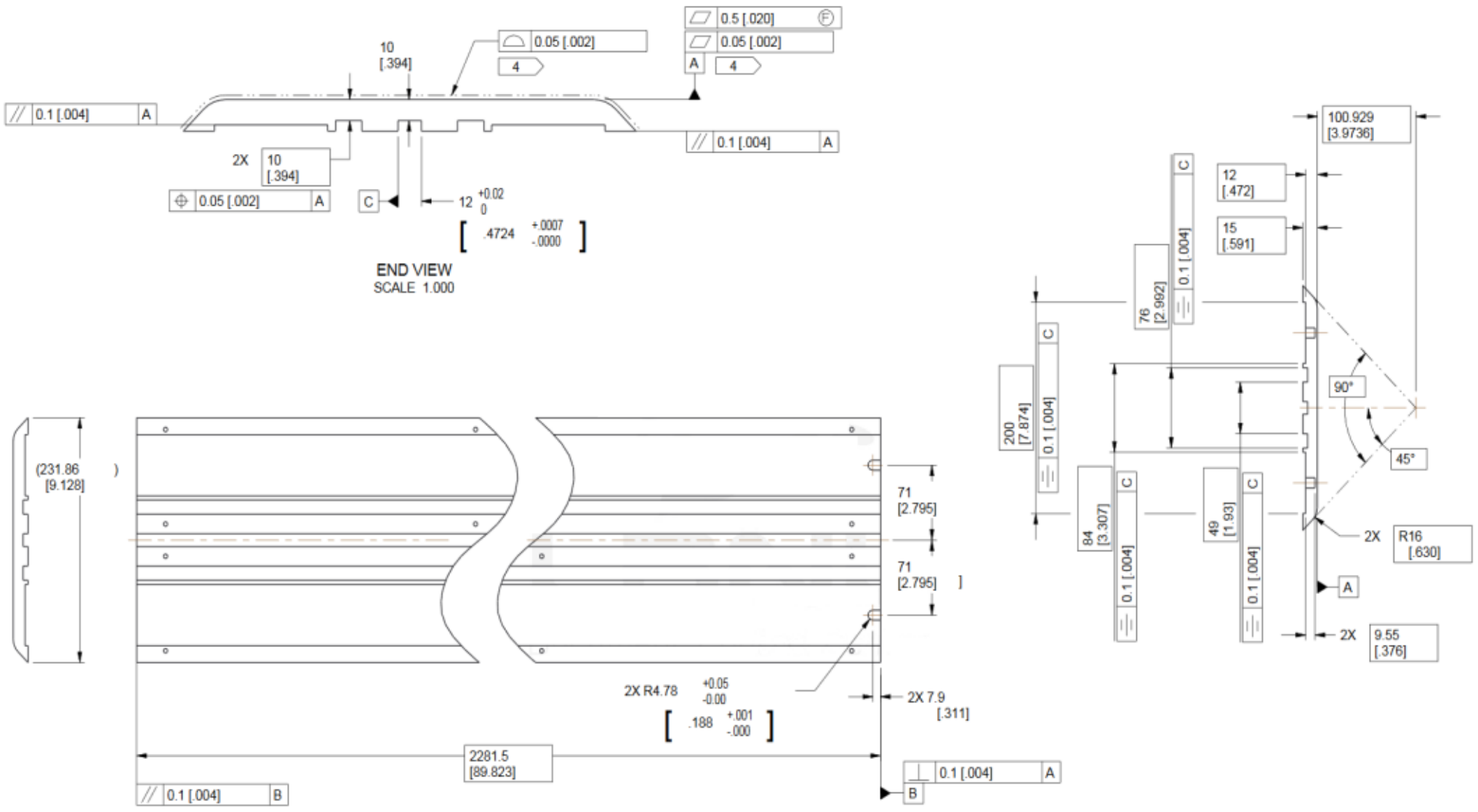

Fig 3.18: Master Key Geometry.

### 3.3.1. Master Key Features

The details of the master key design are shown in Fig. 3.18 and described in the following sections.

#### 3.3.1.1. Outer profile

The master key outer profile surface (see Fig. 3.19) is a high precision machined feature. The aperture profile is designed with a 0.05 mm tolerance. The precise nature of this surface is to allow for two, 2.3 m long master keys configured in a master package assembly to be slid between the yoke master key interface surfaces and the load pad master key interface surface of the coil pack assembly.

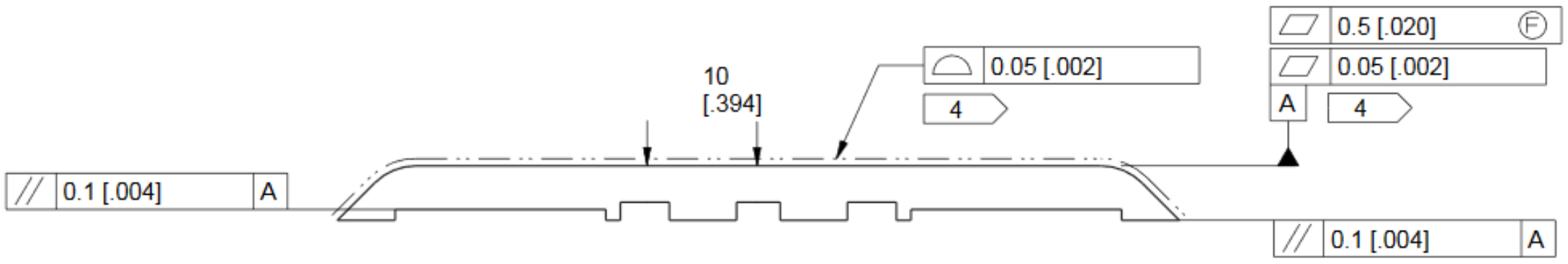

Fig 3.19: Master key outer surface profile.

<u>For the master key outer profile, the following specifications are set:</u>

- ***The master key outer profile scan shall maintain a .05 [.002"] tolerance.***

#### 3.3.1.2. Bladder Slots

The two 58 mm wide channels in the master keys (see Fig. 3.20) are slots for the high pressure bladders (and later for magnetic shims, if needed). When the master keys are mated in a master package assembly the high pressure bladders can be inserted between the master keys down the 58 mm wide slots.

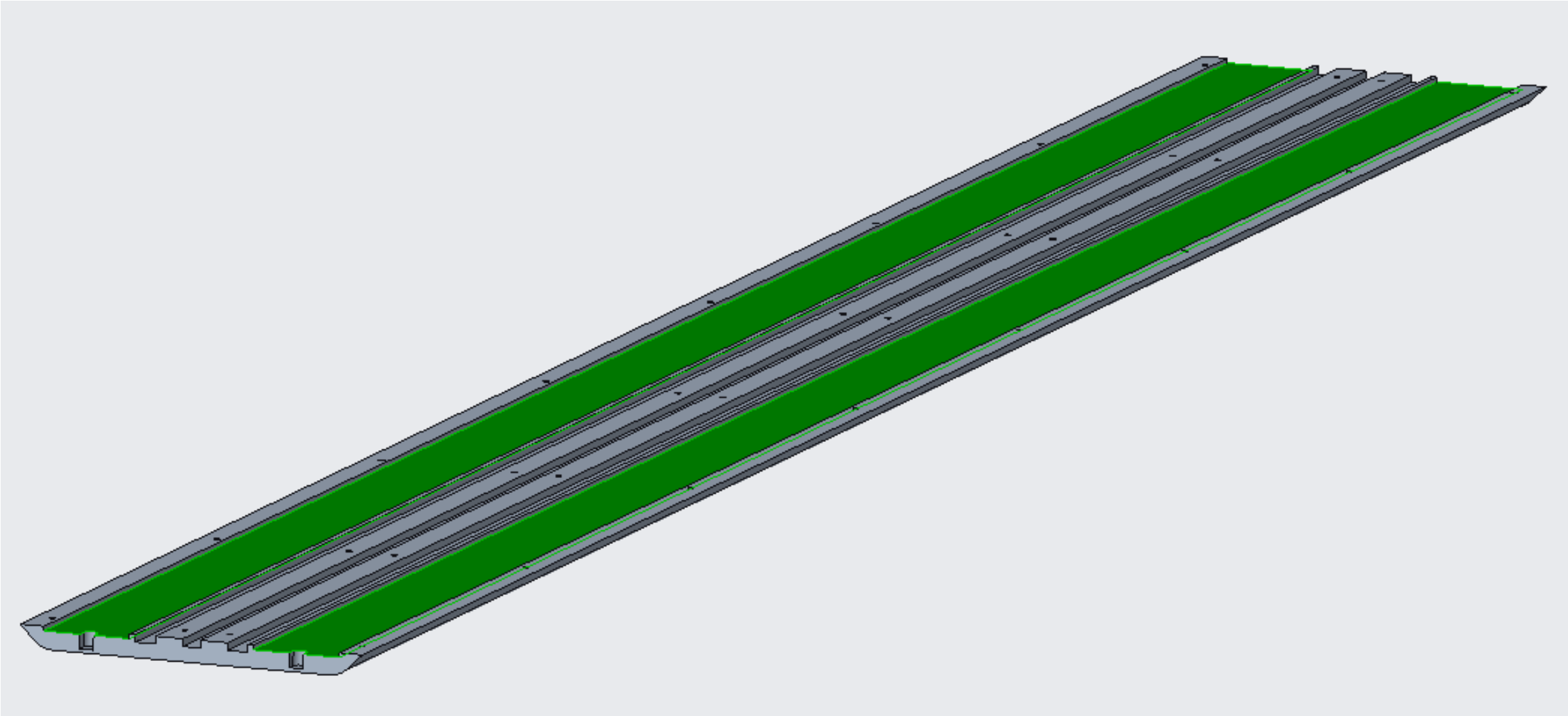

Fig 3.20: Master keys with the bladder slots highlighted.

#### 3.3.1.3. Alignment Key Groove

The single 12 mm wide by 5 mm deep channel in the middle of the master key is a slot for the Master key package alignment key (see Fig. 3.21). This is a precise groove that mates with the bronze alignment key when assembled in the package (see Fig 3.18 for alignment key groove geometry and tolerances). This key ensures the master keys are aligned for the length of the master key package subassembly.

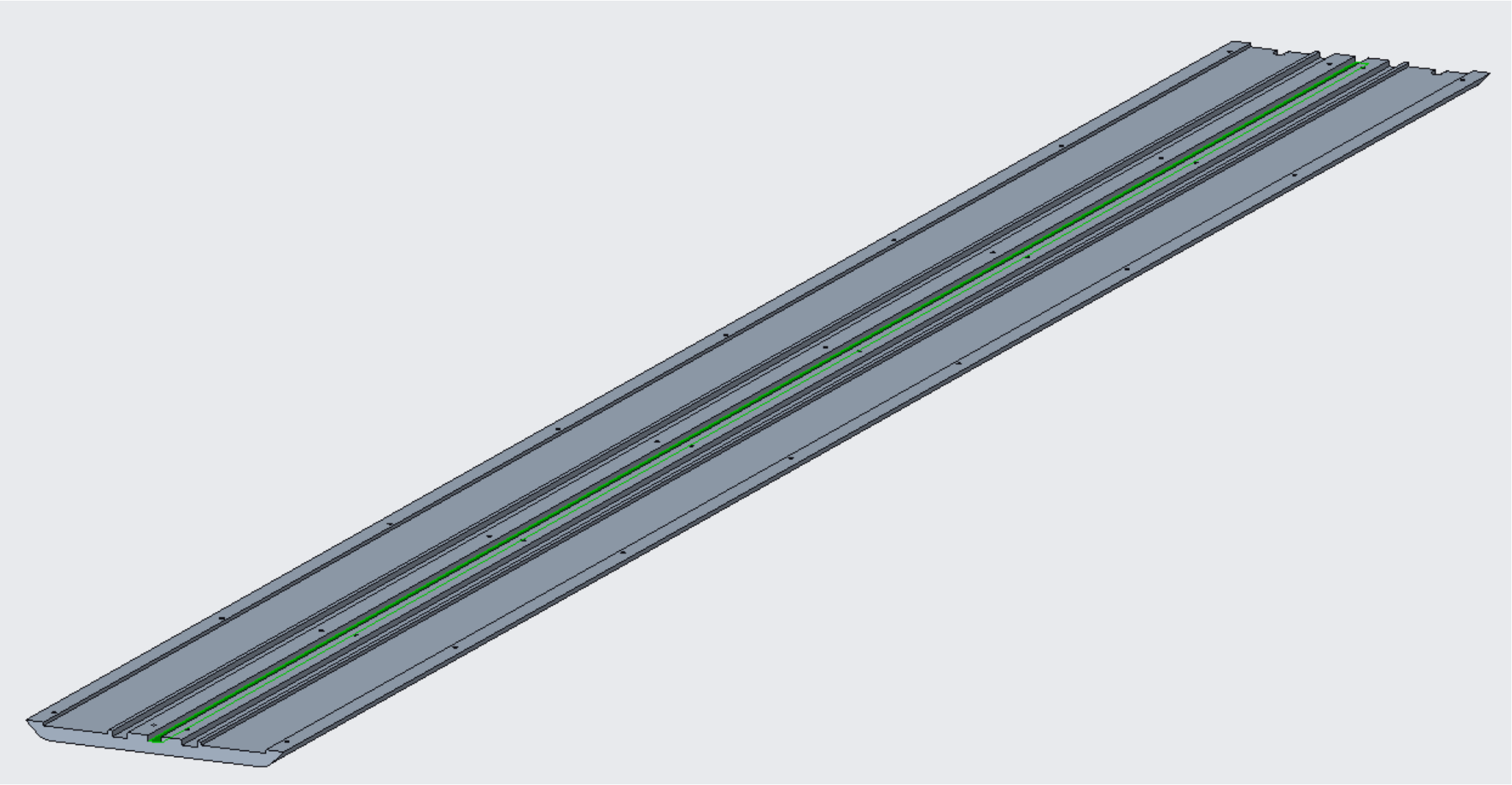

Fig 3.21: Master Key with the alignment key channel highlighted.

#### 3.3.1.4. Load Key Grooves

The two precise 13.5 mm wide by 5 mm deep master key channels to either side of the alignment channel (see Fig. 3.22) are the load key grooves (see Fig 3.18 for load key groove geometry and tolerances). The precision drawn shimmed load key sets are inserted into these grooves during the high pressure bladder operations and serve as interference keys when the bladder pressure is released. The load key/shim stack sets are what maintain the pre-load forces in the assembled magnet.

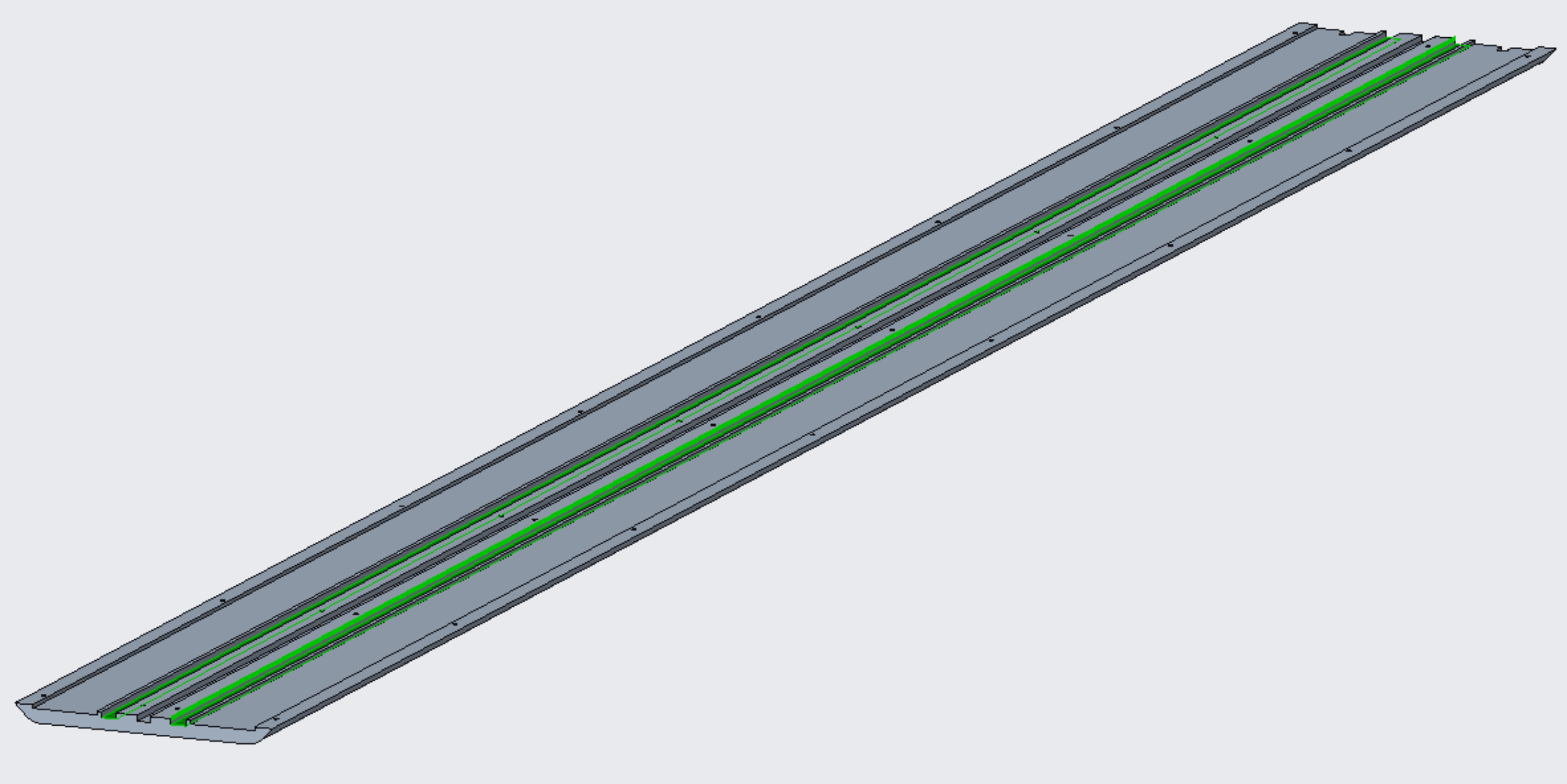

Fig 3.22: Master key with the load key channels highlighted.

### 3.3.2. QC Plan

The quality control of specifications for the master keys are documented in the QC Plan for MQXFA Structures Parts Document [4].

## 3.4. Load Pads

A 3D view of a load pad lamination is shown in Fig. 3.23.

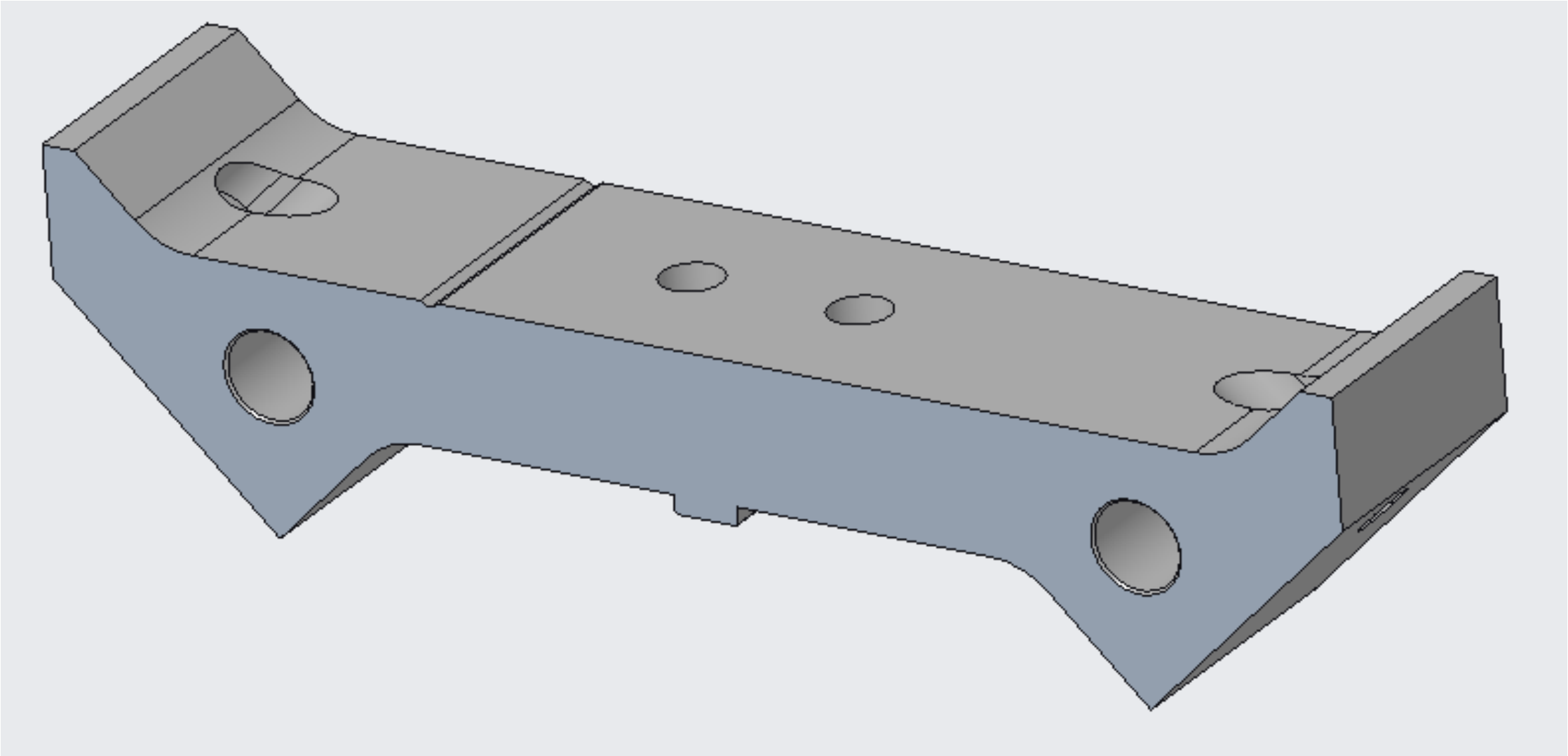

Fig. 3.23: Example of load pad type-3.

### 3.4.1. Load Pad Features

The details of the load pad design are described in the following sections.

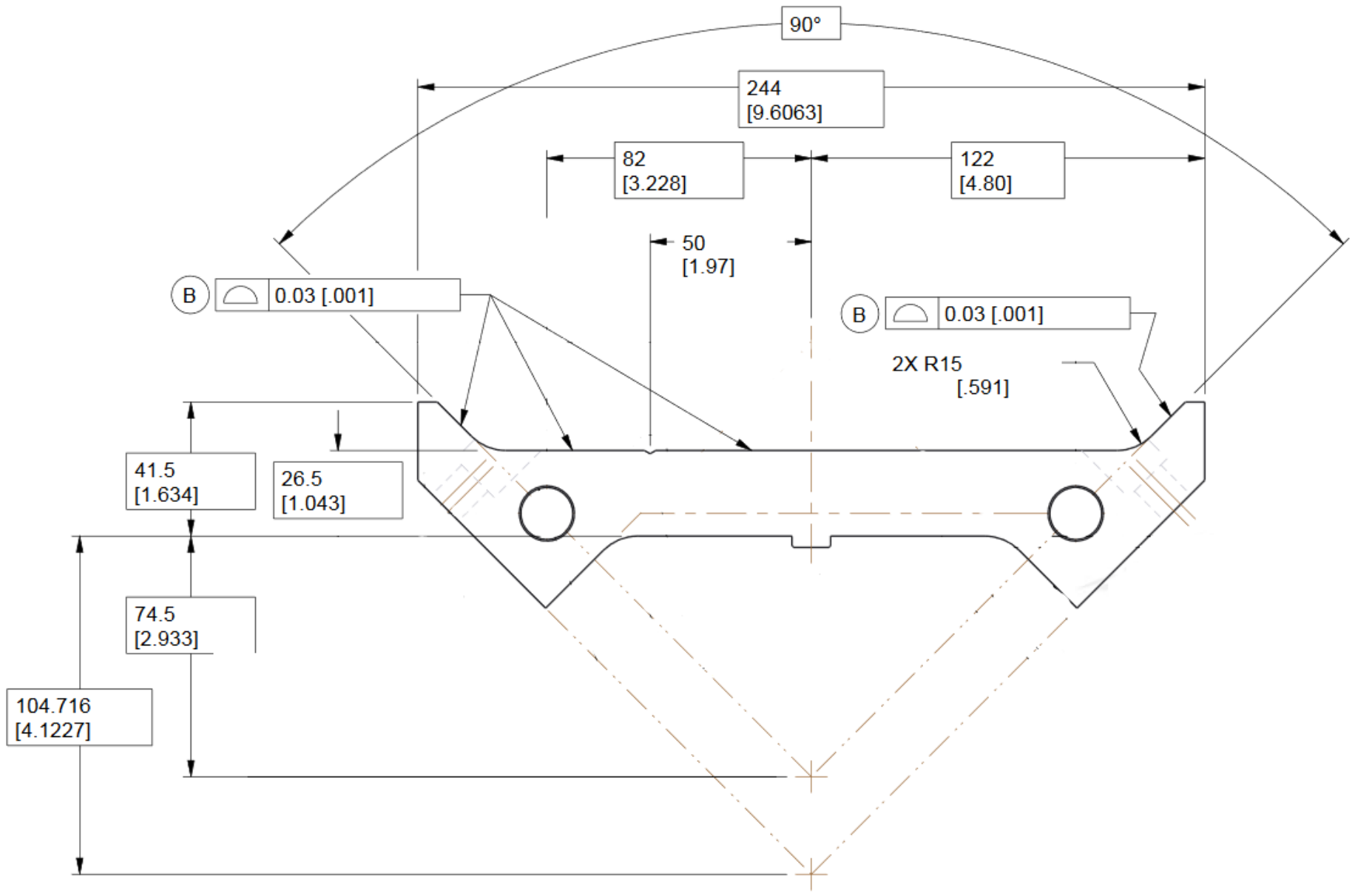

Fig 3.24: Load pad master key interface profile.

#### 3.4.1.1. Master Key interface

The load pad master key interface surface carries a profile tolerance of 0.03 mm (see Fig. 3.24). The features precision is critical. Due to the way in which the load pad are stacked, it is important for adjacent load pads to maintain a uniform smooth surface down the length of load pad stack assemblies. The smooth surface is critical to magnet assembly and applying uniform forces when the magnet is energized.

For the load pad master key interface profile, the following specifications are set:

- ***The Load Pad master key interface profile shall maintain a 0.03 [.001"] tolerance.***

#### 3.4.1.2. Collar Interface Profile

The load pad collar interface surface carries a profile tolerance of 0.03 mm (see Fig. 3.25). The features precision is critical. Due to the nature in the manner in which the load pad laminations are stacked, is important for adjacent load pads to maintain a uniform smooth surface down the length of load pad stack assemblies. The smooth surface is critical to magnet assembly and applying uniform forces when the magnet is energized.

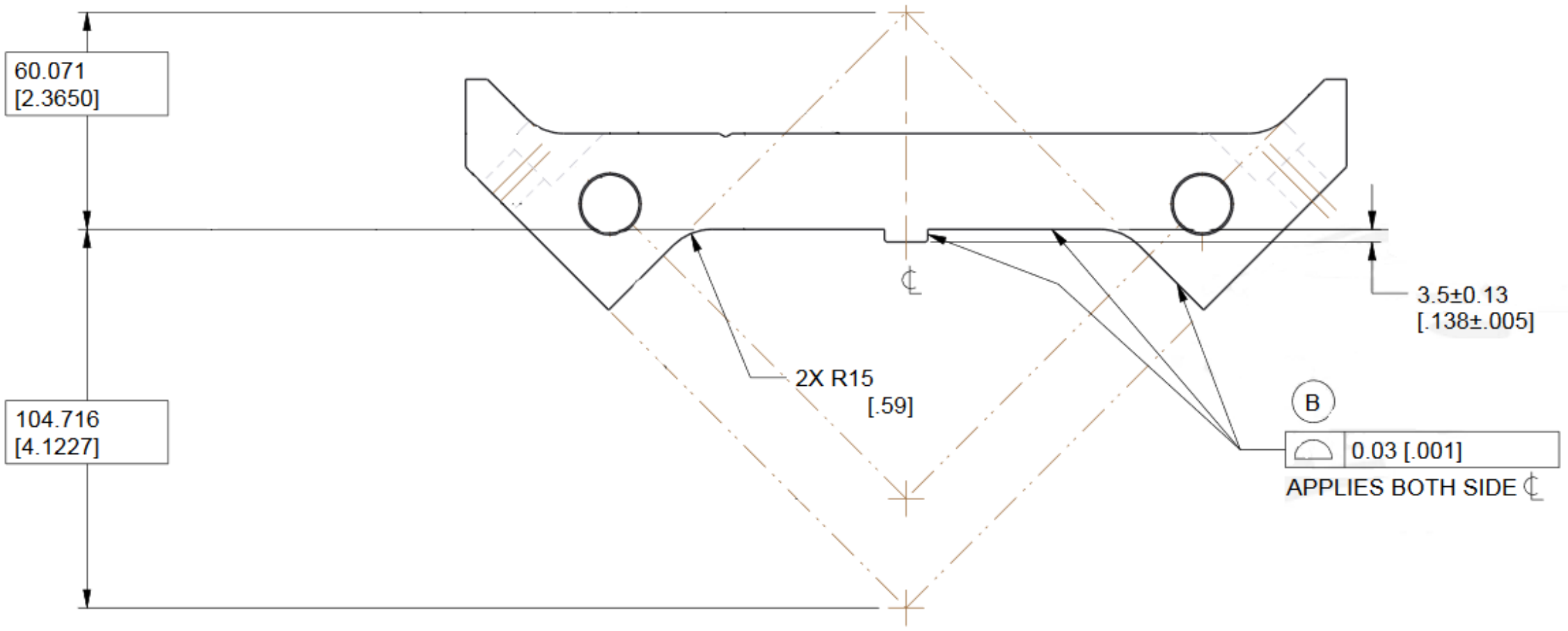

Fig 3.25: Load pad collar interface profile.

For the load pad collar key interface profile, the following specifications are set:

- ***The Load Pad collar interface profile shall maintain a 0.03 [.001"] tolerance.***

#### 3.4.1.3. Tie Rod/ Bushing Holes

The two 16.1 mm holes symmetric to the center line are for securing stacks in position and precise alignment by the use of close-fit bushings (see Fig. 3.26). The bushings lock the lamination stacks into one complete unit when exposed to forces perpendicular to the direction of the holes such as stack lifting operations.

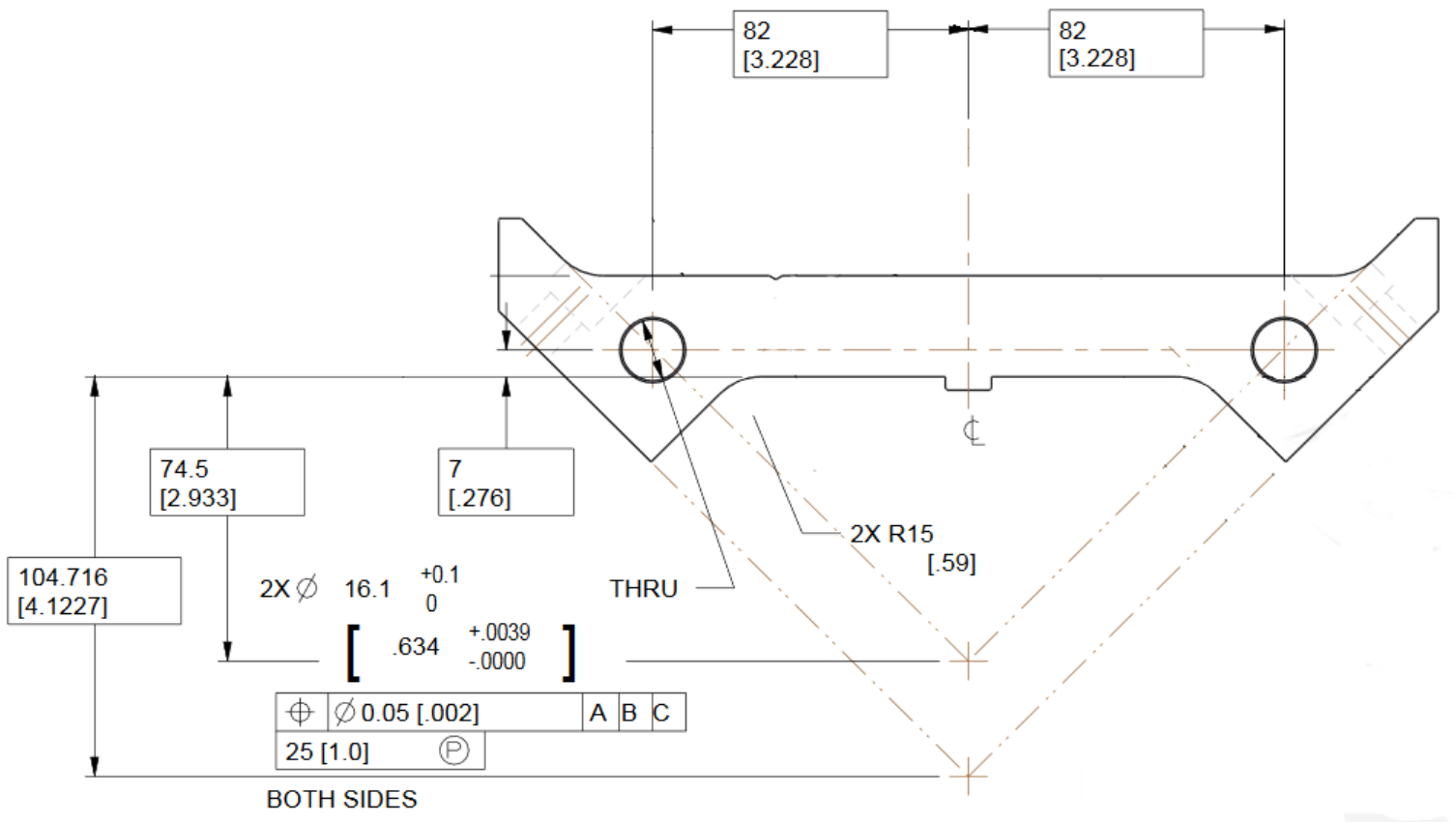

Fig 3.26: Load pad 16.1mm tie rod/bushing holes.

### 3.4.2. Pre-Stacks

The load pad pre-stacks are an intermediary step in the load pad stacking process. The load pads are integrated into full-length assembly (4.563 m long) stacks (see Fig. 3.27). The full-length stacks are comprised of 92 load pad laminations stacked in series. In Fig. 3.27, the pre-stacks are shown: the return end pre-stack 1088 mm long (left), the lead end pre-stack 1050 mm long (middle), and the center pre-stack 1100 long mm (right). The stainless-steel section of the stack was included in the design to reduce the peak magnetic field in the coil ends. Assembling the pre-stacks without additional machining and matching is possible because the suppliers have already managed the tolerance build up by measuring and correcting stack lengths at the pre-stack assembly level.

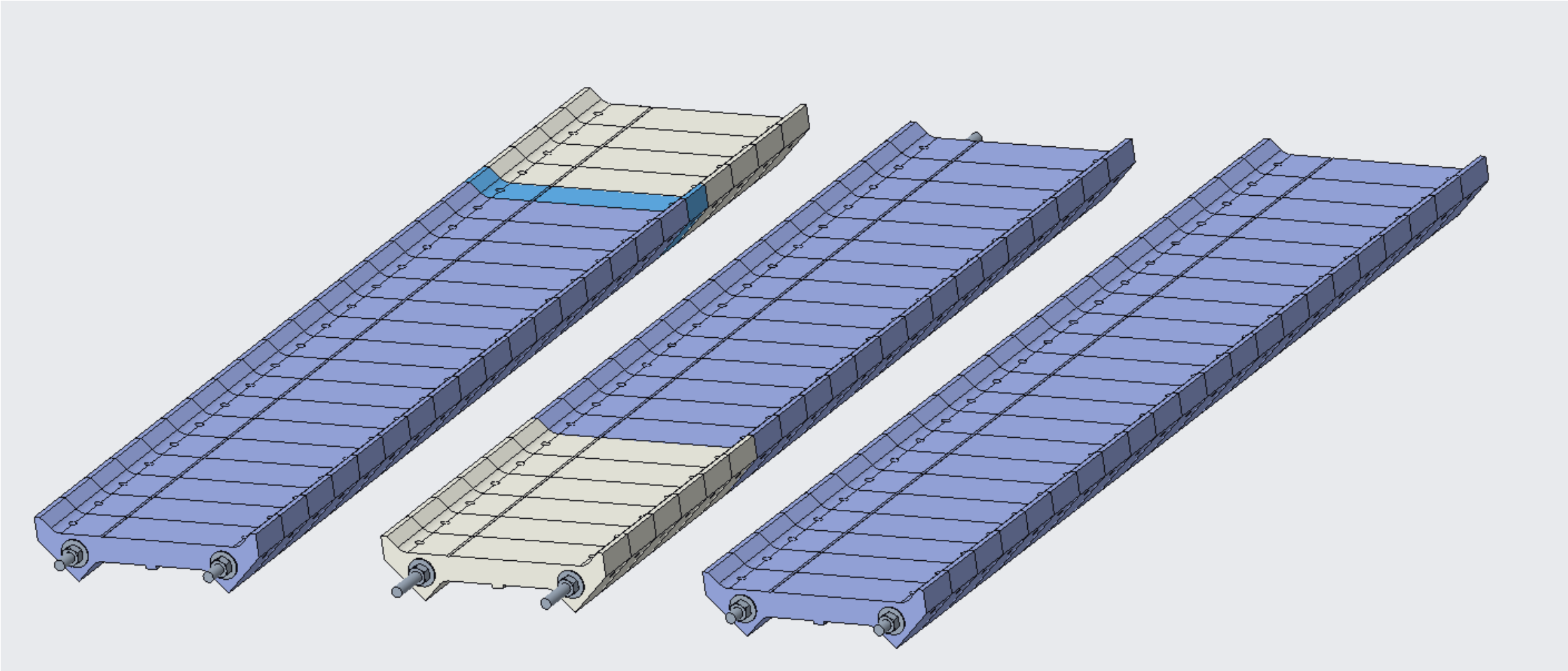

Fig. 3.27: Left: 1088 mm pre-stack (RE), Middle: 1050 mm pre-stack (LE), Right: 1100 mm pre-stack (center).

<u>For the stainless-steel section of the pad pre-stack, the following specifications are set:</u>

- ***On the LE pre-stack, the LE pads shall have a stainless-steel section of 300 ± 0.6 mm and an overall pre-stack length of 1050 ±0.13 mm***
- ***On the RE pre-stack, the RE pads shall have a stainless-steel section of 200 ± 0.4 mm and an overall pre-stack length of 1088 ±0.13 mm***
- ***The 1100 pre-stack shall have an overall length of 1100 ±0.13 mm***

### 3.4.3. QC Plan

The quality control of specifications for the load pads are documented in the QC Plan for MQXFA Structures Parts Document [4].

## 3.5. Collars

A 3D view of a load pad lamination is shown in Fig. 3.28.

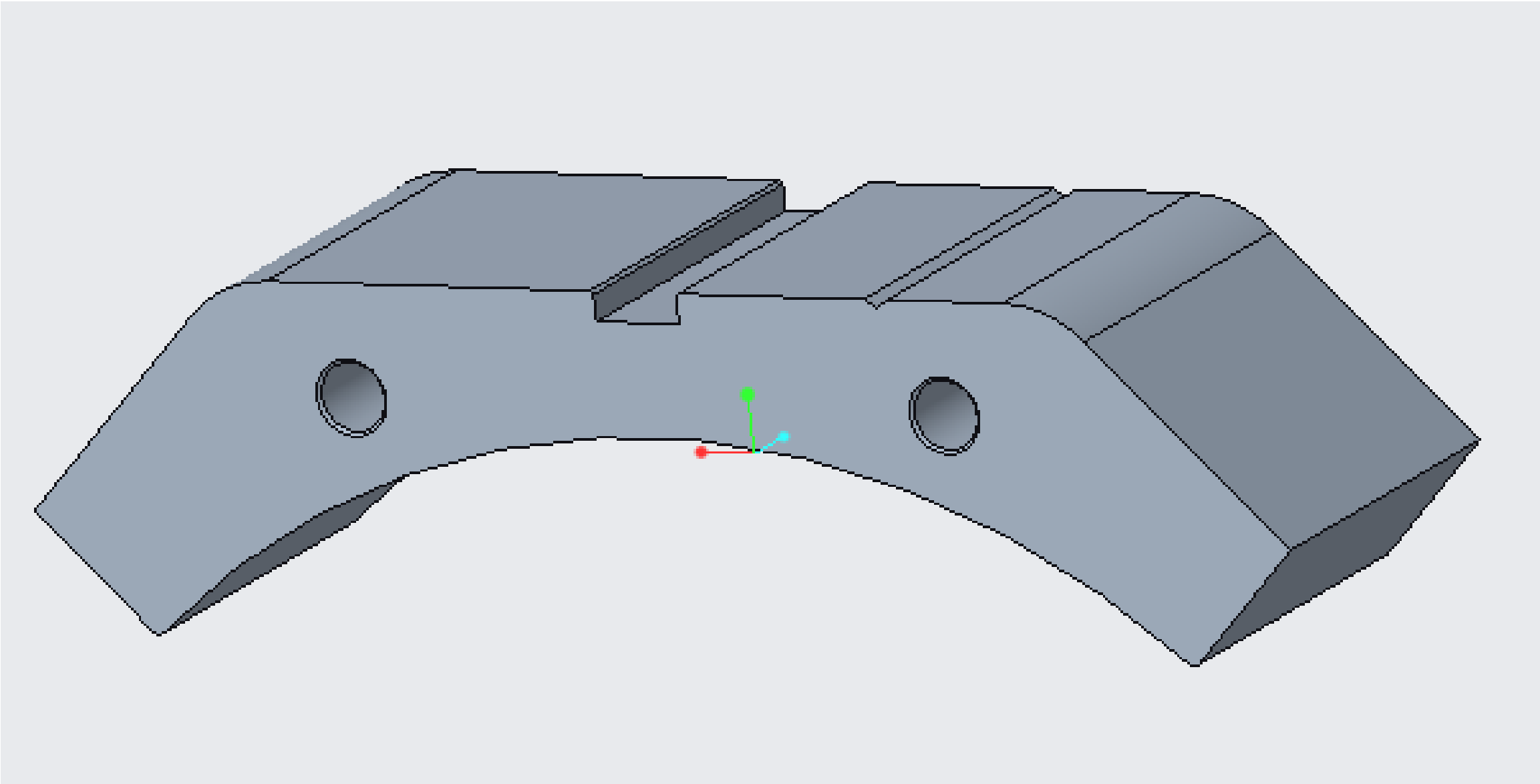

Fig. 3.28: Example of collar type-3.

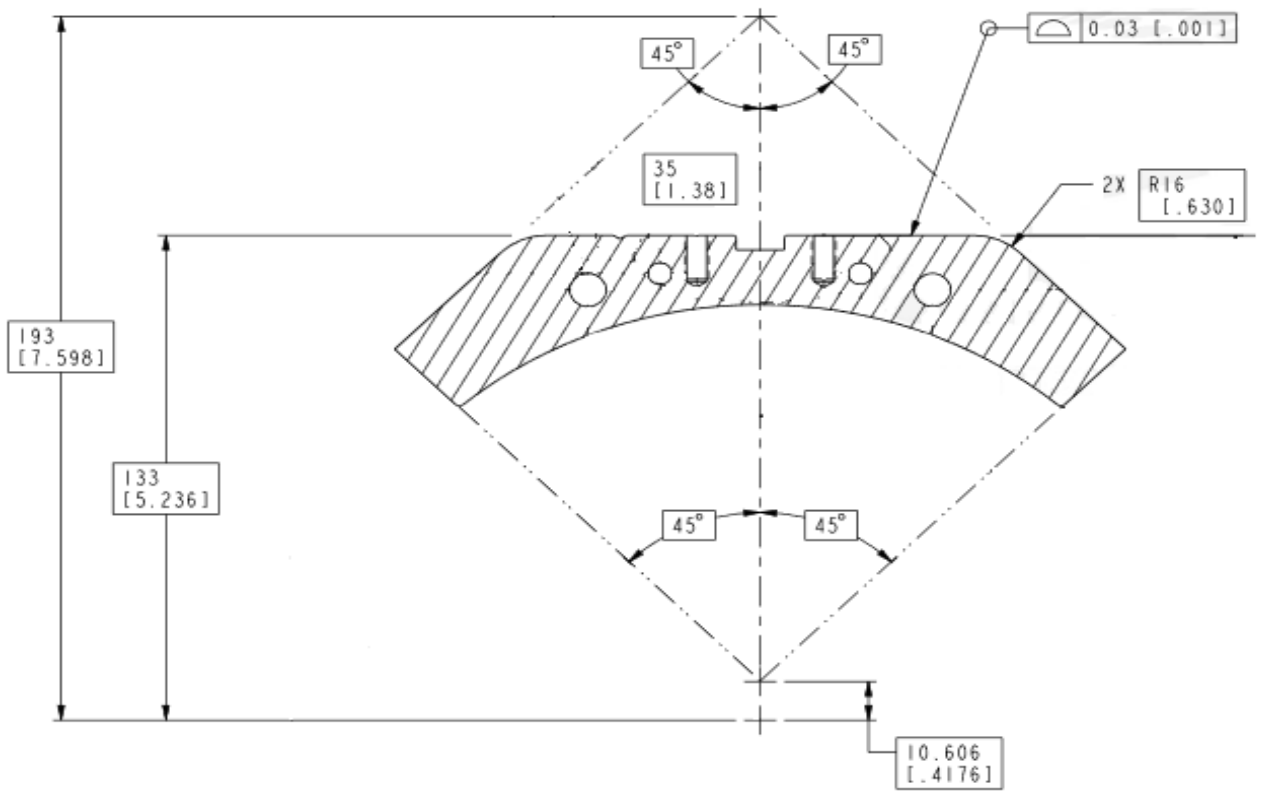

Fig 3.29: Collar load pad interface surface profile.

### 3.5.1. Features of Collars

The details of the load pad design are described in the following sections.

#### 3.5.1.1. Load Pad Interface Profile

The collar load pad interface surface carries a profile tolerance of 0.03 mm (see Fig. 3.29). The features precision is critical. Due to the nature of the interface between the load pad and collars in the assembly, it is important for adjacent collars to maintain a uniform smooth surface down the length of collar stack assemblies. The smooth surface is critical to magnet assembly and applying uniform forces when the magnet is energized.

For the collar to load pad interface profile, the following specifications are set:

- ***The entire collar to load pad interface profile shall maintain a 0.03 [.001"] tolerance all around.***

#### 3.5.1.2. Coil Interfaces

The collar coil interface surface carries a profile tolerance of 0.03 mm (see Fig. 3.30). These surface features' precision is critical. The individual collars are assembled directly on the load pad stack to form the pad-collar subassembly, is important for adjacent collars to maintain a uniform smooth surface down the length of collar stack assemblies. The smooth surface is critical to magnet assembly and applying uniform forces when the magnet is energized.

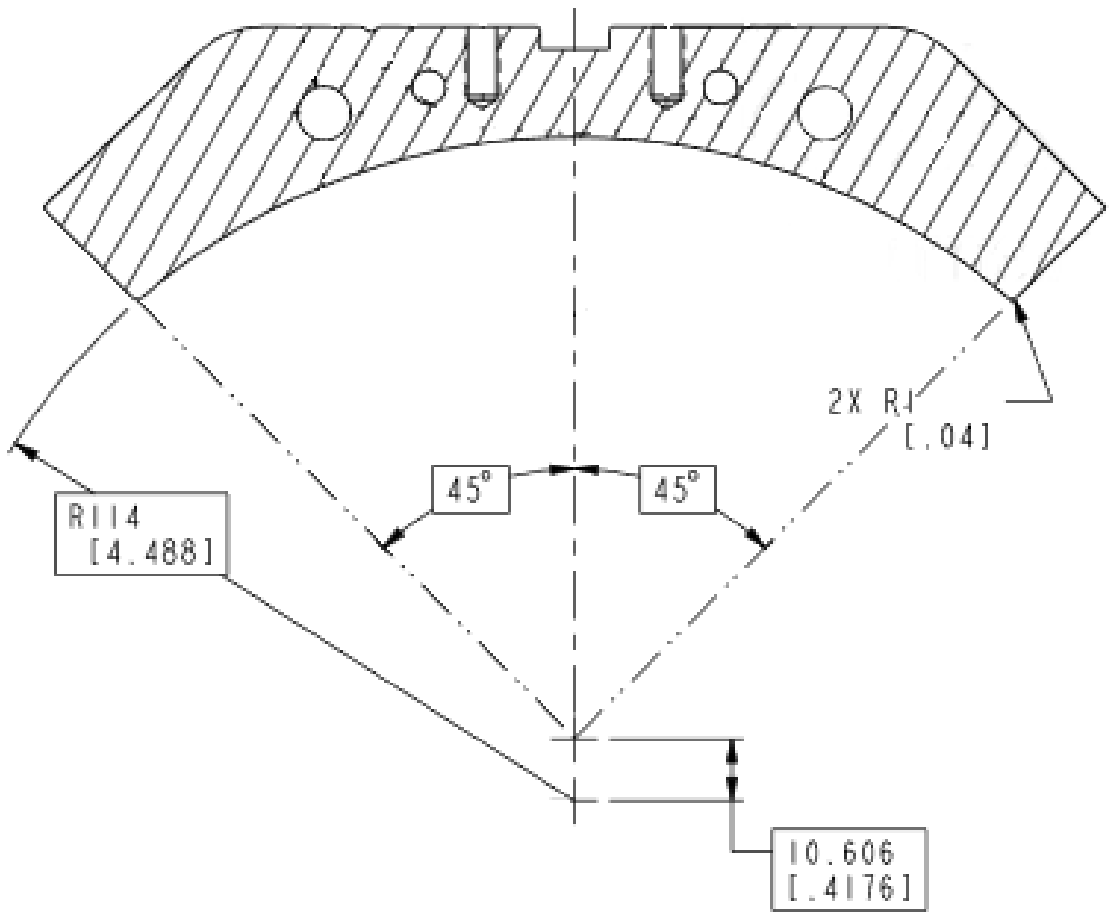

Fig. 3.30: Collar coil interface surface profile.

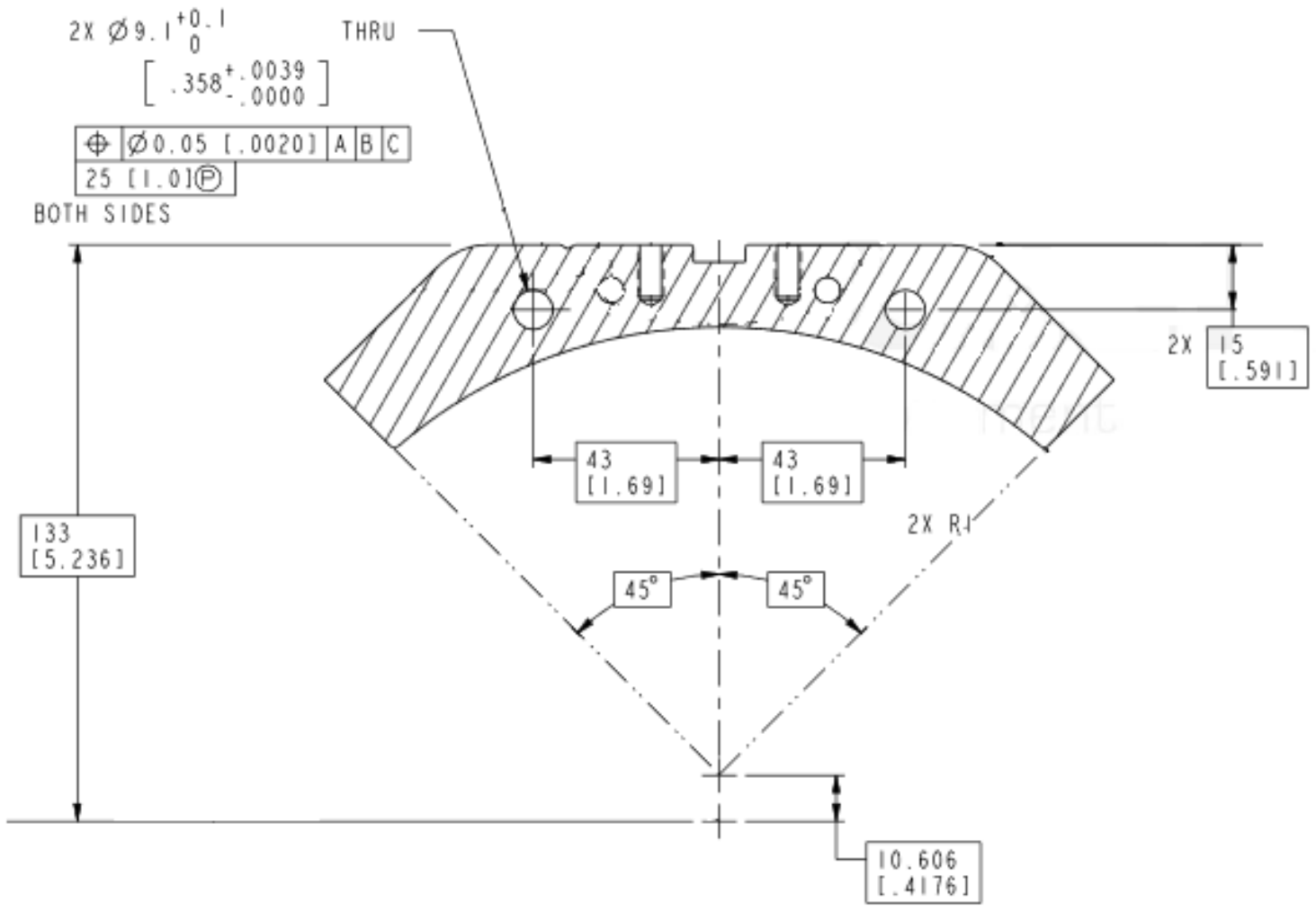

Fig. 3.31: Collar bushing/tie rod holes.

For the collar to coil interface profile, the following specifications are set:

- ***The collar to coil interface profile shall maintain a 0.03 [.001"] tolerance all around.***

#### 3.5.1.3. Tie Rod/ Bushing Holes

The two 9.1 mm holes symmetric to the center line are for securing stacks in position and precise alignment by the use of close-fit bushings (se Fig. 3.31). The bushings lock the lamination stacks into one complete unit when exposed to forces perpendicular to the direction of the holes such as stack lifting operations.

### 3.5.2. QC Plan

The quality control of specifications for the collars are documented in the QC Plan for MQXFA Structures Parts Document [4].

# 4. Magnet Sub-assembly Specifications

The design of the MQXF magnet, is based on two main sub-assemblies: the shell-yoke and the coil-pack. In the next section we will briefly describe the assembly process for both sub-assemblies, the measurements performed and the specifications defined.

## 4.1. Shell-Yoke sub-assembly

The assembly of the shell-yoke sub-assembly is performed by combining two half-length shell-yoke modules: the lead end (LE) module and the return end (RE) module. During the mounting of each module, four shells (three long shells, one short shell) are initially stacked vertically on the assembly stand (see Fig. 4.1, left). Pins are inserted only in the "Top" quadrant slots to align the shells prior to the insertion of the half-length yoke stacks. Four yoke stacks are then inserted vertically inside the stack of shells (see Fig 4.1, right). A bladder operation is performed to insert the gap keys and constrain the yokes against the shells. The cooling holes in the yokes are utilized as the bladder locations to open up the gaps, a solution chosen since it was considered more efficient than the operating the bladders at the internal cross support geometry.

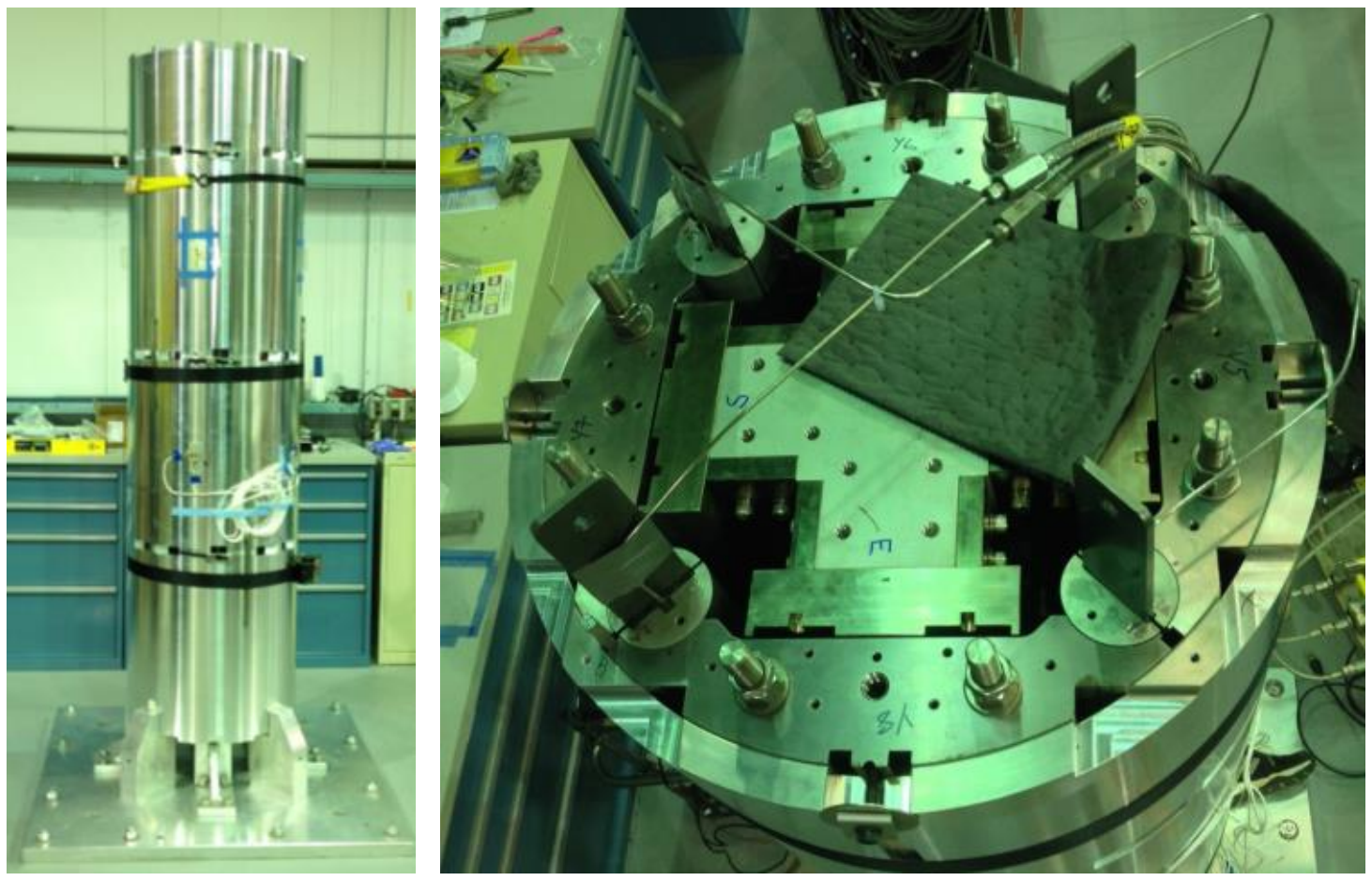

Fig. 4.1. (L) Half-length shell stack on the assembly stand. (R) View from the top, with the yoke quadrants ready to be preloaded.

The nominal yoke gap, i.e. the gap between the yokes close to the shells where the yoke keys are inserted, is 12 mm (see Fig. 4.2 for the yoke key location). However, the gap keys are shimmed to 12.1 mm total. The resulting interference produce azimuthal tension in the shells, which is measured by stain gauge mounted on the outer surface of the shells. Fig. 4.3 show the strain gauge axial locations. Two shells are instrumented in the LE module (shell 2 and 4), while only one shell is instrumented in the RE module (shell 7). In each axial location, both the axial and azimuthal shell strain are measured, in all four quadrants and in the azimuthal position as indicated in Fig. 2.1.

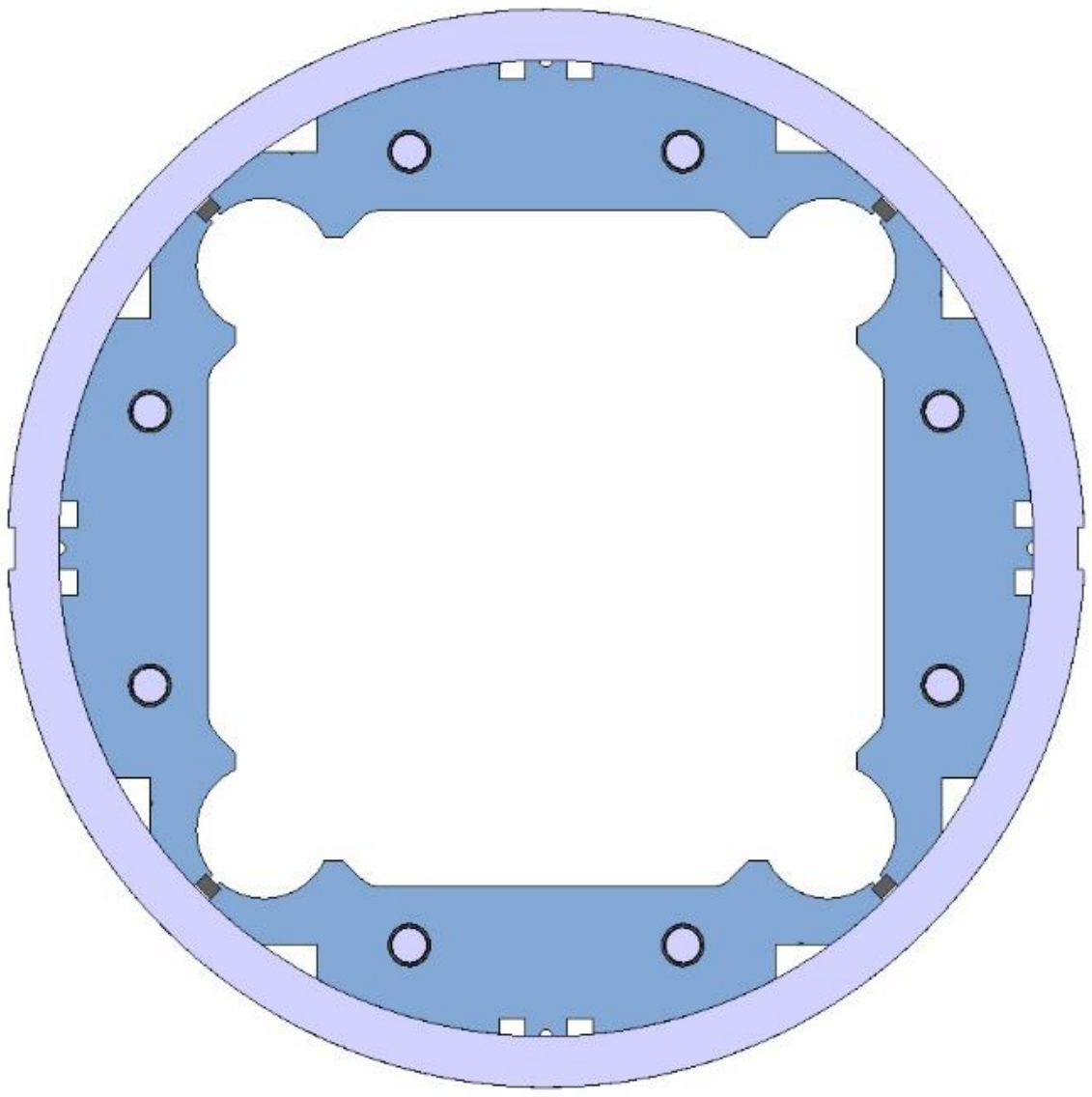

Fig. 4.2. Cross-section of the shell-yoke sub-assembly, with yoke keys inserted in between the yokes.

As can be seen in Fig. 4.4 and Fig. 4.5, where the shell azimuthal and axial micro-strain measured after assembly of the shell-yoke sub-assembly modules are plotted for shells 2,4 and 7 in magnets MQXFA03, A04, A05, A06 (the square solid markers represent the average among the four quadrants, with ±1 σ error bars, the tringle and round markers represent the maximum and minimum values among the four quadrants), the 0.100 mm interference of the yoke key produces a shell strain of about +150 micro-strain in the azimuthal direction, and of about -50 micro-strain in the axial direction.

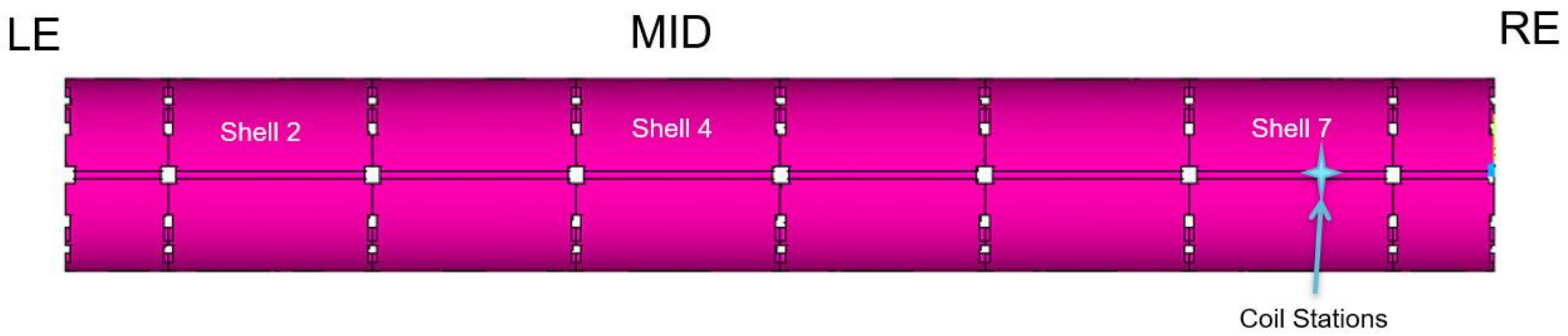

Fig. 4.3. Axial locations of the shell and coil strain gauges.

<u>For the shell azimuthal and axial strain after shell-yoke sub-assembly, the following specifications are set:</u>

- ***The shell strain in the azimuthal direction after shell-yoke sub-assembly*** *shall be within* ***+50/+250 micro-strain***
- ***The shell strain in the axial direction after shell-yoke sub-assembly*** *shall be within* ***-150/+50 micro-strain***

After the vertical loading operation is carried out for both the half-length shell-yoke sub-assemblies, the dimensions of the yoke cavity vertical and horizontal sizes are taken. As can be seen in Fig. 4.6, two horizontal and two vertical measurements are taken on both side of the extremities of the two modules (for a total of 16 measurements). It is important to notice that the vertical bladder operation and the insertion of the yoke keys with 0.100 mm interference determine an increase of the yoke cavity with respect to the nominal values.

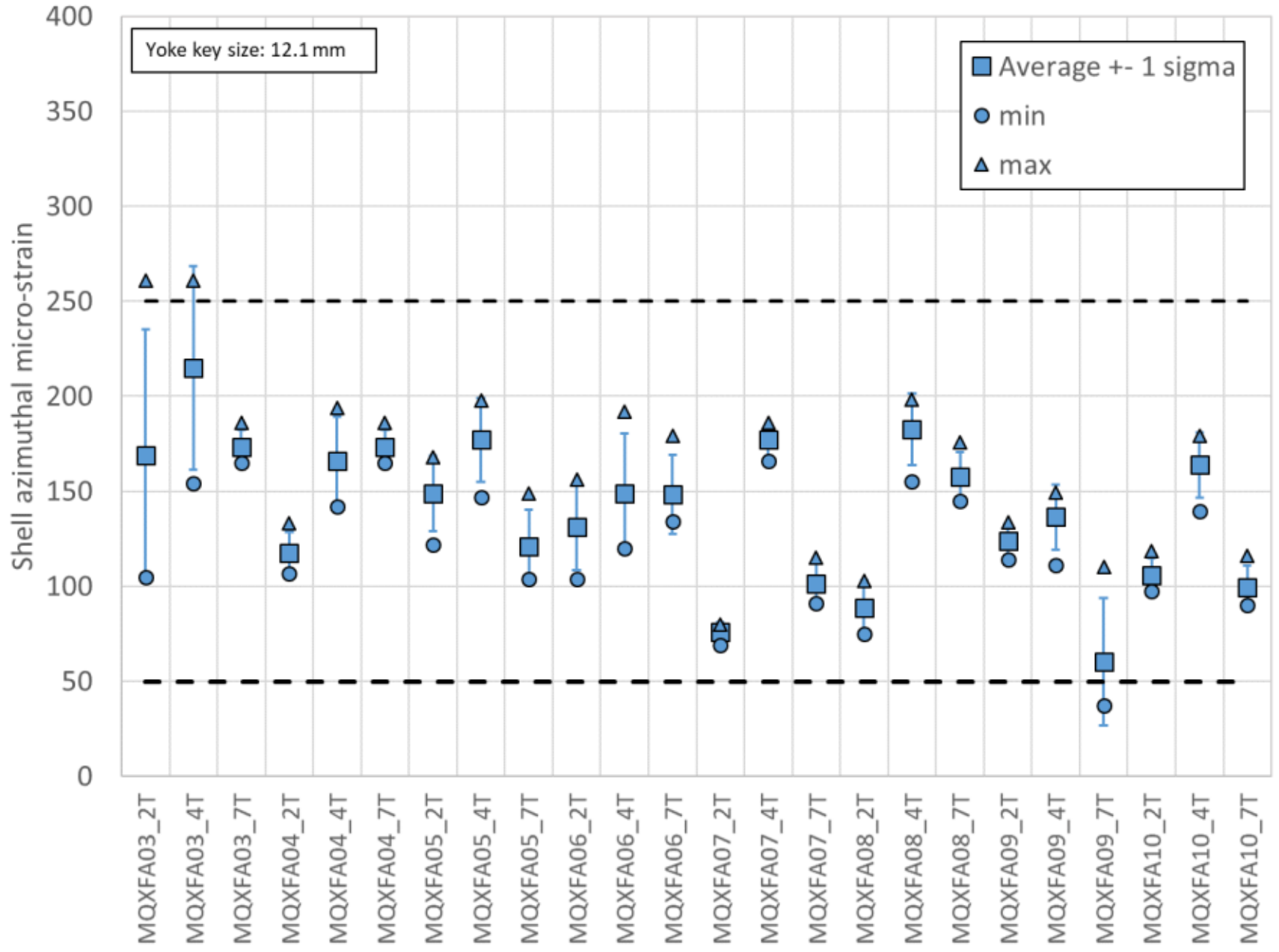

Fig. 4.4. Shell azimuthal micro-strain measured after assembly of shell-yoke sub-assembly, for shells 2,4 and 7 in magnets A03 to A10; the square solid markers represent the average among the four quadrants, with ±1 σ error bars; the triangle and round markers represent the maximum and minimum values among the four quadrants.

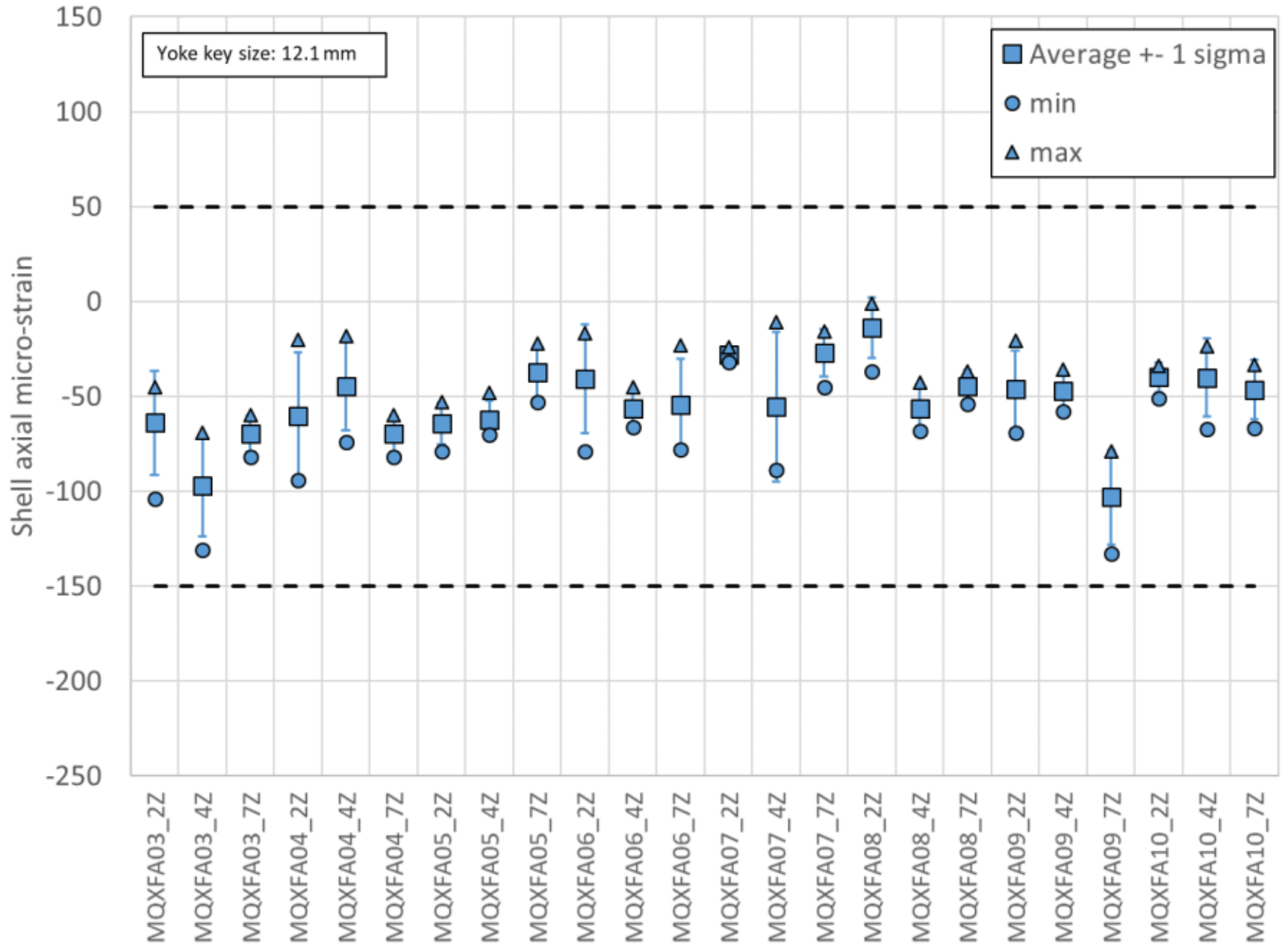

Fig. 4.5. Shell axial micro-strain measured after assembly of shell-yoke sub-assembly, for shells 2,4 and 7 in magnets A03 to A10; the square solid markers represent the average among the four quadrants, with ±1 σ error bars; the triangle and round markers represent the maximum and minimum values among the four quadrants.

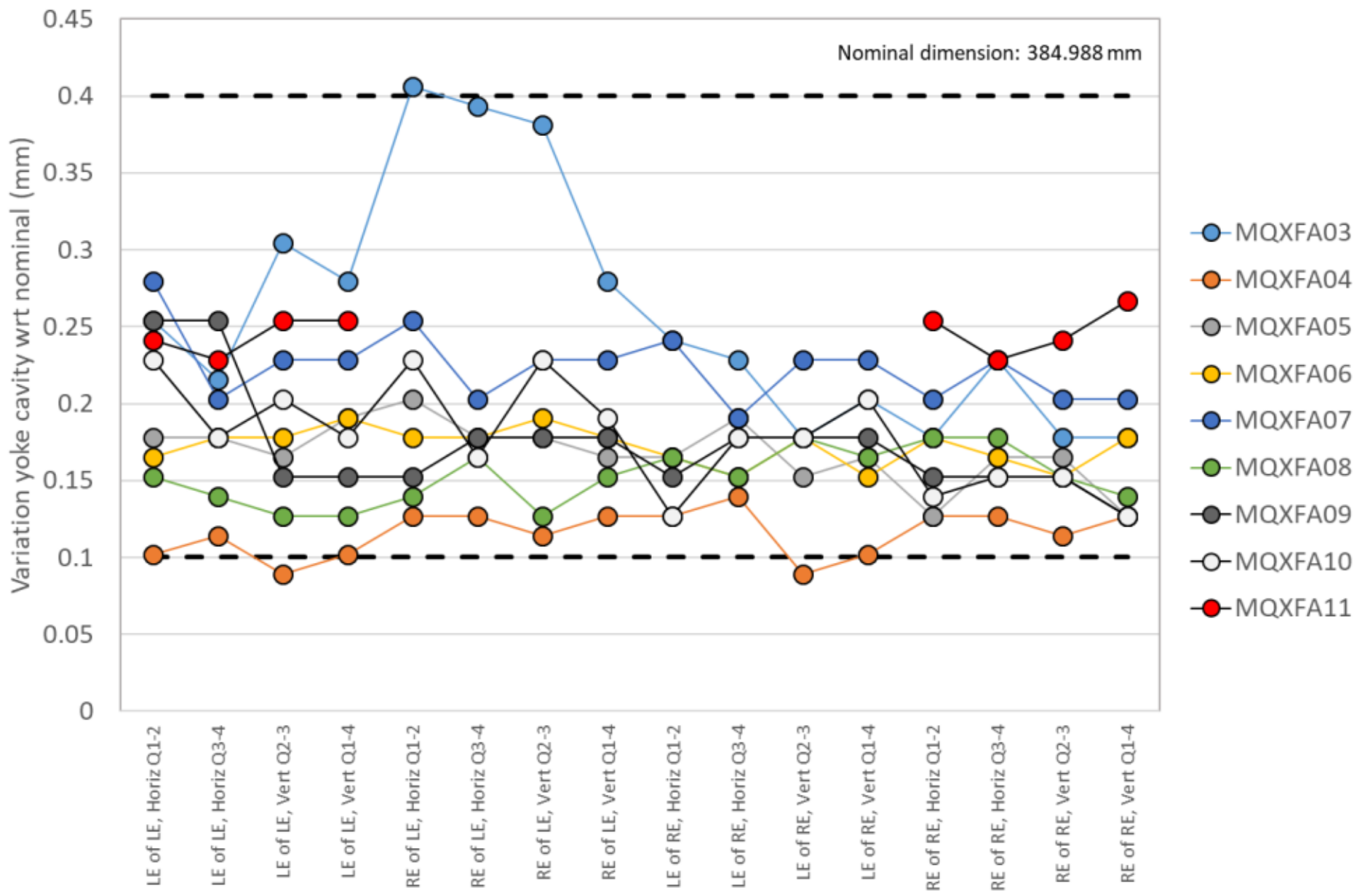

Fig. 4.6. Measurements of variation of yoke cavity dimension with respect to nominal: two horizontal and two vertical measurements taken on both side of the two sub-assemblies in magnets A03 to A11.

<u>For the yoke cavity size after shell-yoke sub-assembly, the following specifications are set:</u>

- ***The variation of yoke cavity dimension with respect to nominal shall be within +0.100/+0.400 mm***

When both half-length sub-assembly modules are joined together, the short yoke tension rods are replaced with full-length ones. A hydraulic rig is used to pre-tension these full-length yoke tension rods to carry 9000 lbs. each (see Fig. 4.7). Each of the four pistons (RSM 200 cylinder with effective area of 4.43 in$^2$), placed in between two tie rods on each quadrant, are pressurized to 4100 PSI, corresponding to 18000 lbs. Then, the nuts are turned to contact and the pressure released.

<u>For the pre-tension of the full-length tie rods of the yoke, the following specifications are set:</u>

- ***The piston (RSM 200 cylinder with effective area of 4.43 in$^2$) shall be pressurized to 4100 ± 200 PSI and the nuts shall be turned to contact.***

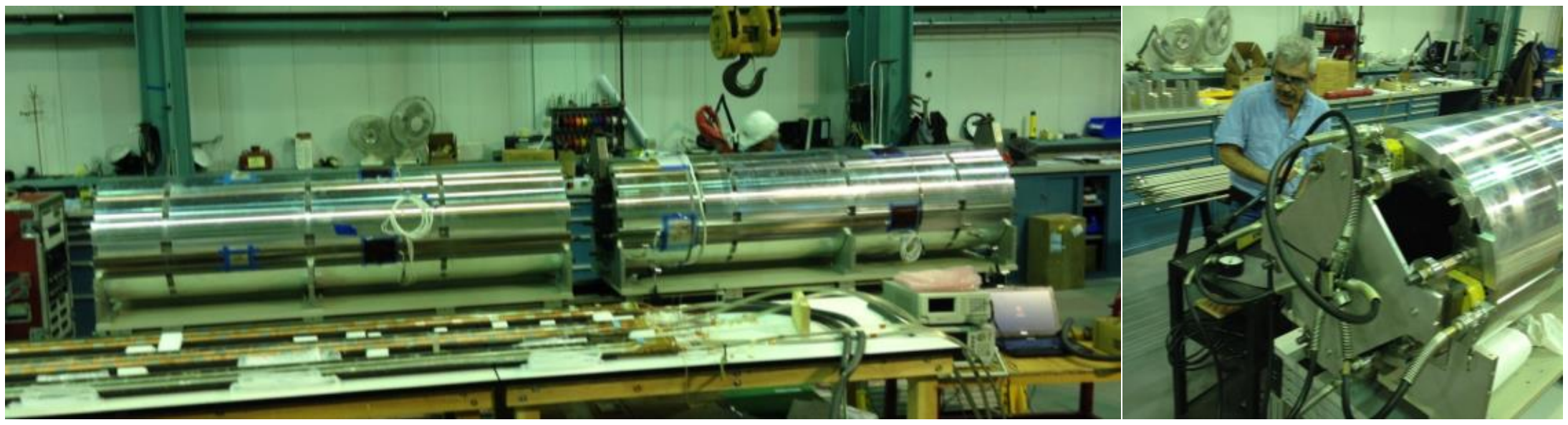

Fig. 4.7. Left: Two half-length yoke-shell subassemblies, ready to be joined. Right: hydraulic rig for pre-tensioning the yoke tie rods.

Once the shell-yoke sub-assembly is completed, the protrusions of the yoke stacks at the LE and RE are measured, being the minimal value 1 mm. Two measurement per yoke quadrant are taken and plotted in Fig. 4.8.

For the yoke protrusion with respect to the shell, the following specifications are set:

- ***The yoke protrusion with respect to the shell shall be within +0.600/+1.200 mm***

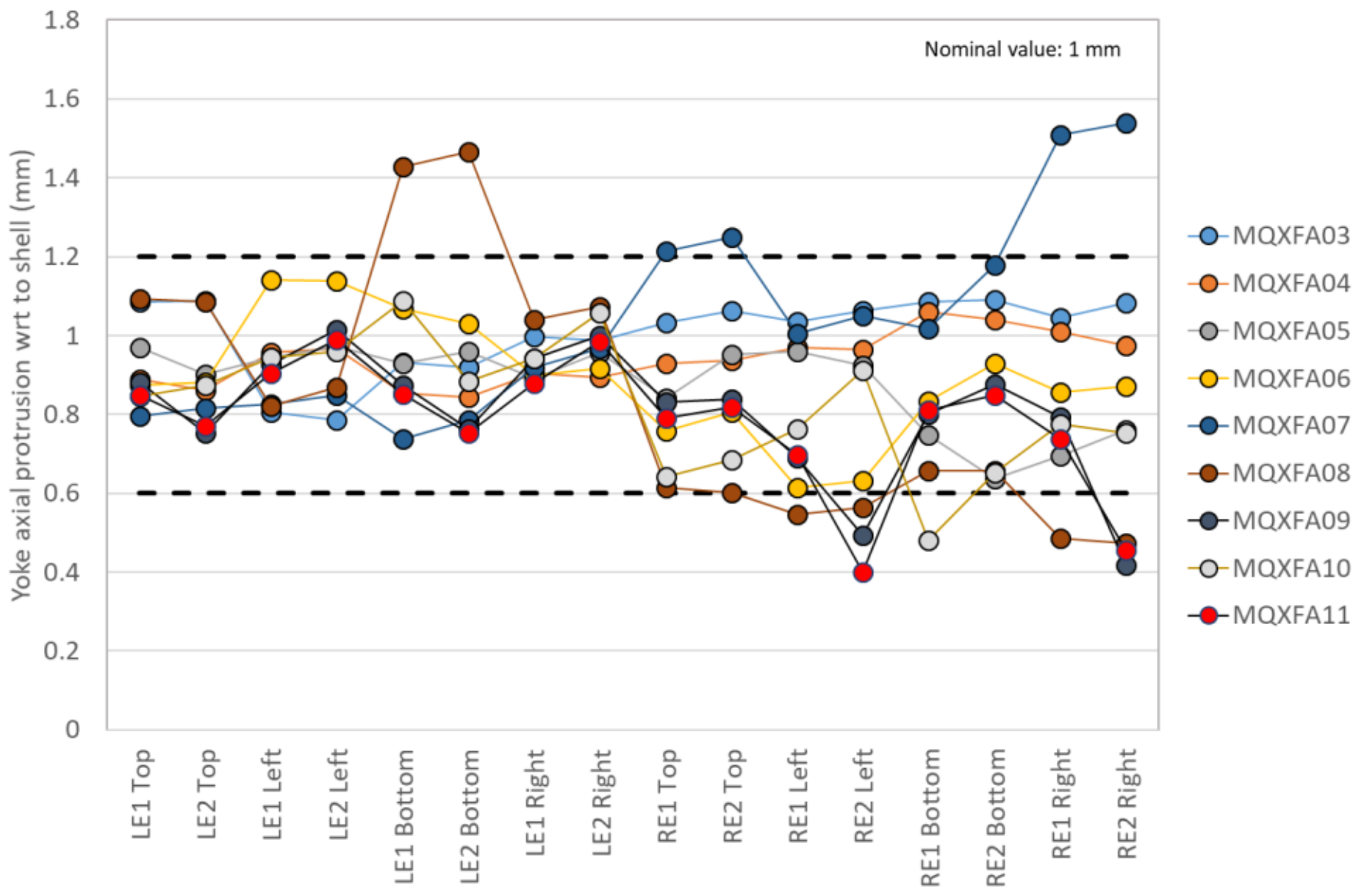

Fig. 4.8. Measurements of yoke protrusion with respect to shell: two measurements are taken per quadrant in MQXFA03-04- 05-06.

It is important to point out that the 1 mm design protrusion of the yokes stack with respect to the shells in the center of the magnet, which is required to guarantee full contact yoke-to-yoke when the LE and the RE modules are combined to form the full shell-yoke sub-assembly, is imposed by shimming the shells with respect to the yokes on the bottom of the vertical assembly shown in Fig. 4.1. Therefore, the protrusion on the bottom of the vertical assembly, which is then the mating face between the modules in the center of the magnet (see Fig. 4.6 left), does not depend on the yoke stack length.

The shell-yoke sub-assembly straightness is monitored by measuring the vertical and horizontal absolute position of fiducial in the mid-plane shell cut-outs (one side, either left of right from the lead end view) and in the top shell cut-outs. Measurements are taken in 7 longitudinal locations (in between shell segments). The fiducials are placed in contact with the yoke laminations, which are accessible through the cut-outs.

For the straightness of the shell-yoke sub-assembly, the following specifications are set:

- ***The maximum spread (max – min) of the vertical and horizontal position of the fiducials in the top shell cut-outs along 7 longitudinal locations shall be ≤ 0.250 mm.***
- ***The maximum spread (max – min) of the vertical and horizontal position of the fiducials in one side of the mid-plane shell cut-outs along 7 longitudinal locations shall be ≤ 0.250 mm.***

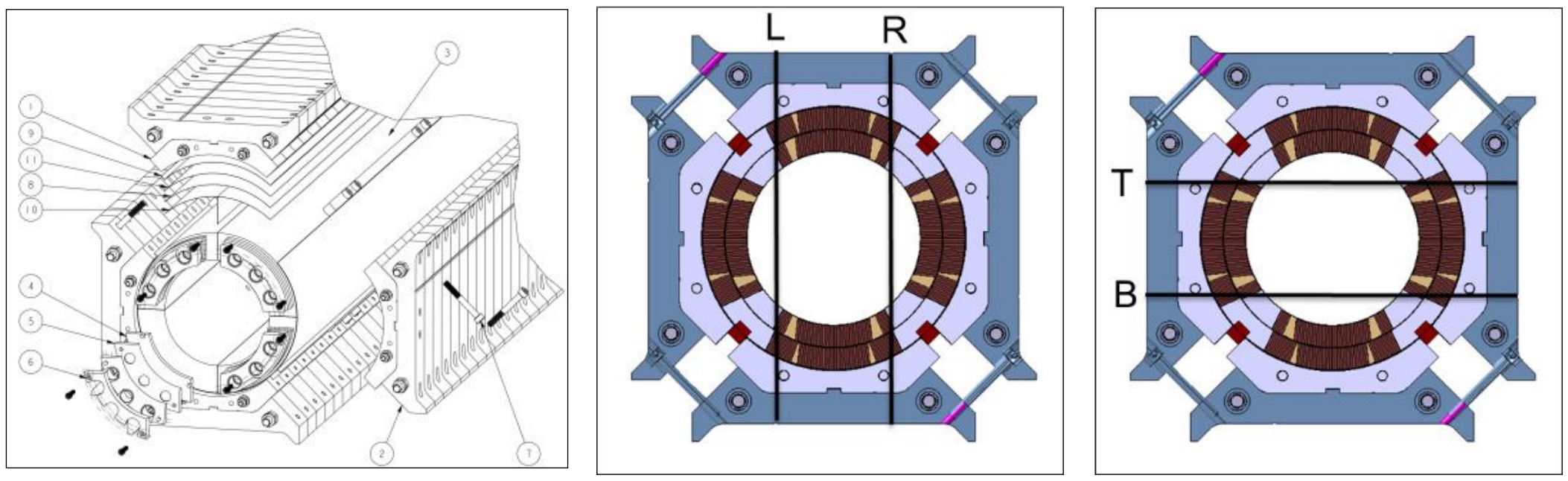

Fig. 4.9. Left: 3D view of coil-pack subassembly and insulation/shim layers. Center: location of the of the left and right vertical dimension measurements. Right: location of the of the top and bottom vertical dimension measurements.

### 4.2. Coil-pack sub-assembly

The assembly of the coil-pack sub-assembly, which is composed by coils, collars and pads, starts by placing the bottom pad-collar stack on the coil pack assembly table. Then, the two-side pad-collar stacks are assembled on the two pivot tables beside the primary assembly table. Based on the earlier CMM measurements performed on the coils, the inner curved surface of the collar stacks is lined with layers of G11 and/or polyimide as radial shims (see Fig. 4.9, left). The bottom pair of dressed coils is then moved onto the bottom pad-collar assembly. The upper pair of coils follows this operation, being secured on the assembly before the side pad-collar assemblies are rotated into place. Finally, the upper pad-collar assembly is placed on top, and all the pads are bolted together.

At this stage, with un-shimmed alignment pole keys inserted into the coils the gaps between these keys and the collars are measured. The function of the pole keys is to ensure the collar-key contact conditions after cool-down. According to FEM computations, described in [5], and shown in Fig. 4.10, this condition is achieved by assembly the coil pack at room temperature with a collar to pole-key gap of 0.200 mm per side (0.400 in the Fig. 4.10, where the total gap left+right is considered). This gap will ensure both proper alignment and minimum interception of the force from the shell at 1.9 K.

Once the coil-pack is assembled, dimensional measurements are taken, and magnetic measurements and electrical test are performed. The related specifications will be described in the next sections.

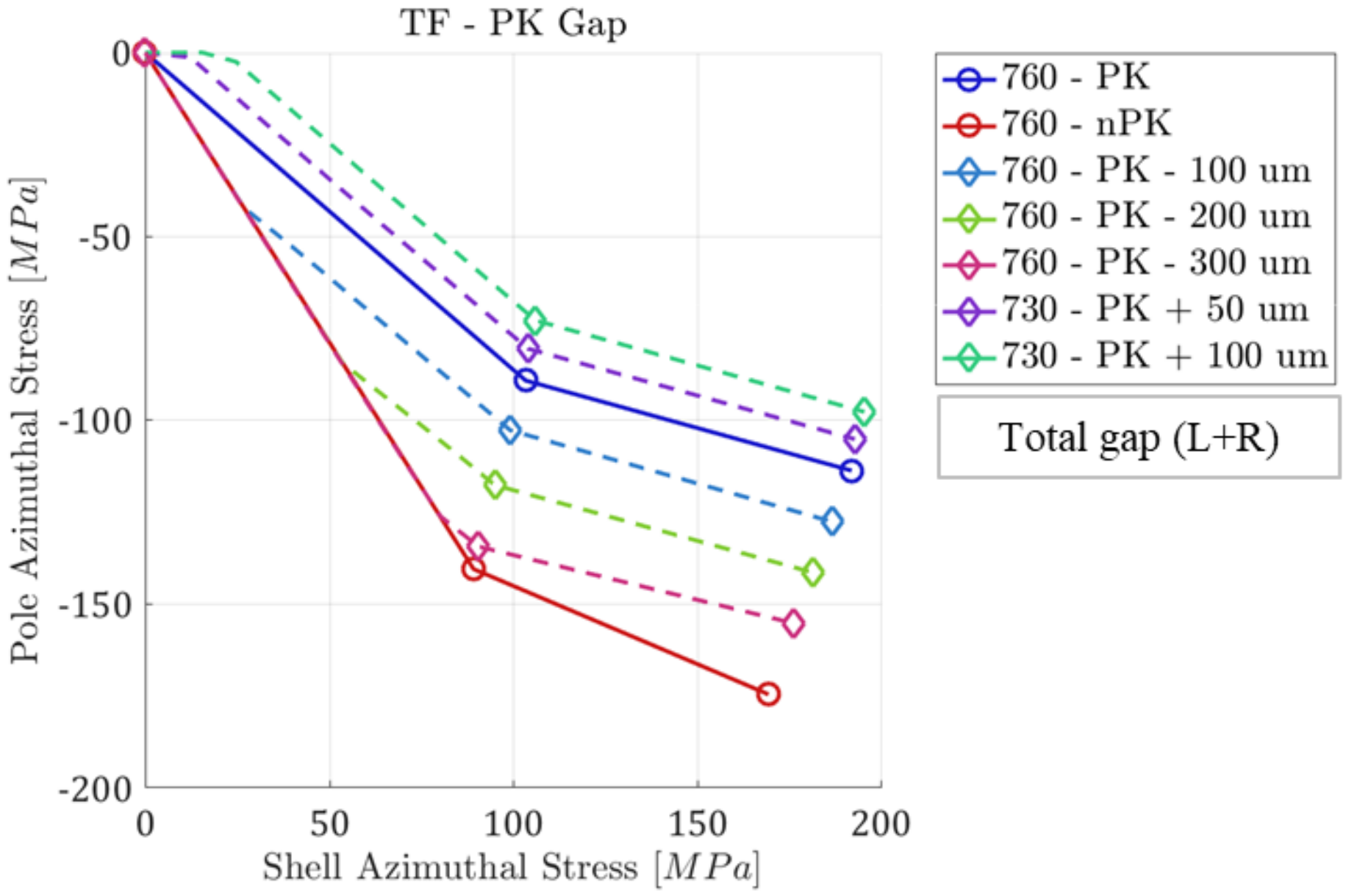

Fig. 4.10. Computed transfer function coil stress vs shell stress as a function of the applied pole-key (PK) interference or gap (total gap, i.e. left + right). The no PK line represents a loading condition where the pole alignment key does not manage to go in contact with the collar sides. The PK line is obtained when the pole key shimming is just enough to put the key sides in contact with the collars at the start of the loading. The dashed lines represent instead shimming conditions creating gap/interference between the key and the collars.

### 4.2.1. Radial gap between insulated coil and collar

The radial shimming of the coil, i.e. the layer of ground plane insulation and G11 and polyimide shims places in between coils and collars are chosen according to the coil dimensional measurements. In the nominal conditions, the main parameters are

- Nominal coil OR: 113.376 mm
- Nominal collar IR: 114 mm
- Nominal radial gap coil to collars: 0.624 mm

From LARP experience, it has become a common practice to remove 0.125 mm of insulation from the nominal radial shim in order to improve the pre-load and stress distribution of the coil. This effect was observed for the first time in the LQ magnets, where it was noticed that the collar had the tendency to contact the coil predominantly on the pole area rather than on the coil outer radius [6]. This effect, called "LQ effect", resulted in excessive coil bending and a lower coil azimuthal pre-load. A significant improvement on the coil stress distribution was achieved by removing radial shims between coil and collar. A mechanical analysis demonstrating the effectiveness of the shim removal was presented in [7].
The same solution is adopted in MQXFA. As a result,

- Nominal radial gap coil to collars after "LQ effect" shim removal: 0.500 mm

From the coil dimensional measurements, and in particular from the arc length excess, it is possible to estimate the effective coil outer radius after mid-plane shims are included, that is

- $\Delta R_{coil}$ = (arc_length_excess + 2*midplane_shim) *2/$\pi$

So, it is possible to compute the effective gap between the insulated coils and the collars, considering both the missing radial shim, and the coil azimuthal dimensions. In Fig. 4.11 we plotted the variation along the longitudinal direction of the gap between collars and insulated coils, considering for each z location the average among the four coils, for magnets MQXFA03-04-05-06-70. In Fig. 4.12, the same plot is done but this time considering the average over the full length, with ±1 $\sigma$ error bars.

For the radial gap between collars and insulated coils, the following specifications are set:

- ***The variation along the longitudinal direction of the radial gap between collars and insulated coils, considering for each z location the average among the four coils, shall be -0.125 mm -0.100 / +0.050 mm.***

*MQXF magnets may be assembled using previously tested coils or a mix of new and previously tested coils. When a coil-pack is done with previously tested coils, it should be taken into account that typically coils experience after pre-load and test a permanent reduction in arc length of the order of 0.100 mm. However, it is expected that at cold and under pre-load conditions consistent with the specification (about 100 MPa at 1.9 K), the coils will have a similar size as during the previous test. Therefore, the shimming of the previously tested coils shall be done assuming the dimensions of the virgin "pre-test" conditions.*

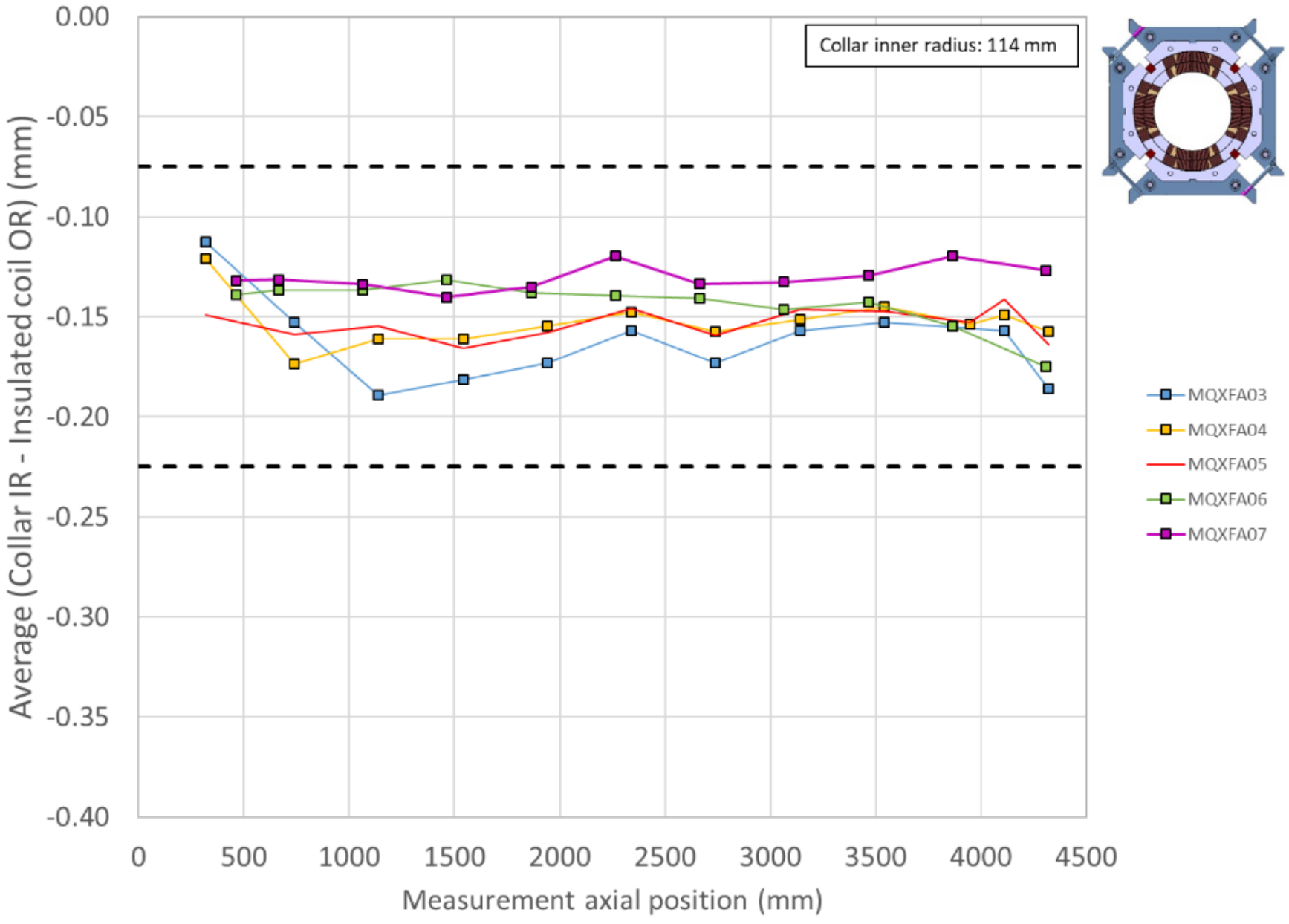

Fig. 4.11. Variation along the longitudinal direction of the radial gap between collars and insulated coils, considering for each z location the average among the four coils.

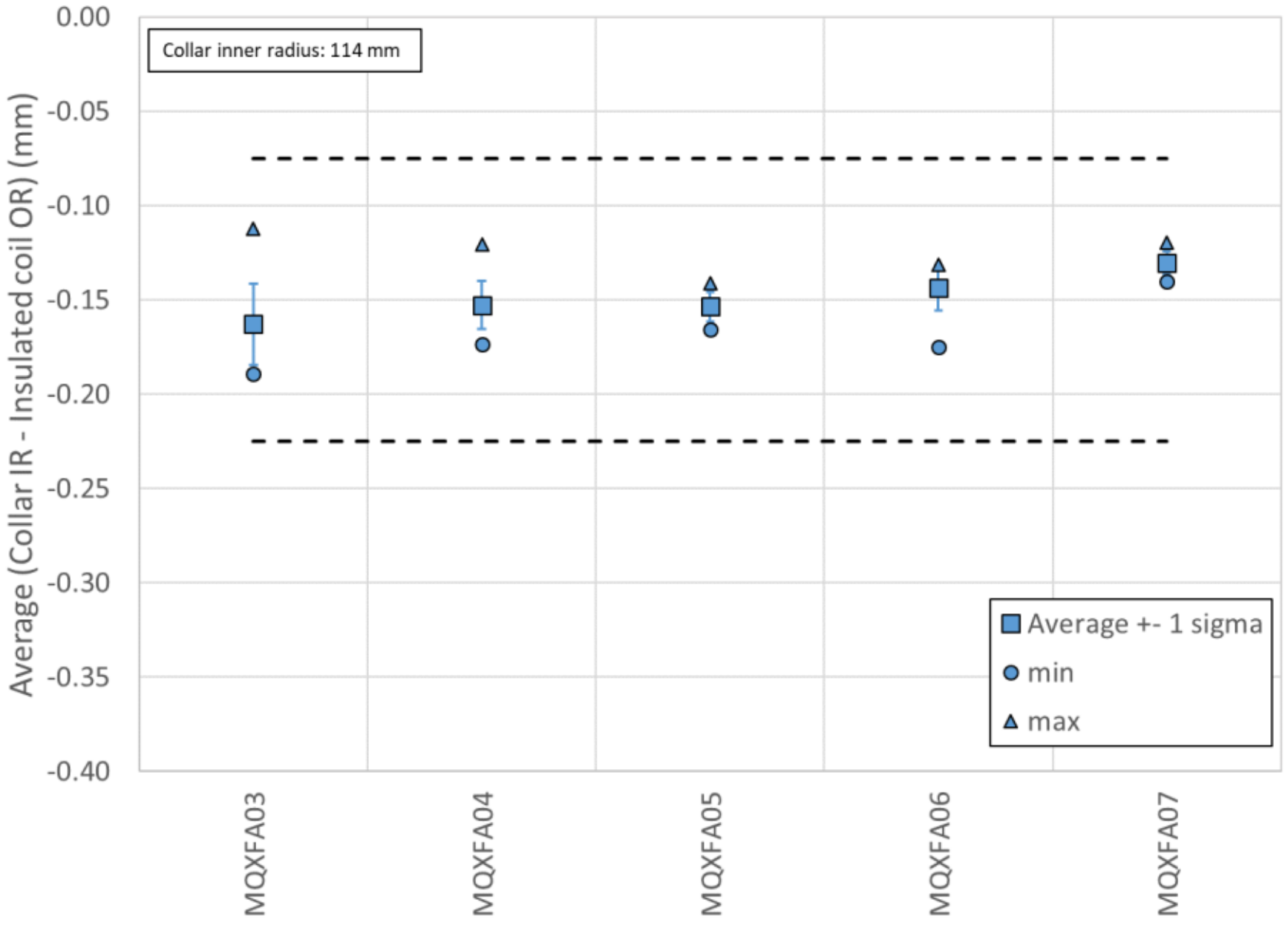

Fig. 4.12. Variation along the longitudinal direction of the radial gap between collars and insulated coils, considering for each z location the average among the four coils. The square solid markers represent the average over the full length, with ±1 σ error bars; the triangle and round markers represent the maximum and minimum values over the full length.

### 4.2.2. Coil-pack dimension

Once the bolting operation is completed, the coil pack vertical dimensions are measured along the length on 11 longitudinal locations. In each cross-section, two horizontal (top and bottom) and two vertical (left and right) dimensions are taken, as shown in the center and right pictures of Fig. 4.9. Two main parameters are monitored

- The **uniformity** of the vertical and horizontal dimensions along the z axis given by
    - The difference of the average of the left and right (top and bottom) vertical (horizontal) dimensions in each cross-section with respect to the measured vertical (horizontal) average dimension of the entire coil pack
- The **squareness** in the vertical and horizontal directions along the z axis, given by
    - The difference of the left and right (top and bottom) vertical (horizontal) dimensions in each cross-section.

The results are shown in Fig. 4.13 to Fig 4.20 for magnets MQXFA03-04-05-06.

For the coil-pack uniformity and squareness the following specifications are set:

- ***The uniformity of the vertical and horizontal dimensions along the z axis shall be within ±0.200 mm***
- ***The squareness of the vertical and horizontal dimensions along the z axis shall be within +0.900 mm***

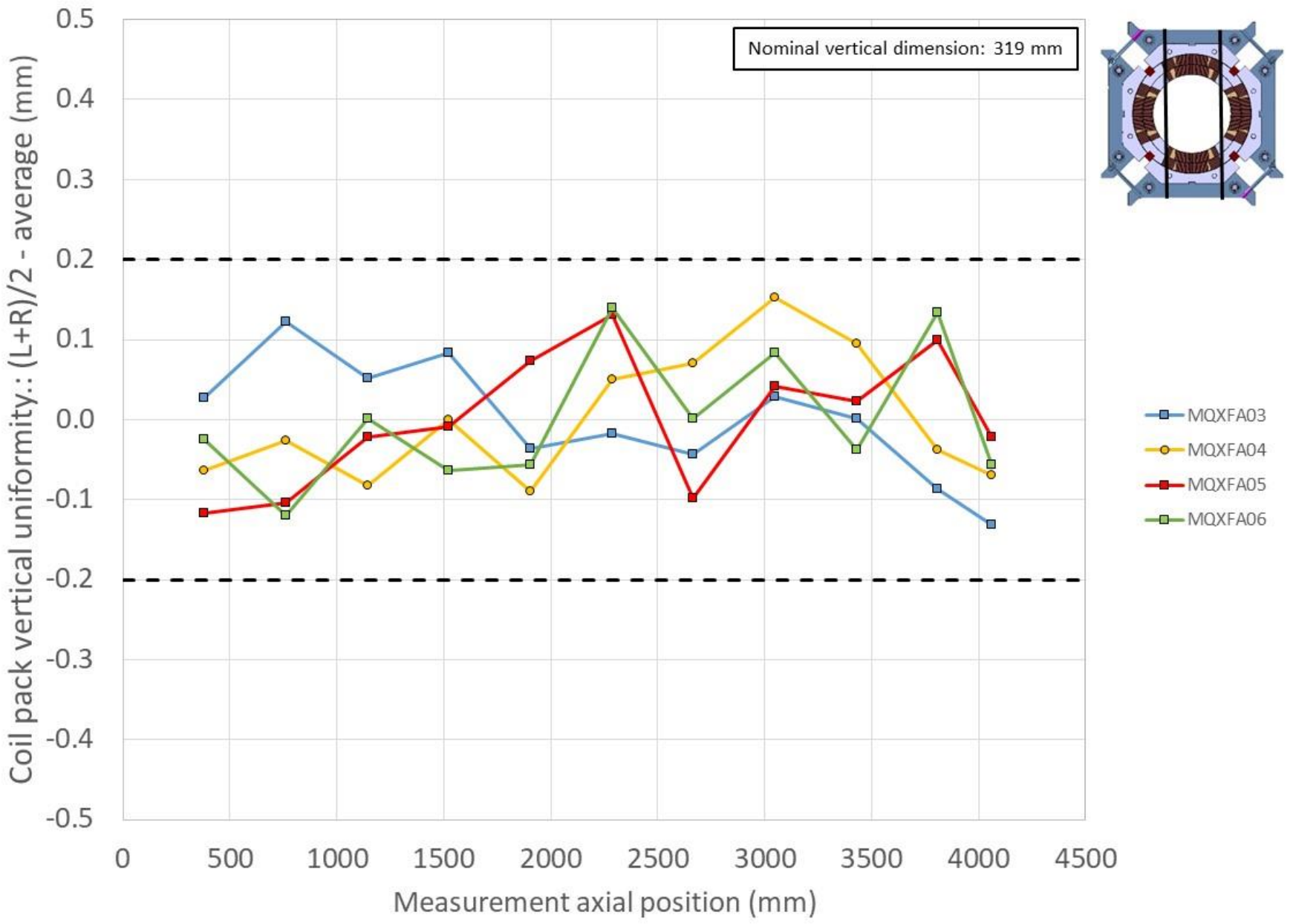

Fig. 4.13. Measured coil-pack vertical uniformity along the z axis, given by the difference of the average of the left and right vertical dimensions in each cross-section with respect to the measured vertical average dimensions of the entire coil pack.

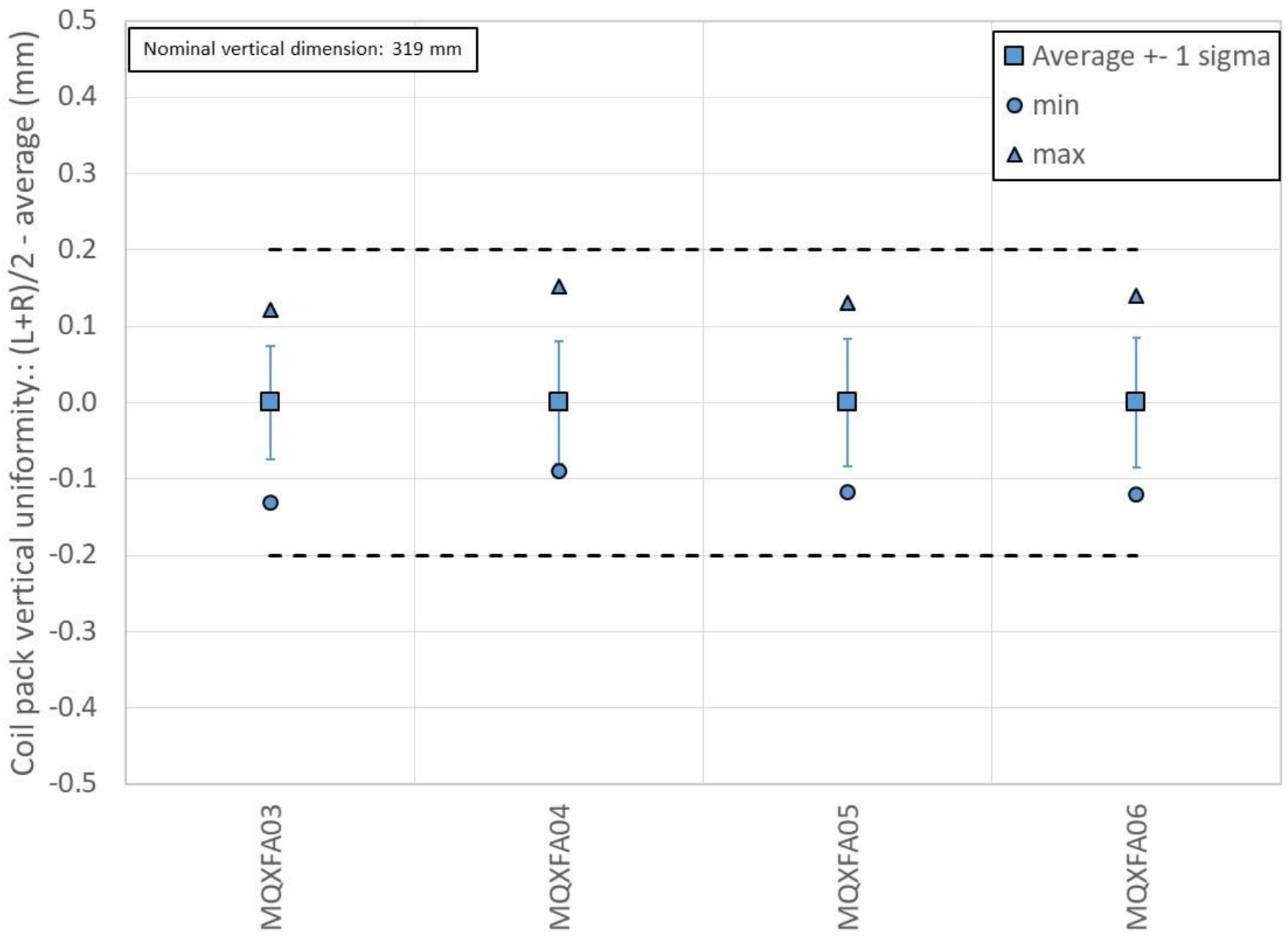

Fig. 4.14. Measured coil-pack vertical uniformity along the z axis, given by the difference of the average of the left and right vertical dimensions in each cross-section with respect to the measured vertical average dimensions of the entire coil pack. The square solid markers represent the average over the full length, with ±1 σ error bars; the triangle and round markers represent the maximum and minimum values over the full length.

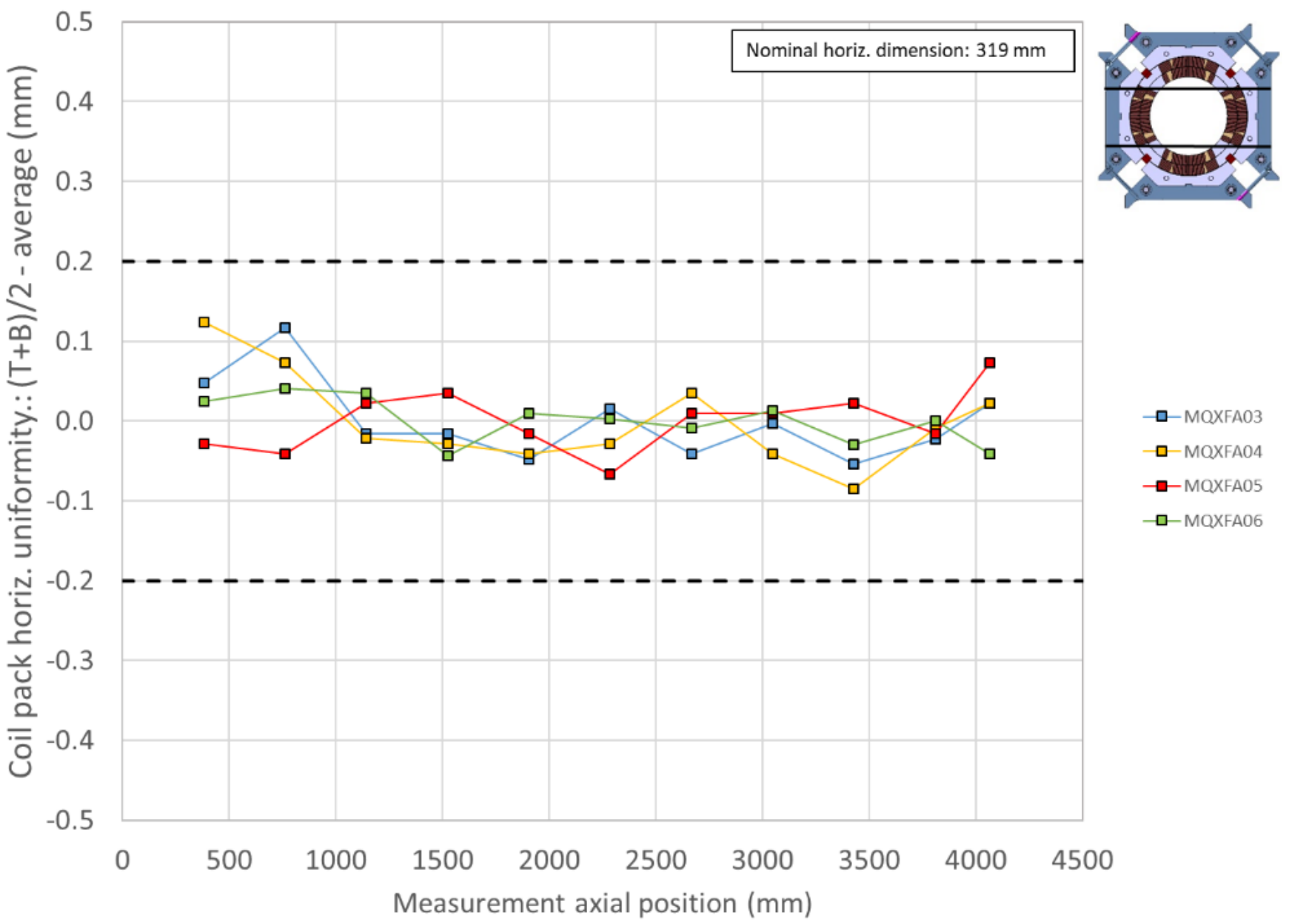

Fig. 4.15. Measured coil-pack horizontal uniformity along the z axis, given by the difference of the average of the top and bottom horizontal dimensions in each cross-section with respect to the measured horizontal average dimensions of the entire coil pack.

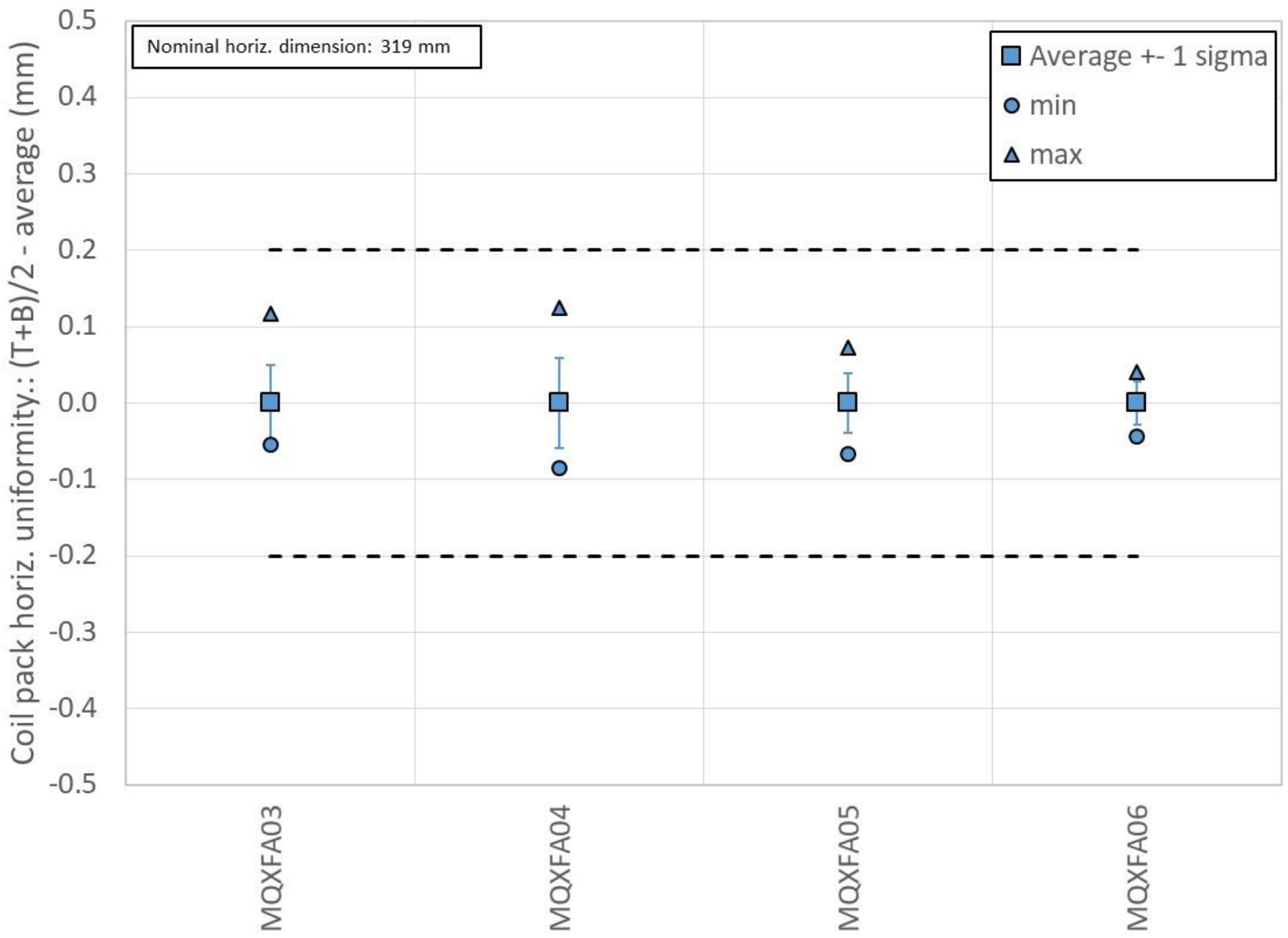

Fig. 4.16. Measured coil-pack horizontal uniformity along the z axis, given by the difference of the average of the top and bottom horizontal dimensions in each cross-section with respect to the measured horizontal average dimensions of the entire coil pack. The square solid markers represent the average over the full length, with ±1 σ error bars; the triangle and round markers represent the maximum and minimum values over the full length.

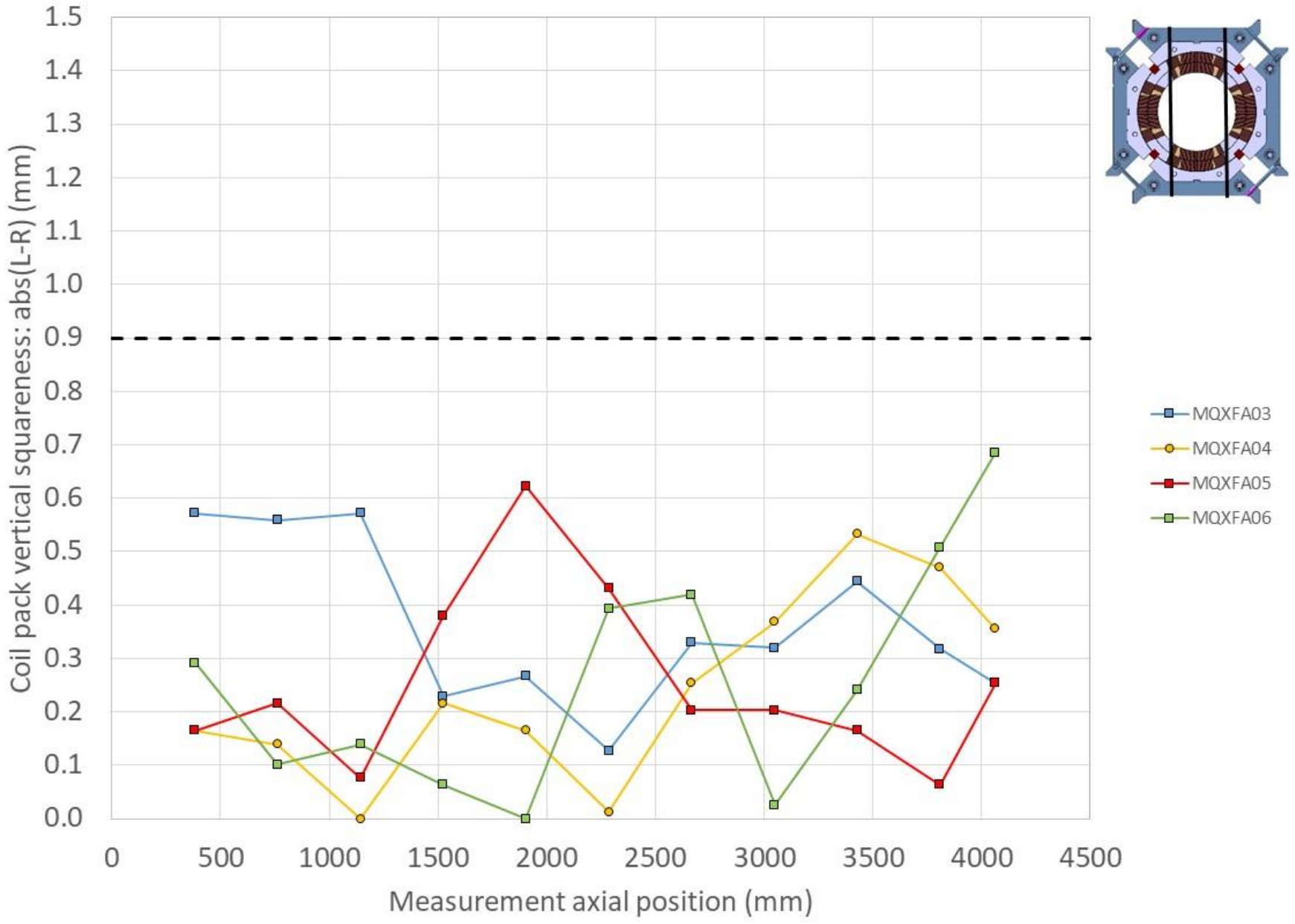

Fig. 4.17. Measured coil-pack squareness in the vertical direction along the z axis, given by the difference of the left and right vertical dimensions in each cross-section.

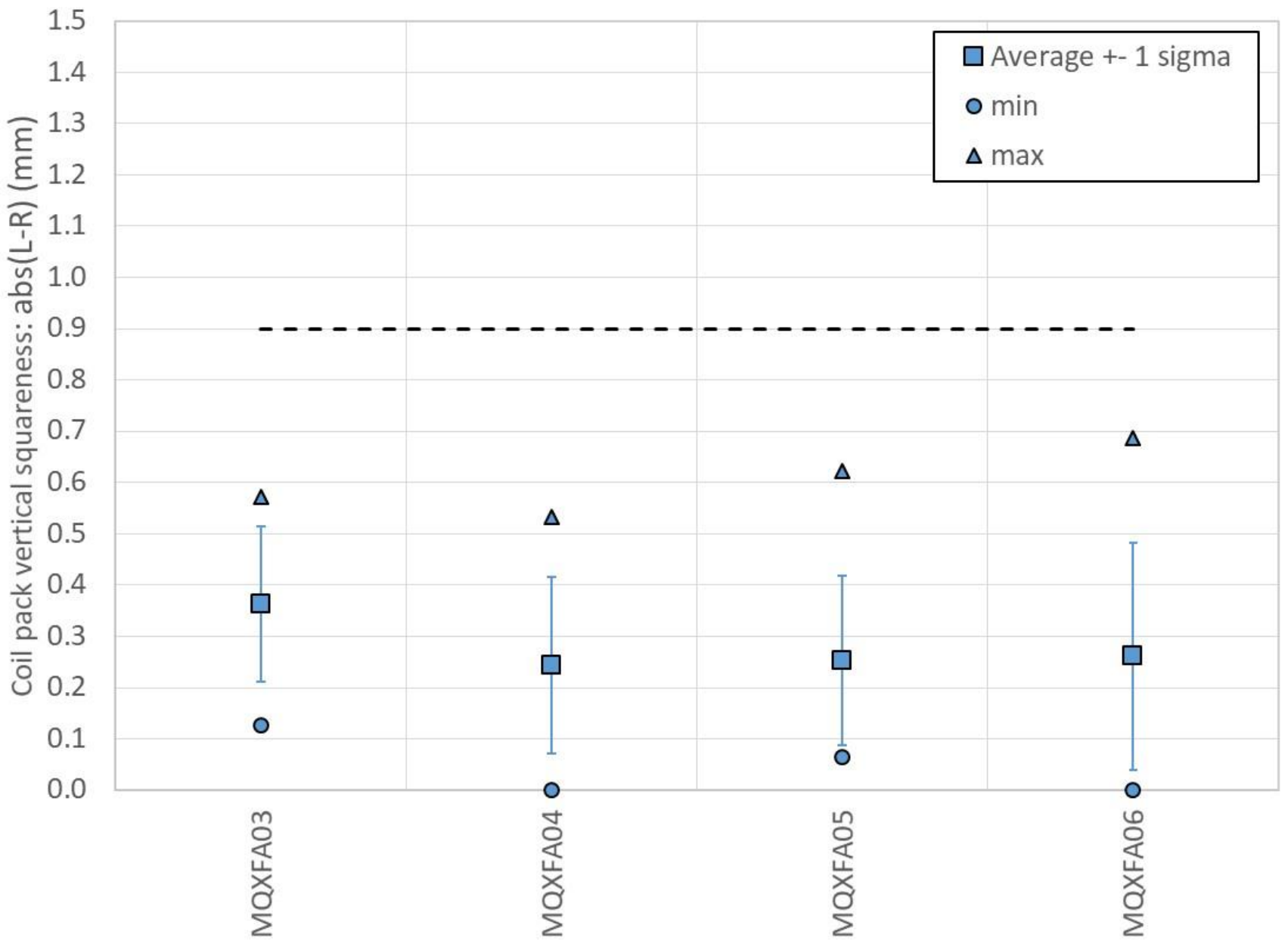

Fig. 4.18. Measured coil-pack squareness in the vertical direction along the z axis, given by the difference of the left and right vertical dimensions in each cross-section. The square solid markers represent the average over the full length, with ±1 σ error bars; the triangle and round markers represent the maximum and minimum values over the full length.

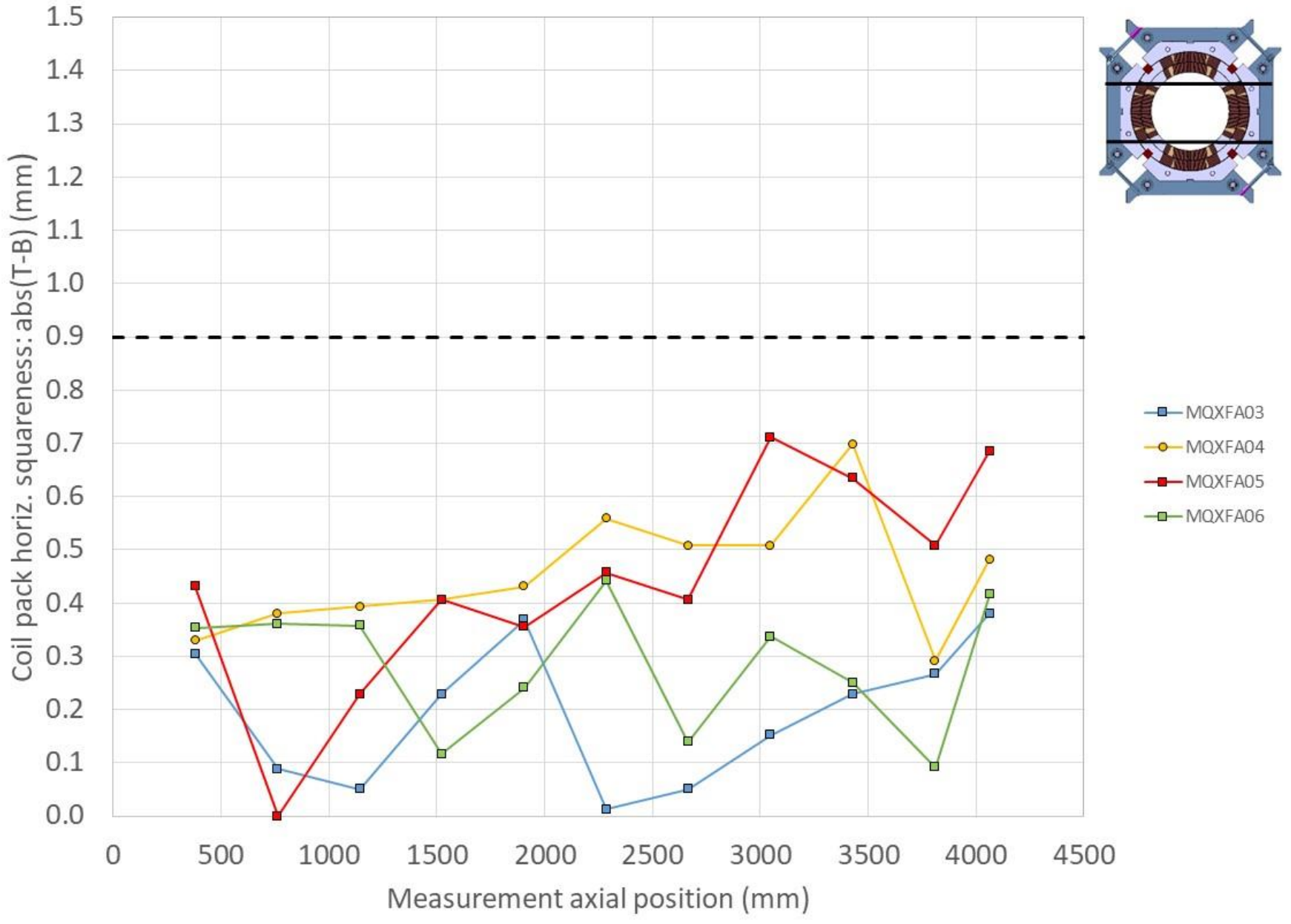

Fig. 4.19. Measured coil-pack squareness in the horizontal direction along the z axis, given by the difference of the top and bottom horizontal dimensions in each cross-section.

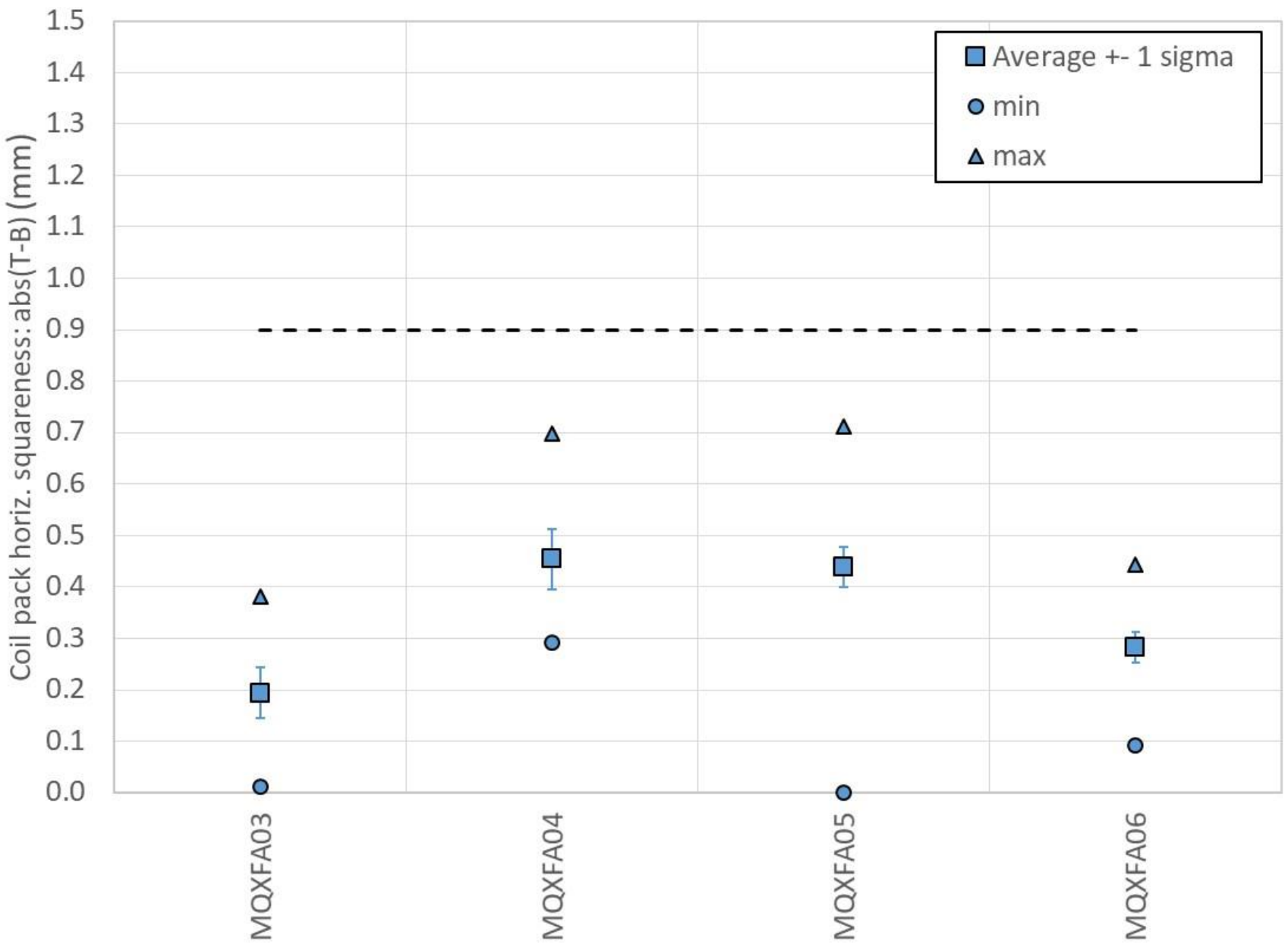

Fig. 4.20. Measured coil-pack squareness in the horizontal direction along the z axis, given by the difference of the top and bottom horizontal dimensions in each cross-section. The square solid markers represent the average over the full length, with ±1 σ error bars; the triangle and round markers represent the maximum and minimum values over the full length.

### 4.2.3. Pole key gap

Once the bolting operation is completed, the gap between the pole key and the collars, including the ground insulation, called the pole key gap (see Fig. 4.21) is measured. According to the finite element computations [5], a gap of 0.200 mm per side at the end of coil-pack bolting results in a contact collar to pole-key at the end of the cool-down. In this ideal scenario, at the end of the cool-down no force interception occurred due to the pole key, but azimuthal alignment between collars and coil is guaranteed.

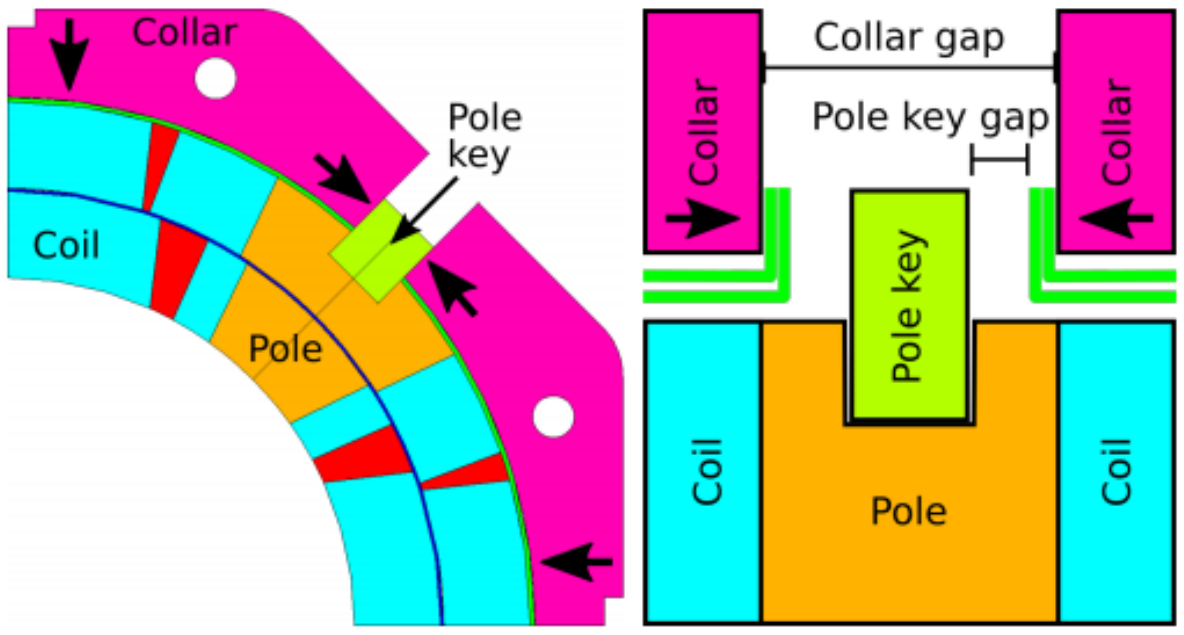

Fig. 4.21. Schematic view of the collared coil (left) and of the pole key (right), courtesy of E. Takala (CERN).

The pole-key gap is measured on all the four coils and along the z axis at 11 longitudinal locations. Fig. 4.22 shows the pole-key gap per side on each quadrant averaged along the magnet length in MQXFA11. Following the inspection of magnet A07 after test, the spec has been updated as follows.

<u>For the pole key gap, the following specifications are set:</u>

- ***The pole key gap, average per side and all quadrants, shall be +0.400 ±0.050 mm in each longitudinal location.***
- ***The minimum pole key gap, average per side, shall be ≥0.300 mm in each quadrant and in each longitudinal location.***

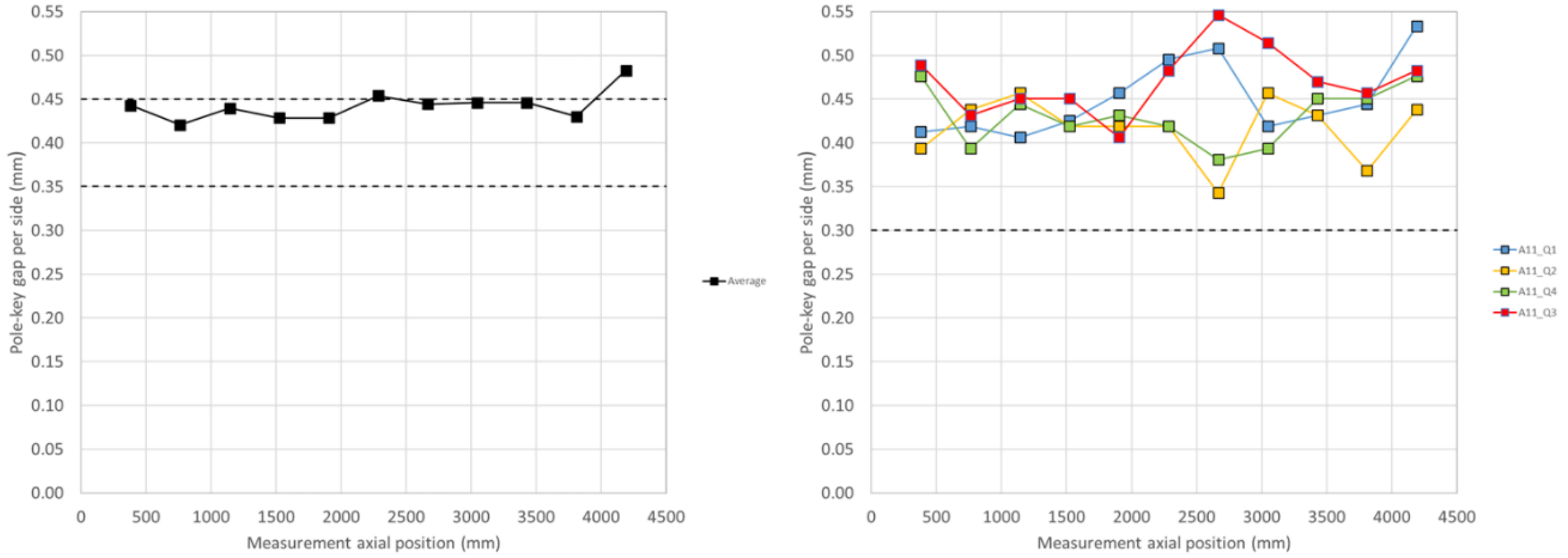

Fig. 4.22. Left: measurements of the pole key gap, average per side and all quadrants, at each longitudinal position of magnet MQXFA11. Right: measurements of the pole key gap, average per side, for each quadrant and at each longitudinal position of magnet MQXFA11. The dashed lines in the figure represent the specs.

### 4.2.4. Coil-pack magnetic measurements

Once the bolting operation is completed, magnetic measurements are performed, focusing on the absolute value of the main field transfer function averaged over the straight part (see Fig. 4.23) and the field errors in units averaged in the straight part (see Fig. 4.24).

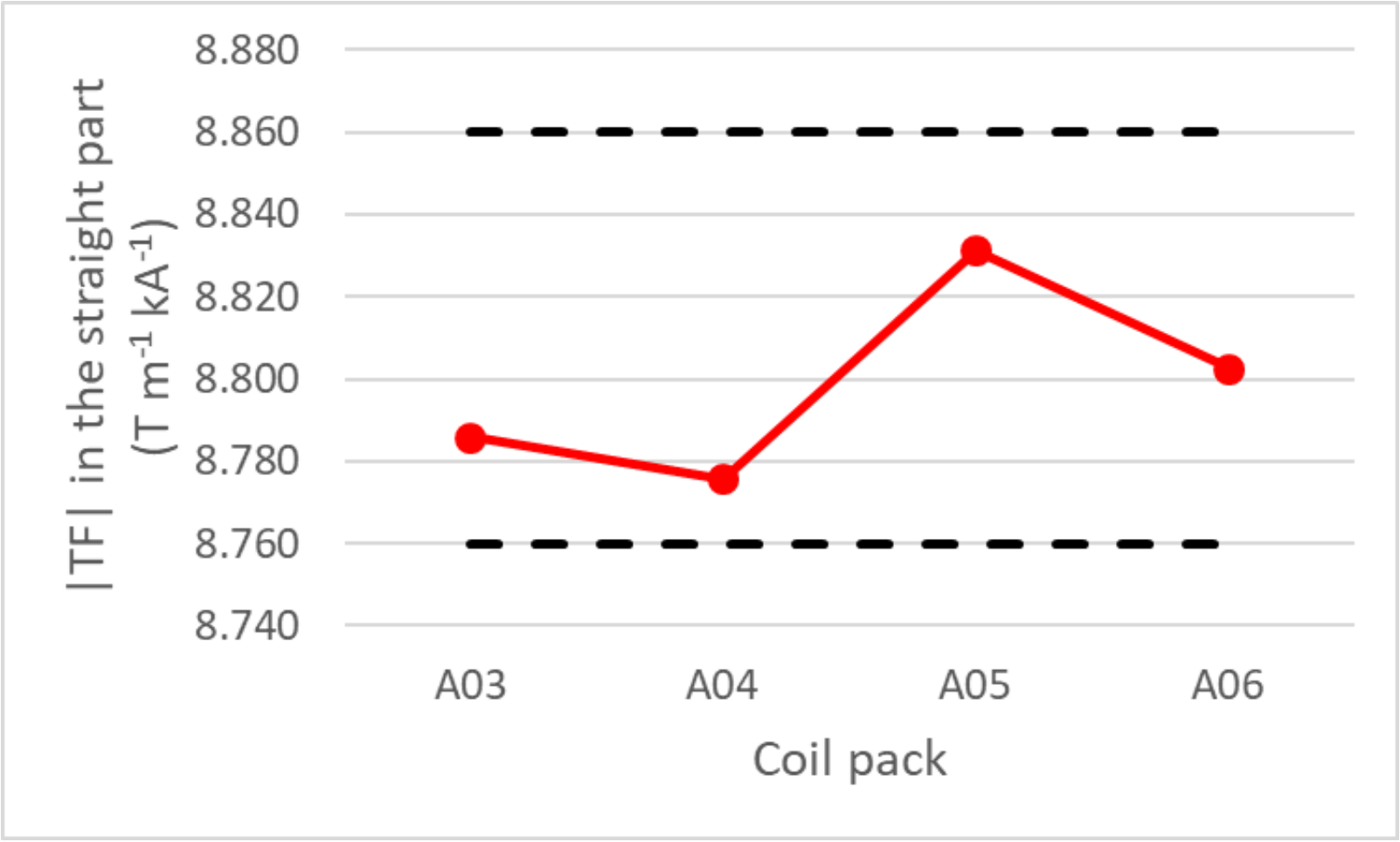

Fig. 4.23: Absolute value of the main field transfer function in the straight part of the coil packs. Data of A03 to A06 were measured using a 110 mm long rotating coil. The lower bound is 8.760 T/m/kA and the upper bound is 8.860 T/m/kA.

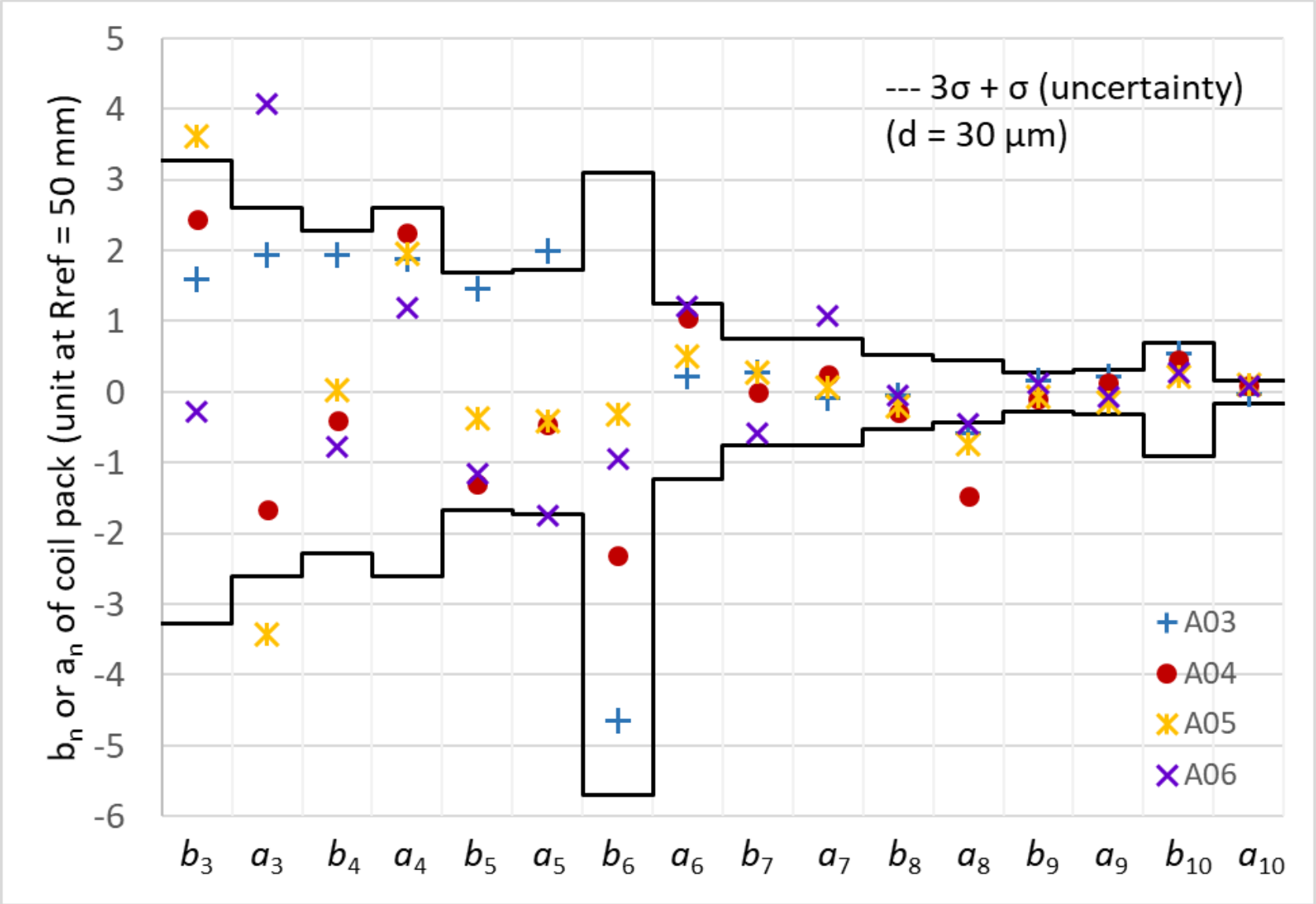

Fig. 4.24: Field errors of the coil pack. Data of A03 to A06 were measured using a 110 mm long rotating coil.

For the main field transfer function, the following specifications are set:

- ***The absolute value of the main field transfer function averaged over the straight part of the coil pack shall be within +8.760 T/(m kA) and +8.860 T/(m kA).***

For the field errors, the following targets are set:

- ***The field errors in units averaged in the straight part of the coil pack with a reference radius of 50 mm should be within the upper and lower bounds determined according to the Triplet Field Quality Table in Table 1 of [8]***
    - *A systematic component (0.9 units) associated with magnet assembly and cooldown is considered for $b_6$.*

### 4.2.5. Coil-pack electrical tests

The electrical tests performed on the coil-pack are described in the document "MQXF magnet Electrical QA at LBNL" [9]. The purpose is to check if there is any electrical issue due to the coil-pack assembly process.

For the electrical tests of the coil pack, the following specifications are set in [9]:

- ***1. Resistance check with hand-held multimeter***
    - ***a. Coil to structure***
    - ***b. Coil to coil***
    - ***c. Coil to PH***
- ***2. Hipot.***
    - ***a. Individual coil to structure, 3680 V.***
- ***3. Impulse test up to 2500 V, Direct- and Reversed-polarity.***
    - ***a. 1000 V, 1500 V, 2000 V, 2500 V***

## 5. Magnet Assembly Specifications

The completion of the shell-yoke sub-assembly and of the coil-pack sub-assembly, with the dimensional, electrical and magnetic measurements associated, is followed by the assembly and pre-load of the full magnet. The coil-pack is slid inside the shell-yoke sub-assembly, and master keys and bladders are inserted in the slots between yoke and pads (see Fig. 5.1, left). The high-pressure pump is attached to the bladders on both ends of the magnet (keys and bladders are half-length), and the axial loading rig is attached to the axial rods to perform the axial pre-load operations (see Fig 5.1 right). By inflating the bladders at progressively higher pressures, the load keys are shimmed to obtain the target strain on the shells and coils. In parallel, the axial rods are pre-tensioned. The strain in shells, coils, and rods is monitored with strain gauges.

Fig. 4.3 shows the shell and coil strain gauge locations. Three shells are instrumented (shell 2, 4, and 7). In each axial location, both the axial and azimuthal shell strain are measured, for all four quadrants in the azimuthal position indicated in Fig. 2.1. All four coils (winding poles) are instrumented with azimuthal and axial strain gauges, in a location close to the RE (see Fig. 4.3 and Fig. 2.1). Finally, each axial rod is equipped with an axial strain gauge.

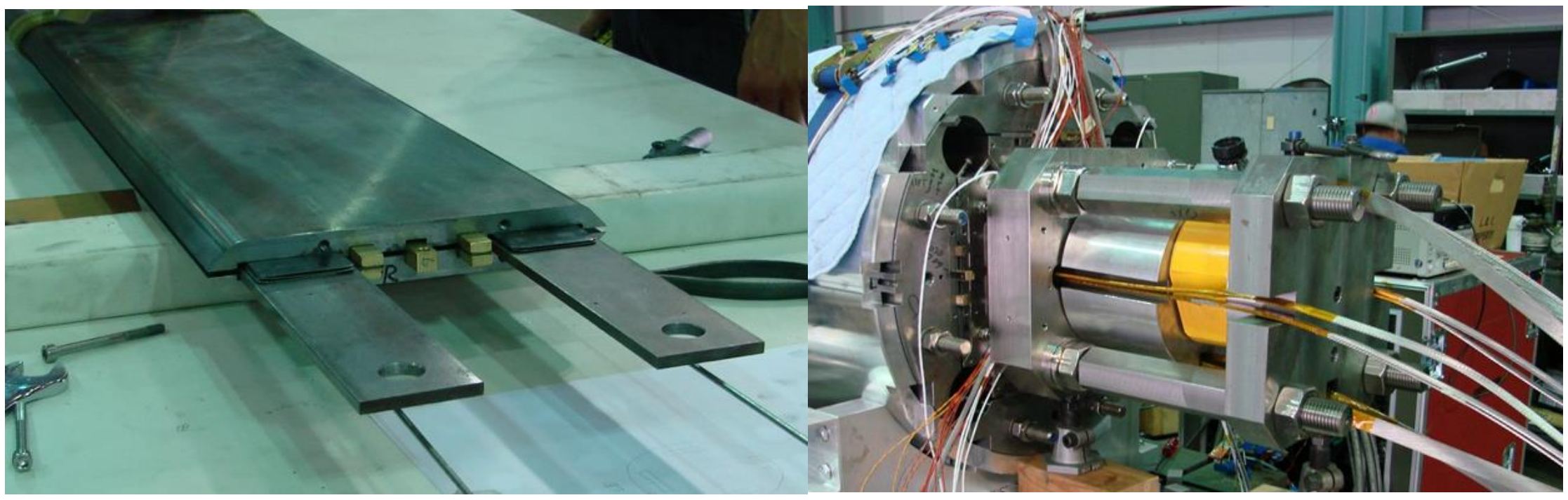

Fig. 5.1: Left: master key assembly with bladders and pulling shims; right: axial pre-load system.

For the pre-load operation sequence, the following specifications are set:

- ***The loading operation shall follow the sequence described below:***
  1. ***Pre-load axial rods with average to a measured of strain ~50 με***
  2. ***Apply 50% azimuthal pre-load***
  3. ***Apply 50% axial pre-load***
  4. ***Apply 100% azimuthal pre-load***
  5. ***Apply 100% axial pre-load***

The pre-load sequence can be seen in Fig. 5.2, where the coil, shell and rod strain during MQXFA06 pre-loading are plot. The vertical brown bands represent the axial loading phase, which, following the specifications, is executed in two steps, in correspondence of 50% and 100% of the coil-shell target.

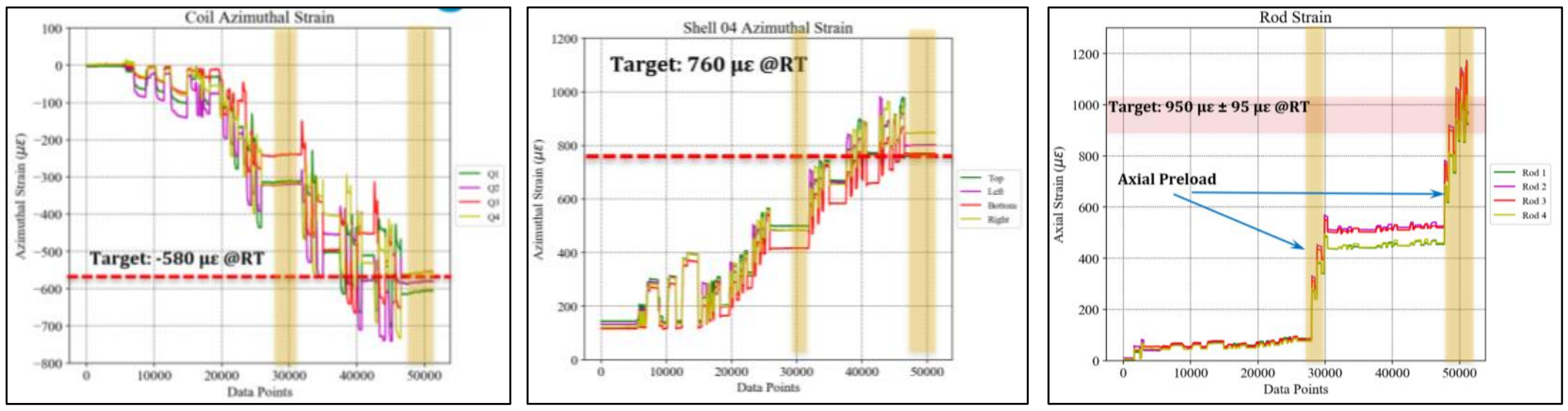

Fig. 5.2: Pre-load sequence of MQXFA06: coil strain (left), shell strain (center), rod strain (right).

The target pre-load levels for shells, coils and rods are chosen based on the experience of successful short and pre-series models, which is summarized in Fig. 5.3, showing the coil azimuthal pre-load vs. the axial pre-load force, both after cool-down. The dashed lines indicate the computed reference levels to guarantee full pre-load (no unloading) in the straight section and in the ends respectively at Nominal and Ultimate field. The specification for the room temperature pre-loads is set in order to achieve after cool-down a pre-compression of 100-110 MPa in the coil and a total axial preload of 1.1 MN in the rods. These values correspond to the ones applied in MQXFS4, which showed the best training performance among the short models (see Fig. 5.4).

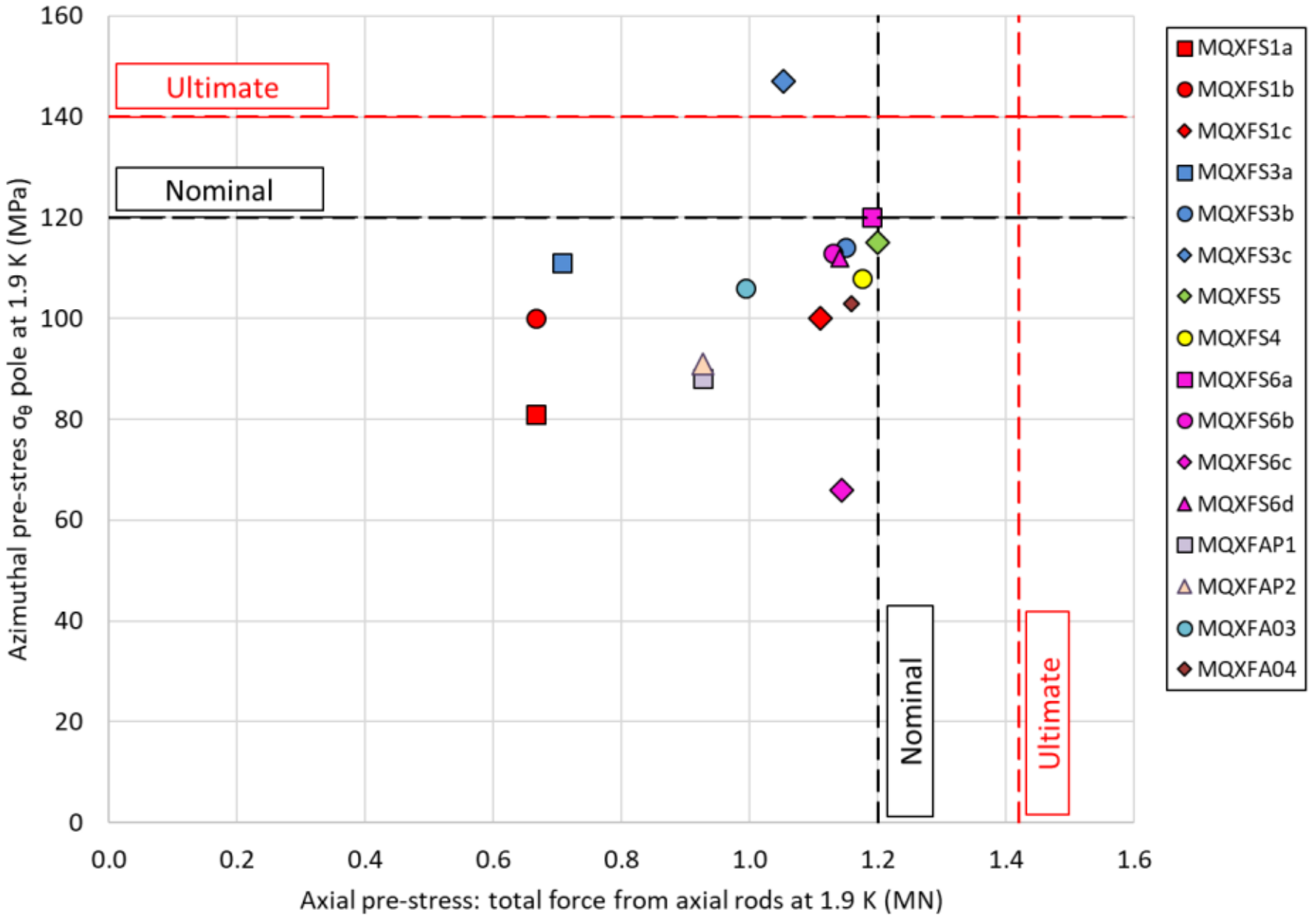

Fig. 5.3: Coil azimuthal pre-load vs. the axial pre-load force, both after cool-down, applied in different MQXF magnets. The dashed lines indicate computed reference levels to guarantee full coil-pole pre-load (no unloading) in the straight section and in the ends respectively at Nominal and Ultimate field.

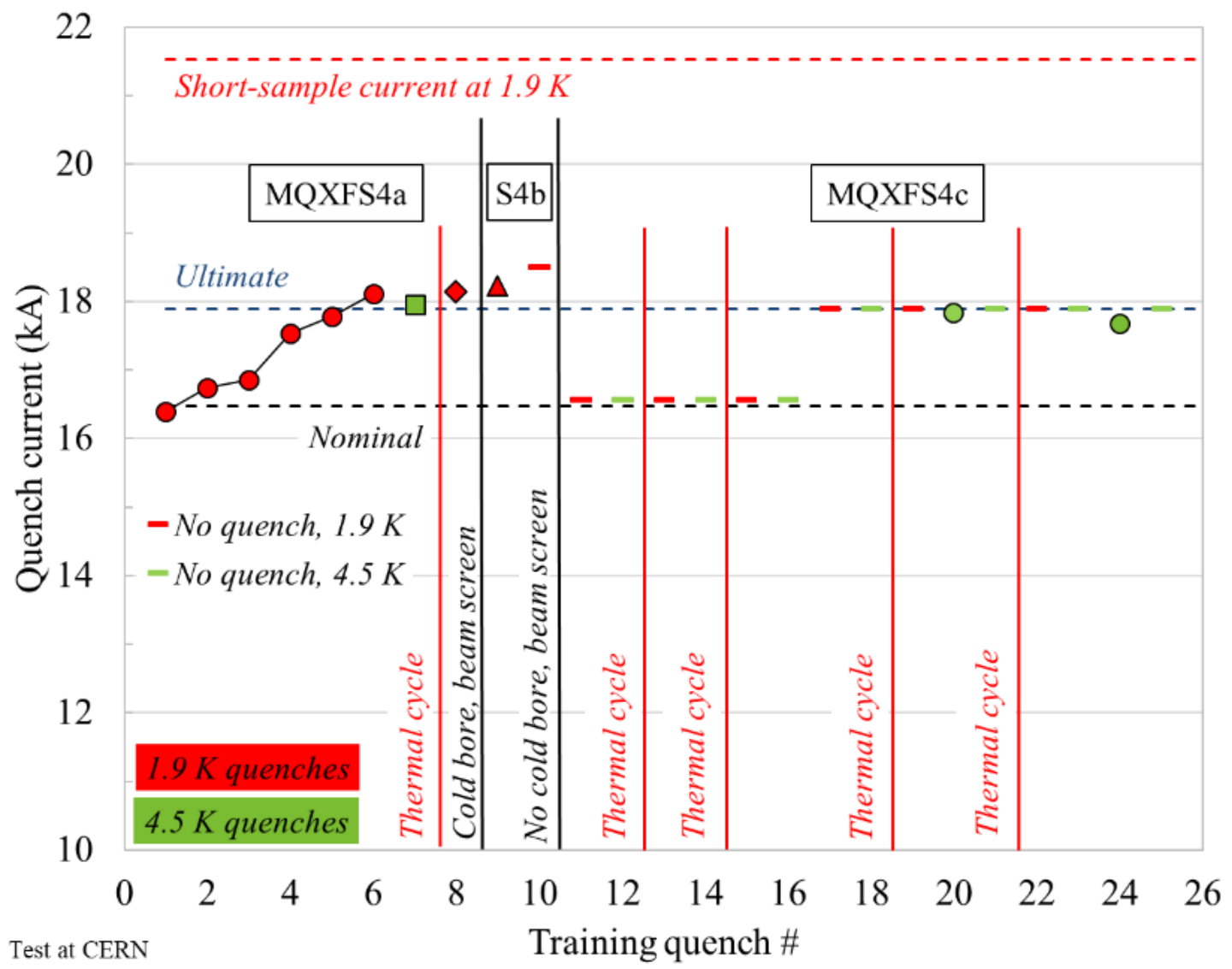

Fig. 5.4: Training performance of MQXFS04.

For the pre-load targets at the end of the pre-load operations, the following specifications are set:

- ***The target average measured stress on coils, shells and rods at the end of the loading (after at least 24 h) shall be***
  - ***Shell average azimuthal stress: +58 ± 6 MPa***
  - ***Coil (winding pole) average azimuthal stress: -80 ± 8 MPa***
  - ***Rod average strain: +950 $\mu\varepsilon$ ± 95 $\mu\varepsilon$***

An example is shown in Fig. 5.4, where the final values of MQXF06 are plotted.

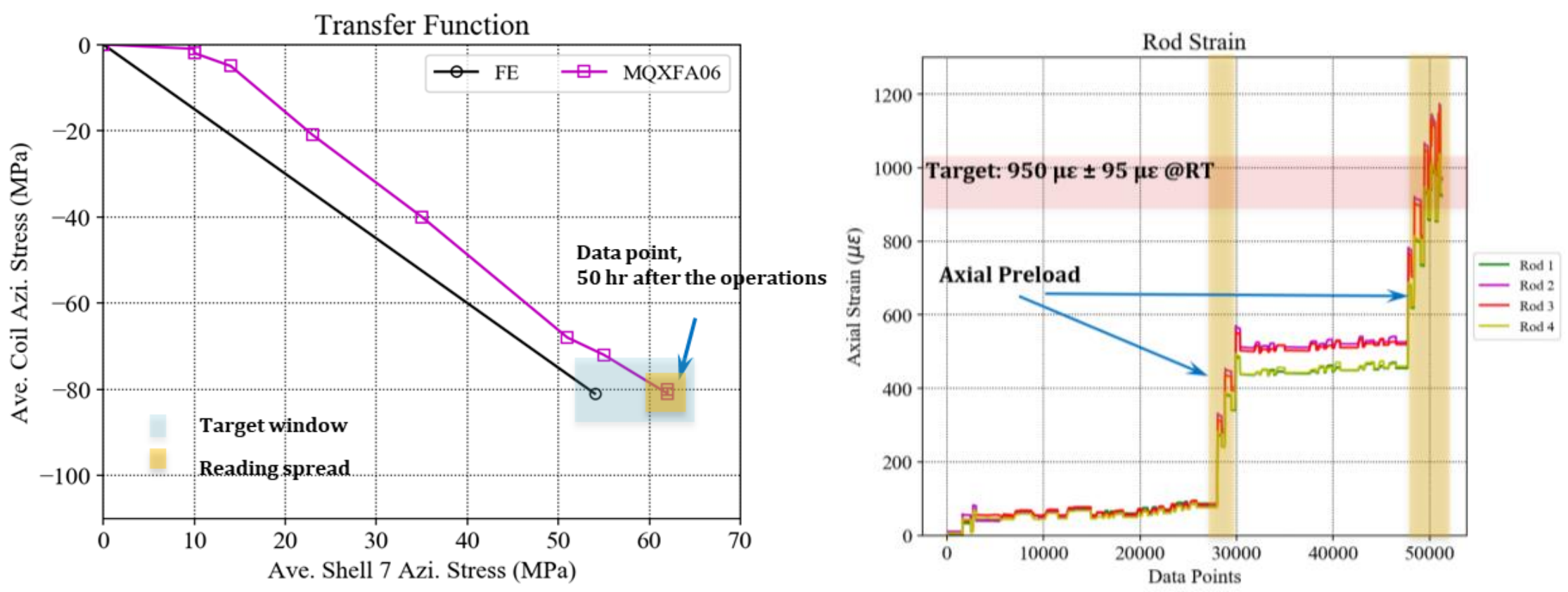

Fig. 5.4: Azimuthal and axial pre-loading of MQXFA06.

In order to minimize the risk of permanent conductor degradation due to high stress during pre-load at room temperature, a maximum stress of -120 MPa has been established in [1]. The value was based on experience from LARP short model magnets. Moving from short models to long coils, a more conservative values is chosen in the specification for MQXFA.

<u>For the maximum stress reached during the pre-load operations, the following specifications are set:</u>

- ***During the entire pre-load operation, the maximum compression measured on each coil shall never exceed -110 MPa***

An example of the coil stress measured during the bladder operations of MQXFA06 is shown in Fig. 5.5.

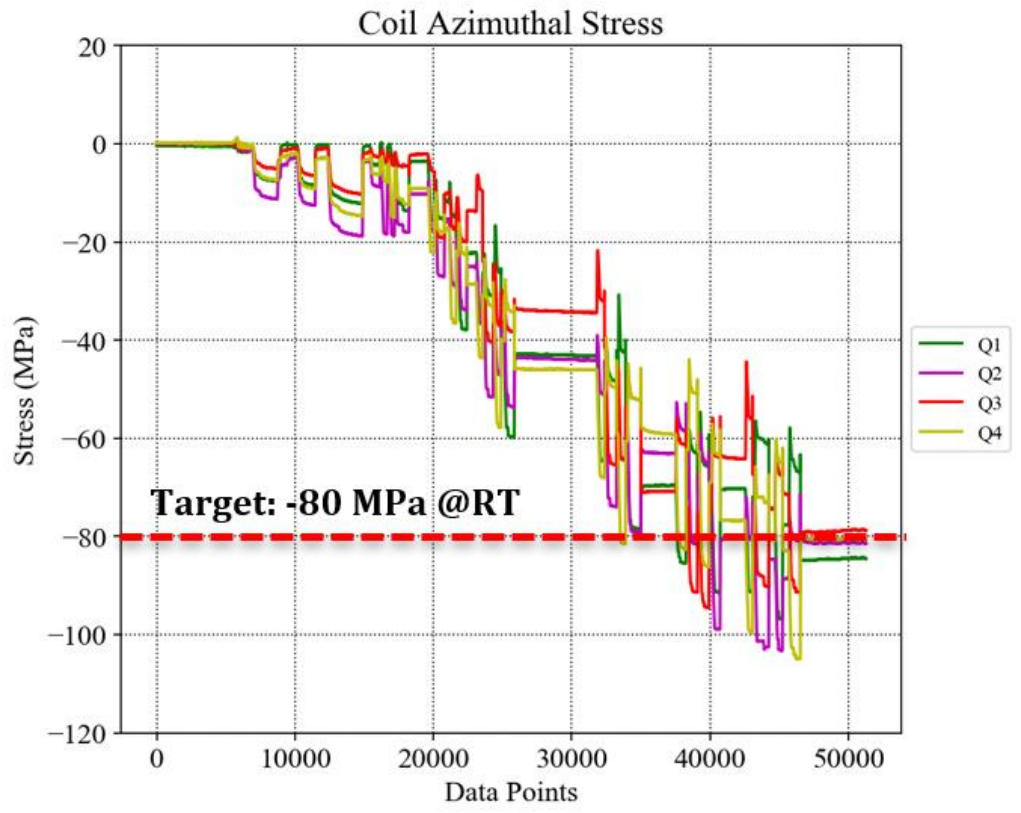

Fig. 5.5: Coil stress measured during bladder operations of MQXFA06.

As stated at the beginning of the section, the coil stress is measured only in one axial location, at 3965 mm from the lead end, as can be seen in Fig. 5.6 (and reported in reference [10]), which indicate the location of the coil dimensional measurements and of the strain gauge station.

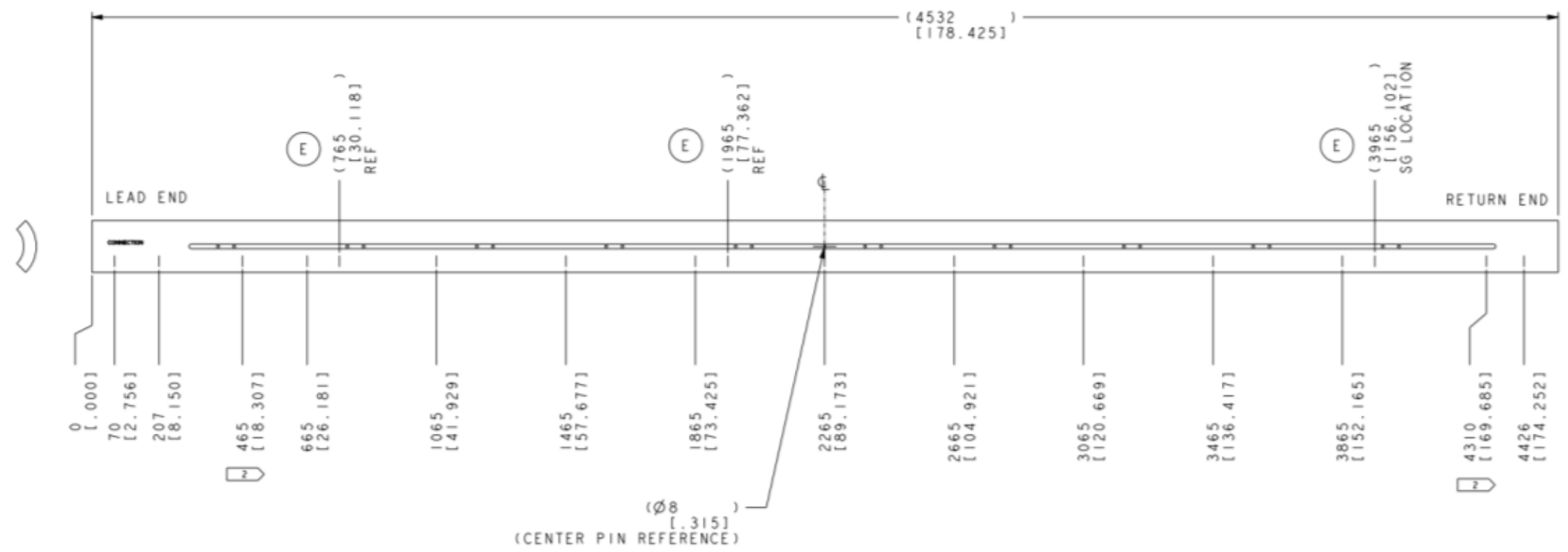

Fig. 5.6: Locations of coil dimensional measurements and of strain gauges.

So, in order to estimate the stress in the coil along the longitudinal location, we rely on a sensitivity analysis of the coil and shell stress vs the load key thickness performed with 2D FE models (see [7], [11], [12]) and confirmed with strain data in short models. The analysis suggests that the computed sensitivity of shell and coil stress to key thickness (which basically correspond to coil radial variations) is respectively 120 MPa/mm and -200 MPa/mm. These values are confirmed in short models with a pole key gap of at least 0.200 mm, like MQXFS4 (see Fig. 5.7).

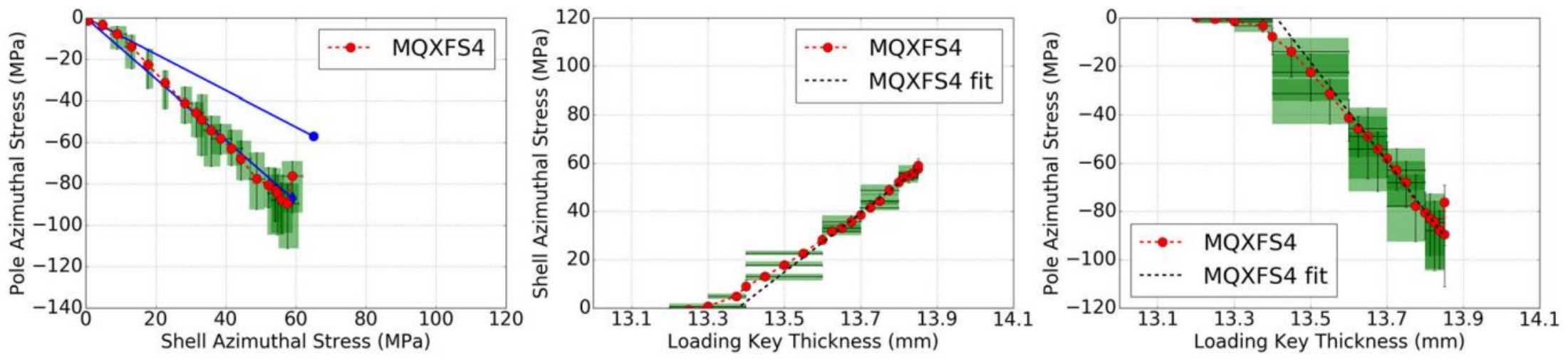

Fig. 5.7: Measured shell and coil stress in MQXFS4 (red markers with variation ranges in green). Left: transfer function coil vs shell, with computed values (solid lines) with pole key and without pole keys (lower line); Center and right: sensitivity to load key thickness.

According to the sensitivity analysis:

- +0.100 mm of coil outer radius increase
  - -20 MPa of coil stress variation
  - +12 MPa of shell stress variation

- +0.100 mm of total coil arc-length increase (0.050 per mid-plane)
  - -13 MPa of coil stress variation
  - +8 MPa of shell stress variation

Therefore, in order to minimize the variation of stress along the length, we define a target range of the coil azimuthal size with respect to the size at the strain gauge station. The range is defined based on the sensitivity analysis after coil shimming and assuming that only the average coil size plays a role in the coil stress.

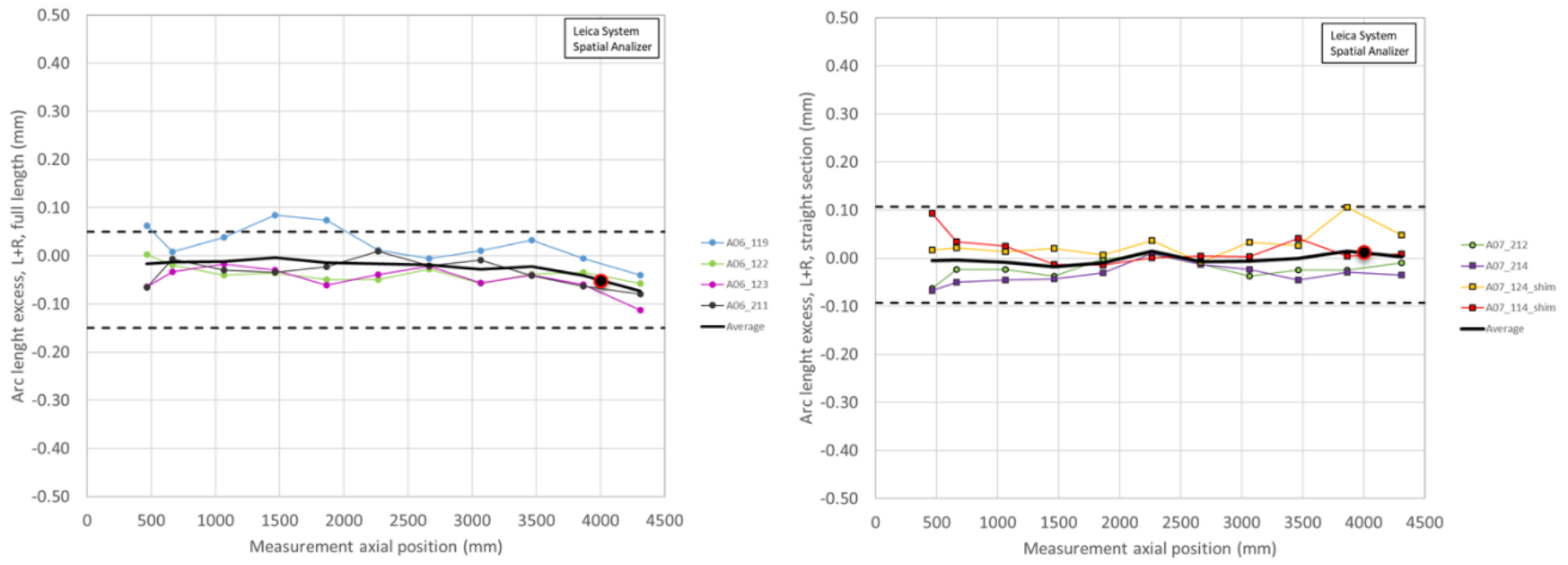

Fig. 5.8: Average shimmed coil size (arc length) for MQXFA6 and MQXFA07. The black round market indicates the location of the strain gauge station.

<u>For the maximum coil size (arc length) variation along the coil length, the following specifications are set:</u>

- ***Average coil size (arc length) after shimming shall range within ±100 µm with respect to average coil size measured in strain gauge location.***

A representation of this specification is show in Fig. 5.8, where the average shimmed coil size (arc length) for MQXFA6 and MQXFA07 is plot, and where the black round marker indicates the location of the strain gauge station.

The range of ±100 µm can be justified as follows.

In regards of the -80 ± 8 MPa specification for the average stress after bladder operation, and considering the aforementioned stress sensitivity values, a ±100 µm average arc length variation with respect to the strain gauge location results in

- A possible maximum average coil stress of -88-(13) MPa= -101 MPa
  - As shown in Fig. 5.3, and according to measured stress data during pre-loading of successful short models (i.e. short model which reached ultimate field), the average coil stress is ranging
    - At warm from -28 MPa (MQXFS6c) to -99 MPa (MQXFS5)
    - At cold from -66 MPa (MQXFS6c) to -118 MPa (MQXFS4-5)
  - Therefore, the estimated maximum average coil stress is marginally higher than the value measured in MQXFS5 without taking into account MQXFS5 coil size variations
- A possible minimum average coil stress of -72+(13) MPa = -59 MPa
  - Therefore, the estimated minimum average coil stress is higher than the -28 MPa measured in MQXFS6c
    - The statement is true also considering the further reduction of coil pre-compression of 5 MPa due to the shell segmentation, as shown in the computations plotted in Fig. 5.9 [13].

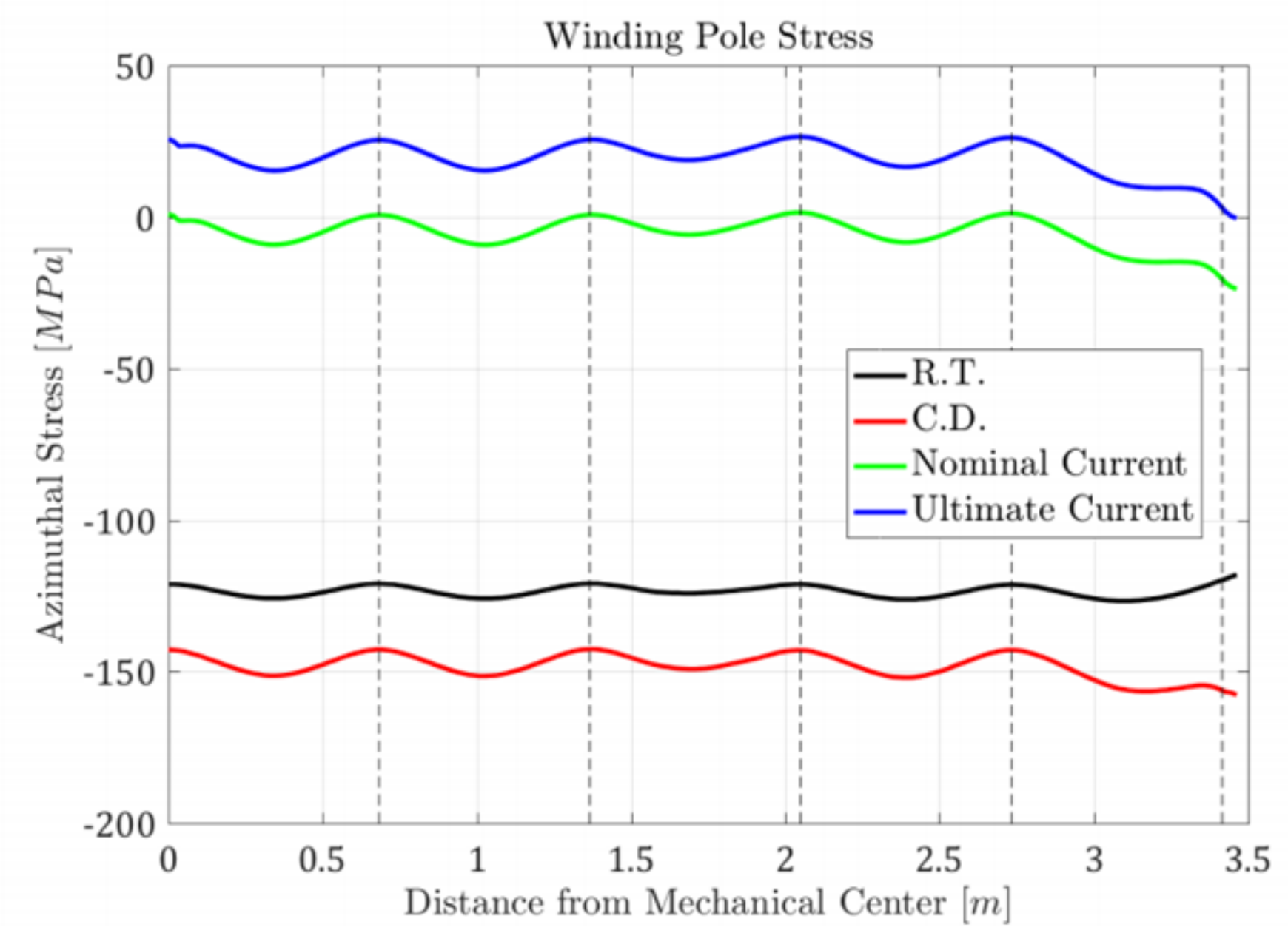

Fig. 5.9: Computed azimuthal winding pole stress along coil length in MQXFB [13].

In regards of the -110 MPa specification for maximum compression measured on each coil during pre-load operations, and considering the aforementioned stress sensitivity values, a ±100 µm average arc length variation with respect to the strain gauge location results in

- A possible maximum coil stress of -110-(13) MPa = -123 MPa
  - According to measured stress data during pre-loading of successful short models (i.e. short model which reached ultimate field), the peak coil stress reached during bladder operations is ranging
    - At warm from -23 MPa (MQXFS6c) to -140 MPa (MQXFS5, see Fig. 5.10)
    - At cold from -64 MPa (MQXFS6c) to -145 MPa (MQXFS4, see Fig. 5.11)
  - Therefore, the estimated maximum stress of -123 MPa would be in between marginally higher than the -120 MPa limit established in [1] and still lower than the -140 MPa reached in MQXFS5.
    - In correspondence of the shell cuts this value could be further reduced by 5 MPa at warm and 10 MPa at cold

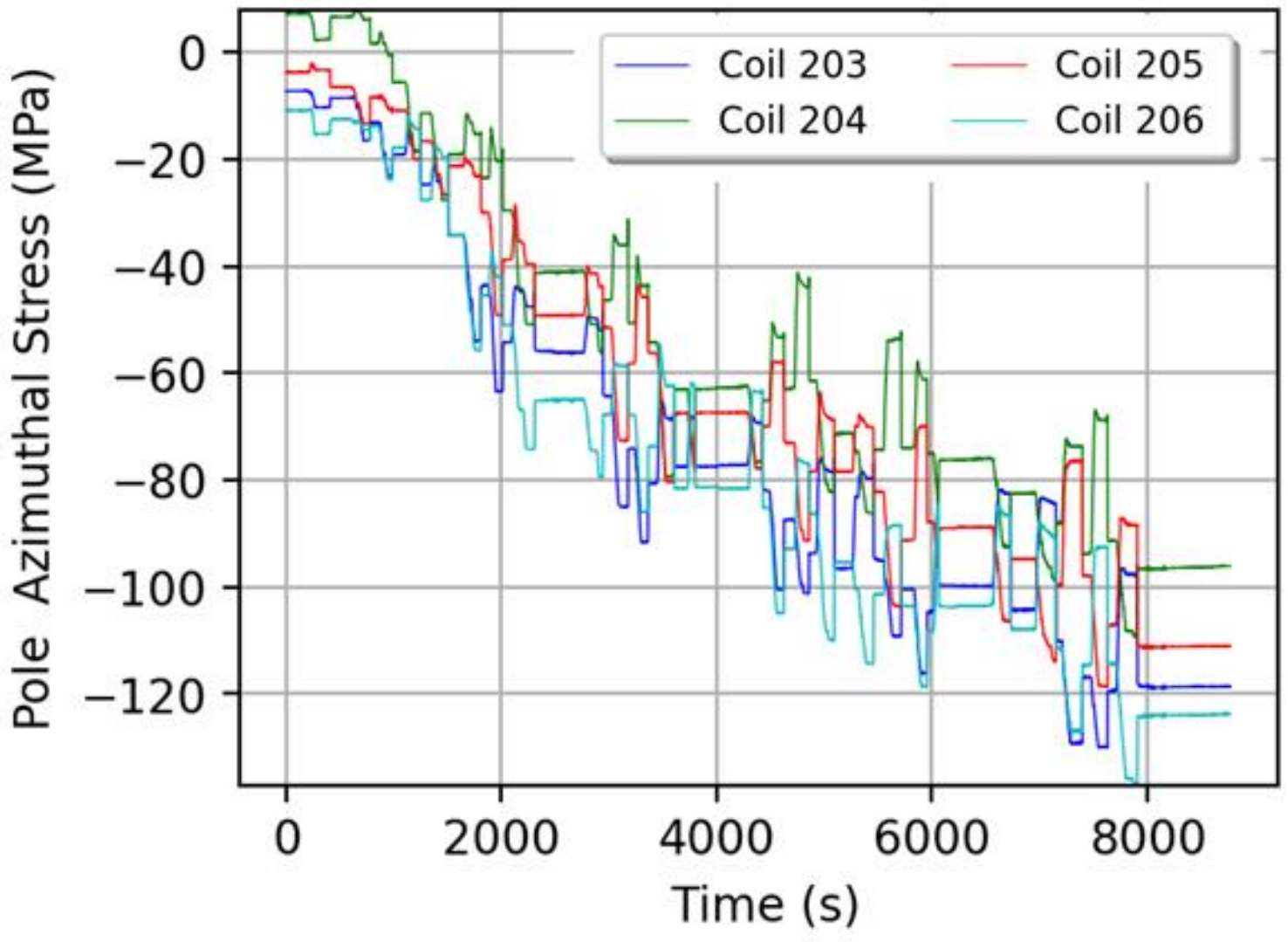

Fig. 5.10: Measured coil tress during bladder operation in MQXFS5.

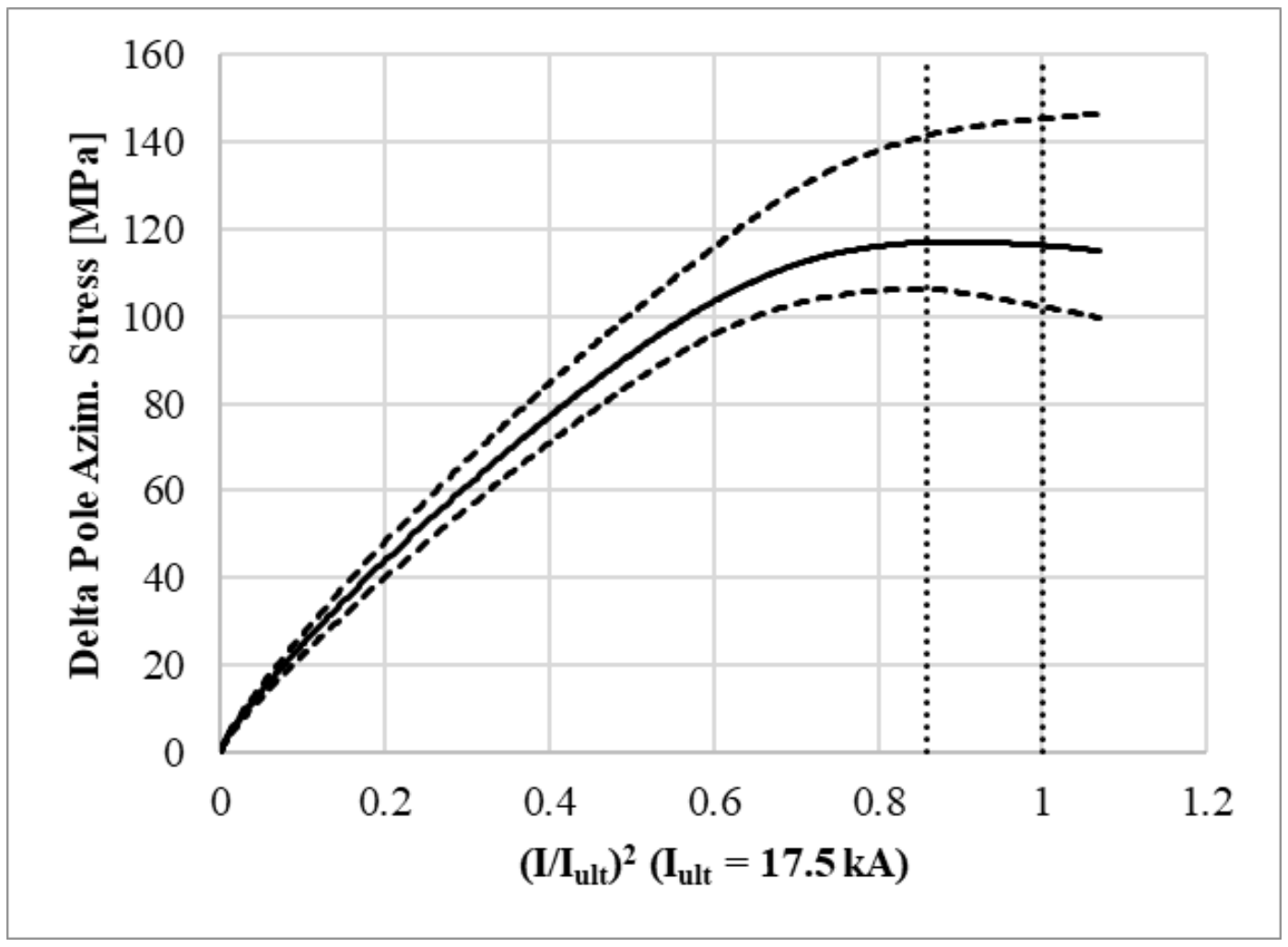

Fig. 5.11: Measured unloading of coil pole during excitation in MQXFS4: average (solid line) and max-min (dashed lines)

# 6. Assembled Magnet Specifications

All MQXFA "assembled magnet" requirements are described in [8]. The requirements are classified into two groups: "Threshold" requirements (R-T) and "Objective" requirements (R-O). Threshold requirements are requirements that contain at least one parameter that the project must achieve, and objective requirements are requirements that the project should achieve and will strive to achieve.

In the following sections we focus on the requirements related to the MQXFA assembled magnet and we address how the requirements are verified.

## 6.1. Magnet aperture

<u>For the magnet aperture, the following specifications are set (requirement R-T-01 in [8])</u>

- ***The MQXFA coil aperture at room temperature without preload is 149.5 mm. Coil bumpers (pions) of 1.2 mm thickness shall be installed on the four coil poles. The minimum free coil aperture at room temperature after coil bumpers installation, magnet assembly and preload shall be 146.7 mm. This guarantees an annulus free for HeII of minimum thickness 1.2 mm, average thickness larger or equal to 1.5 mm.***

The aperture is verified with a carbon fiber cylinder with pads that engage with the coil bumpers. The design of the cylinder and pad is shown in Fig. 6.1 and 6.2. Because of the presence of the strain gauges at the RE of the coils, of the eleven pion bumper locations only ten can be verified, up to approximately 4060 mm from the end of the splice box, as shown in Fig. 6.3.

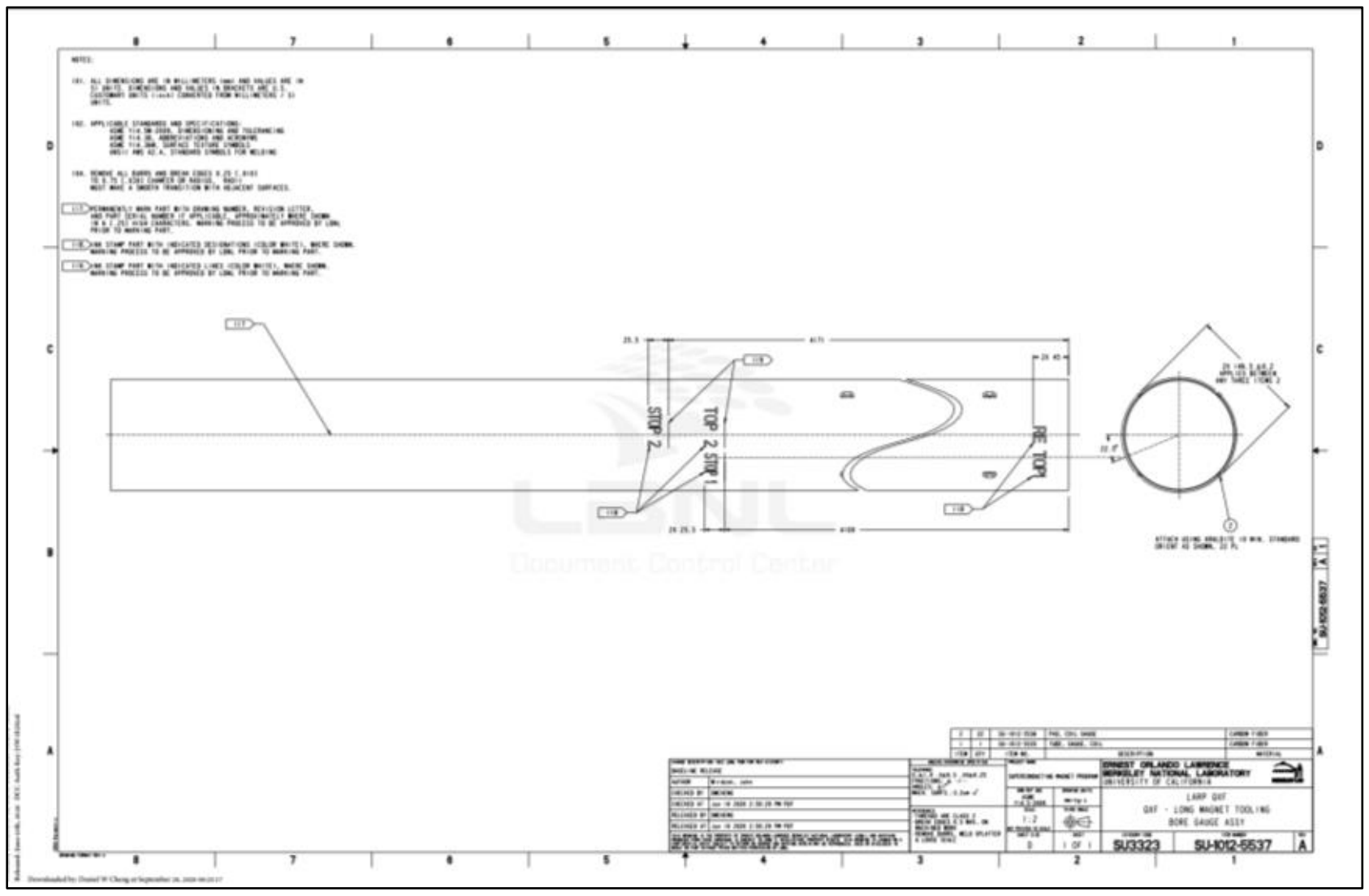

Fig 6.1: Drawing of the cylinder used to verify magnet aperture.

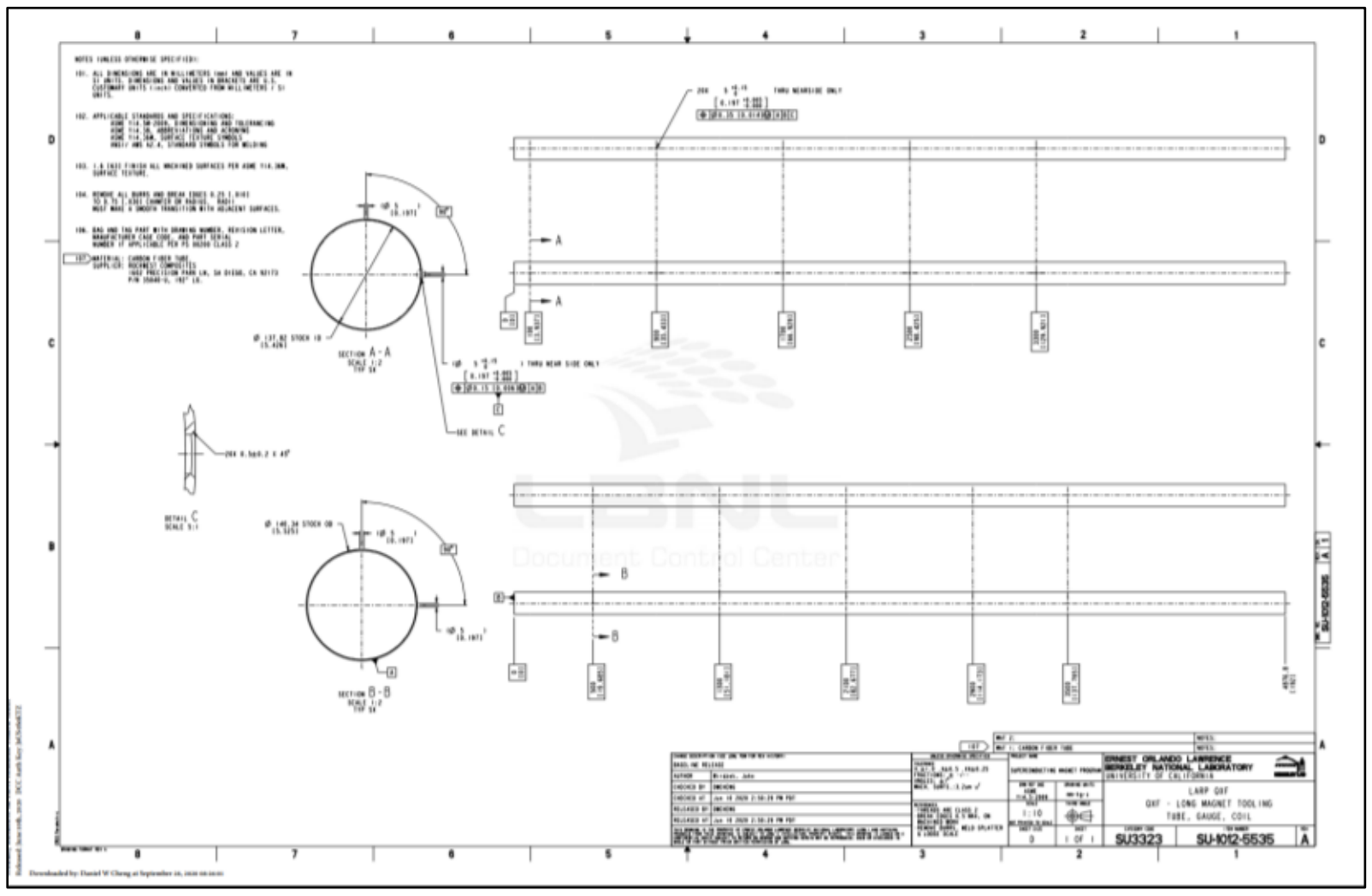

Fig 6.2: Drawing of the pad used to verify magnet aperture.

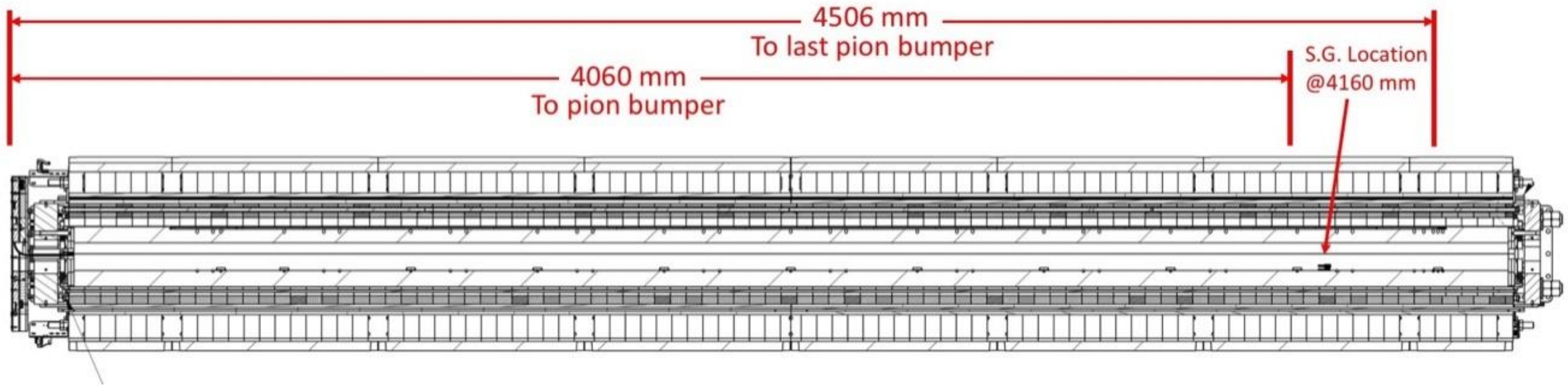

Fig 6.3 Cross-section of the magnet, showing 4060 mm of the magnet aperture that can be verified after magnet assembly.

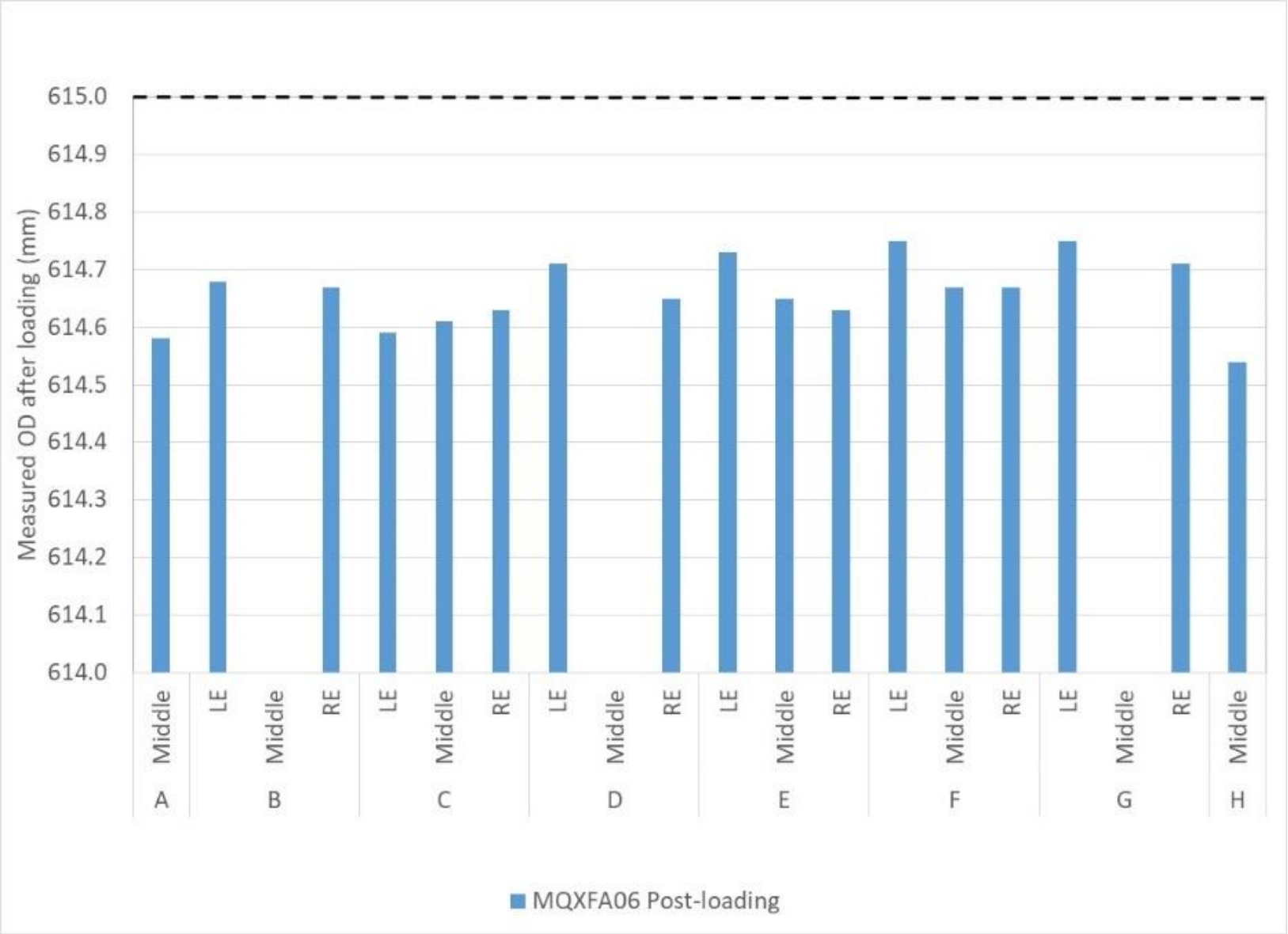

Fig 6.4: Measurements of the outer diameter in MQXFA06.

## 6.2. Magnet outer diameter and straightness

For the magnet outer diameter, the following specifications are set (requirement R-T-02 in [8])

- ***The MQXFA nominal outer diameter without preload is 614 mm. The MQXFA outer diameter best-fit after assembly and preload shall not exceed 615 mm.***

The OD of the magnet is measured using a pi tape after preload operations have been completed. In Fig 6.4, the measurements performed on MQXFA06 are shown.

The magnet straightness is monitored by measuring the vertical and horizontal absolute position of fiducial in the mid-plane shell cut-outs (both sides, left and right from the lead end view) and in the top shell cut-outs. Measurements are taken in 7 longitudinal locations (in between shell segments). The fiducials are placed in contact with the yoke laminations, which are accessible through the cut-outs.

For the straightness of the magnet, the following specifications are set:

- ***The maximum spread (max – min) of the vertical and horizontal position of the fiducials in the top shell cut-outs along 7 longitudinal locations shall be ≤0.250 mm.***

- ***The maximum spread (max – min) of the vertical and horizontal position of the fiducials in both sides of the mid-plane shell cut-outs along 7 longitudinal locations shall be ≤ 0.250 mm.***

## 6.3. Magnet cooling

For the magnet cooling, the following specifications are set (requirement R-T-06 and R-T-08 in [8])

- ***The MQXFA cooling channels shall be capable of accommodating two (2) heat exchanger tubes running along the length of the magnet in the yoke cooling channels. The minimum diameter of the MQXFA yoke cooling channels that will provide an adequate gap around the heat exchanger tubes is 77 mm.***
- ***The MQXFA shall have provisions for the following cooling passages: (1) Free passage through the coil pole and subsequent G-11 alignment key equivalent of 8 mm diameter holes repeated every 50 mm; (2) free helium paths interconnecting the four yoke cooling channels holes; and (3) a free cross-sectional area of at least 150 cm$^2$***

These specifications are addressed by the design of the components described in Sect. 4.

## 6.4. Splices and instrumentation

For the splice and instrumentation, the following specifications are set (requirement R-T-14 and R-T-15 in [5])

- ***Splices are to be soldered with CERN approved materials.***
- ***Voltage Taps: the MQXFA magnet shall be delivered with three redundant (3x2) quench detection voltage taps located on each magnet lead and at the electrical midpoint of the magnet circuit; and two (2) voltage taps for each internal MQXFA Nb3Sn-NbTi splice. Each voltage tap used for critical quench detection shall have a redundant voltage tap.***

The process of making the coil lead solder splices and installation of the Splice Box is described in the Work Instruction document "Splice Box Work Instructions (WI)" [14] and is carried out according to the CERN approved materials.

The configuration of voltage taps and other instrumentation wires is described in the Work Instruction Document "MQXFA Magnet Connectors and Finishing Work Instructions (WI)" [15], and in particular in the documents

- "Lead End connector Skirt Layout" [16].
- "Return End Connector Skirt Layout" [17].
- "MQXFA Series SG Schematic" [18].
- "MQXFA Series VT Schematic" [19].
- "MQXFA Pre-Series VT Schematic" [20].
- "MQXFA Series PH Schematic" [21].
- "MQXFA Integration and Loading" [22].

The process to install instrumentation and connector is given in [23].

All strain gauges, mounted on shells, coils, and rods are used during magnet assembly and during vertical test, and they are removed before cold-mass assembly together with connectors.

All instrumentation wires that will stay with the magnet are to be AXON HH2619-LH (Voltage Tap wires) and AXON HH1819-LH (Quench Heater wires).

### 6.5. Electrical tests

For the coils and quench heaters, the following specifications are set (requirement R-T-16 in [8])

- ***The MQXFA magnet coils and quench protection heaters shall pass the hi-pot test specified in Table 3 [8] before cold test.***

The table 3 of reference [8] is shown below in Fig. 6.5.

Table 3 [2]: Required hi-pot test voltages and leakage current
Based on Electrical Design Criteria for HL-LHC Inner Triplet Magnets [EDMS 1963398]

| Circuit Element | Expected Vmax [V] | V hi-pot | I hi-pot [μA]*** | Minimum time duration [s] |
|---|---|---|---|---|
| Coil to Ground at RT before helium exposure * | n.a. | 3.68 kV | 10 | 30 |
| Coil to Quench Heater at RT before helium exposure * | n.a. | 3.68 kV | 10 | 30 |
| Coil to Ground at cold ** | 670 | 1.84 kV | 10 | 30 |
| Coil to Quench Heater at cold ** | 900 | 2.3 kV | 10 | 30 |
| Coil to Ground at RT after helium exposure * | n.a. | 368 V | 10 | 30 |
| Coil to Quench Heater at RT after helium exposure * | n.a. | 460 V | 10 | 30 |

* Room Temperature conditions refer to air at 20±3 °C and relative humidity lower than 60%

** Cold conditions refer to nominal cryogenic conditions (superfluid helium)

*** Maximum leakage current does not include leakage of the test station.

Fig 6.5: Table of electrical tests.

Electrical QC tests are performed at several stages of magnet assembly and are described in the document "MQXF magnet Electrical QA at LBNL" [9]. The purpose of these tests, in addition to demonstrating that each magnet meets requirements, is to obtain a record of magnet electrical behavior before shipment and perform parameter tracking. They are summarized below.

1. Sequential resistance measurements, including the fixed voltage taps, for quench detection at pigtail connector (the header side) at 1 A.
    a. Current source (+) connected to A Lead, and (-) connected to B Lead.
    b. Verify current being applied to properly calculate resistances from measured voltages.
2. Magnet inductance measurements (L&Q) at 20, 100, and 1000 Hz.
3. Heater circuit resistance measurements.
4. Hipot
    a. Coil-to-Gnd Hipot 3680 V
        i. Heater floating. Cable shielding is tied to structure and thus grounded.
    b. Heater circuit to Coil Hipot 3680 V
        i. Structure is tied to cable shielding and both are floating.
        ii. Tested coils in group (Q1-Q2, Q4-Q3) according to PH circuits
    c. Heater-to-Gnd Hipot 3680 V
        i. Coil floating.
        ii. Outer layer PH circuits, each circuit tested individually.
5. Impulse tests at 1000 V, 1500 V, 2000 V, 2500 V, Direct- and Reversed-polarity for each voltage.

## 6.6. Radiation resistance

For the radiation resistance, the following specifications are set (requirement R-T-20 in [8])

- ***All MQXFA components must withstand a radiation dose of 35 MGy, or shall be approved by CERN for use in a specific location as shown in MQXFA Materials List [EDMS 1786261].***

All components of the structure are included in the "MQXFA Materials List" approved by CERN [24].

## 6.7. Interfaces

For the interfaces, the following specifications are set (requirement R-T-22 in [8])

- ***The MQXFA magnets shall meet the interface specifications with the following systems: (1) other LMQXFA Cold Mass components; (2) the CERN supplied power system; (3) the CERN supplied quench protection system, and (4) the CERN supplied instrumentation system. These interfaces are specified in Interface Control Document [25].***

Interfaces between the MQXFA magnet, the LMQXFA cold-mass and CERN systems are described in "MQXFA Magnet Interface Specification" [26], and are controlled by the Interface Control Documents [25]-[27].

## 6.8. Field quality

### 6.8.1. Positions of the local magnetic center and magnetic field angle

For the Positions of the local magnetic center and magnetic field angle, the following specifications are set (requirement MQXFA-R-O-01in [8])

- ***The positions of the local magnetic center and magnetic field angle are measured along the magnet axis with values obtained by averaging sections ≤ 500 mm. In each section, the local magnetic center is to be within ±0.5 mm from the magnet magnetic axis both in horizontal and in vertical direction. The local magnetic field angle in each section is to be ±2 mrad from the average magnetic field angle of the whole magnet, with the exception of the first measurement in the connection side where field angle is affected by magnet leads.***

The magnetic measurements of the local magnetic center and of the magnetic field angle are shown, with the specification ranges, in Fig. 6.6 to Fig. 6.8.

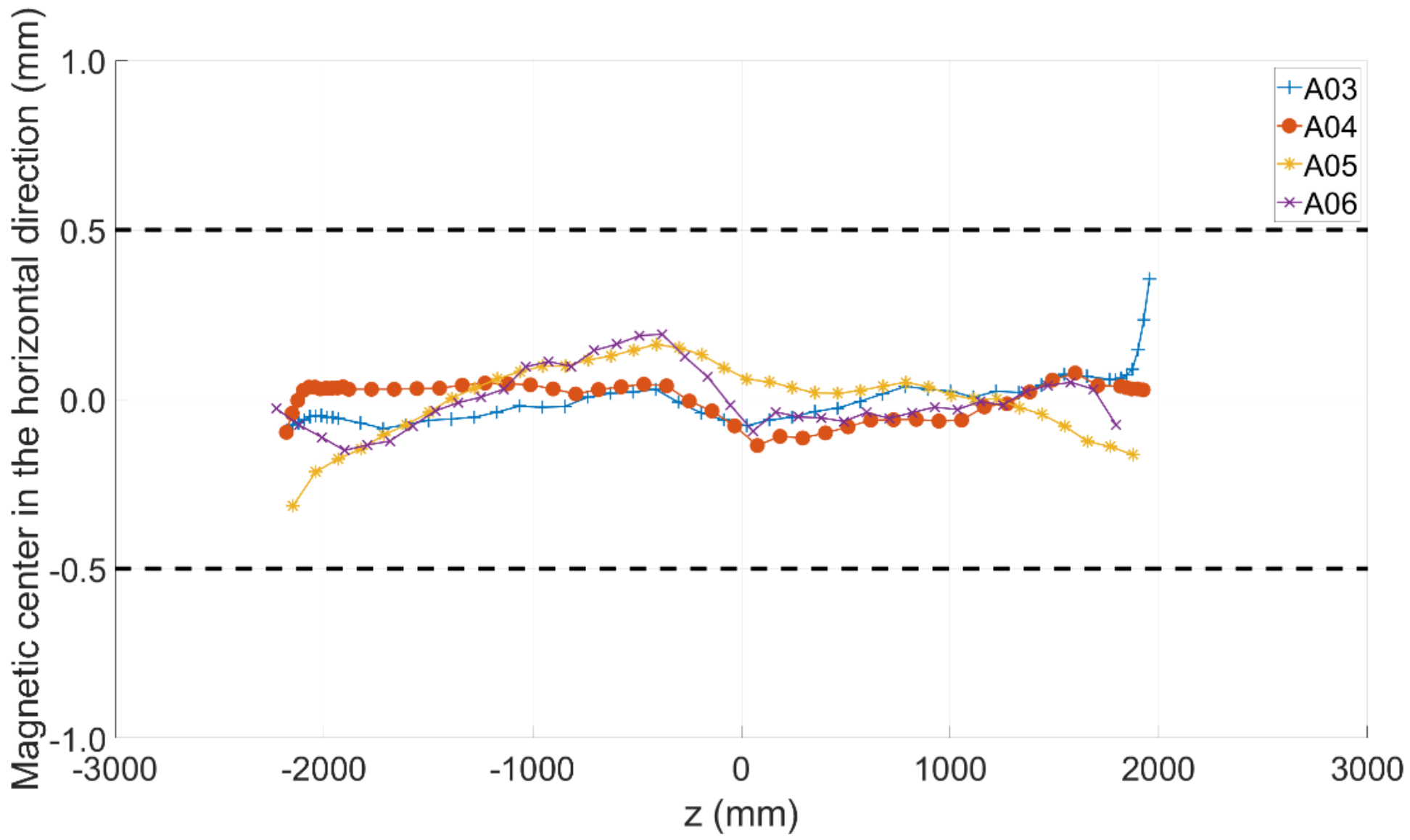

Fig.6.6: Magnetic center in the horizontal direction with respect to the fiducials. Only the region with a local transfer function higher than 6 T/m/kA is considered. The value of the magnetic center averaged over the region is set to zero. Data of A03 to A06 were measured using a 110 mm long rotating coil.

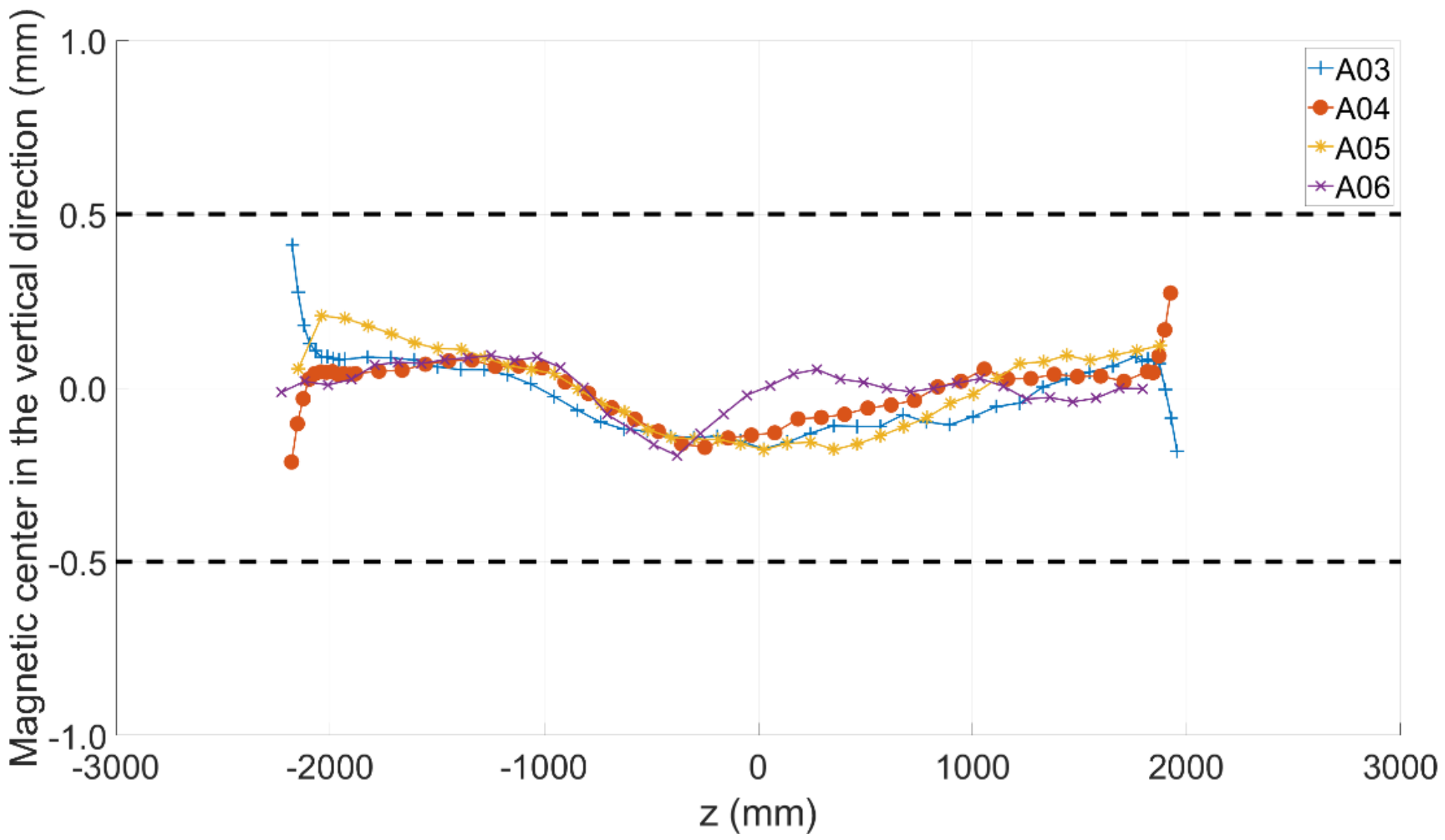

Fig.6.7: Magnetic center in the vertical direction with respect to the fiducials. Only the region with a local transfer function higher than 6 T/m/kA is considered. The value of the magnetic center averaged over the region is set to zero. Data of A03 to A06 were measured using a 110 mm long rotating coil.

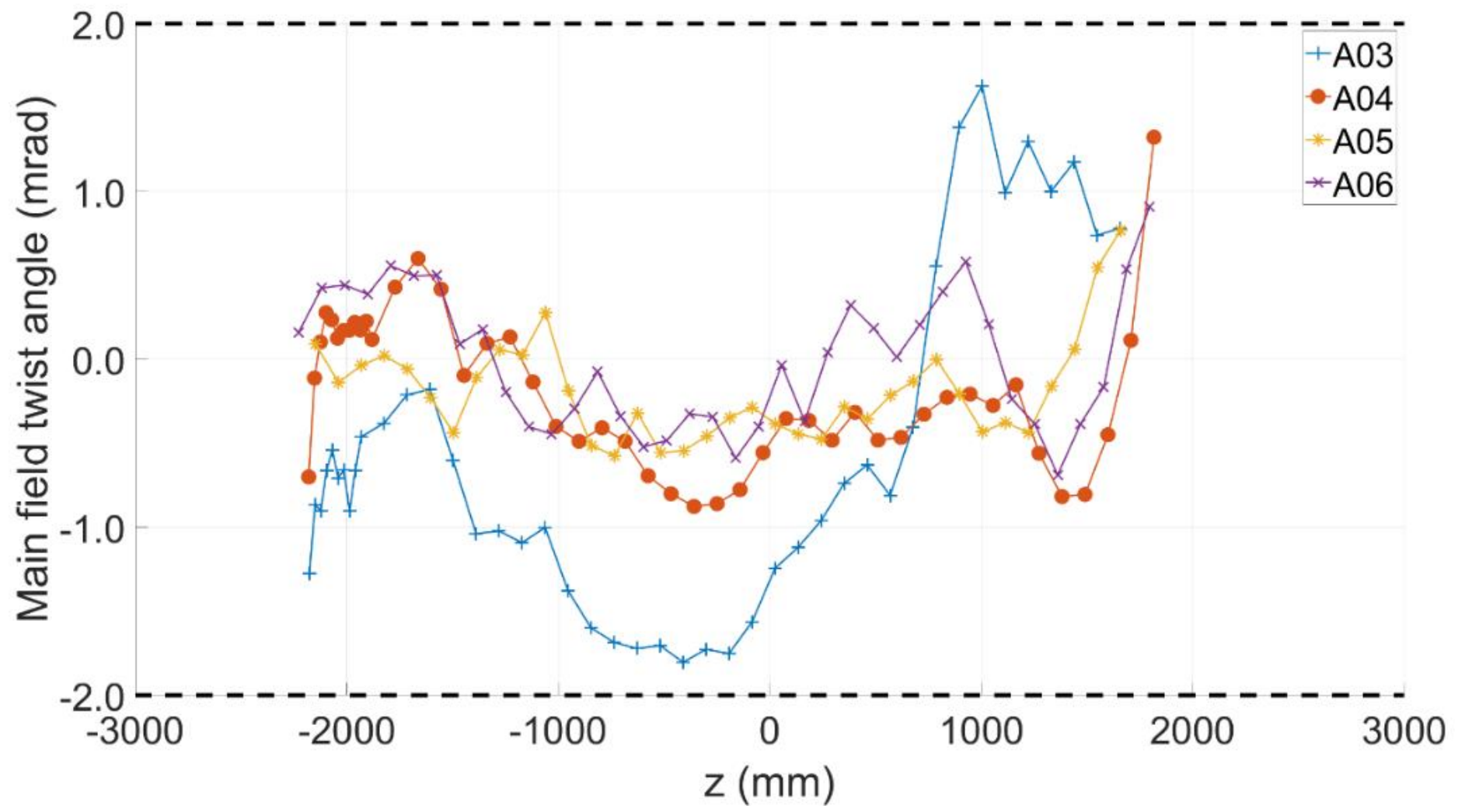

Fig.6.8: Main field twist angle. Only the region with a local transfer function higher than 6 T/m/kA is considered. The value of the twist angle averaged over the region is set to zero. The data points in the lead end were removed due to the effect of current leads. Data of A03 to A06 were measured using a 110 mm long rotating coil.

### 6.8.2. Field errors

For the field errors, the following targets are set (requirement MQXFA-R-O-02 in [8]):

- ***The field errors in units averaged in the straight part of the assembled magnet at a reference radius of 50 mm should be within the upper and lower bounds determined according to the Triplet Field Quality Table in Table 1 of [8]***
  - *A systematic component (0.9 units) associated with magnet assembly and cooldown is considered for $b_6$.*

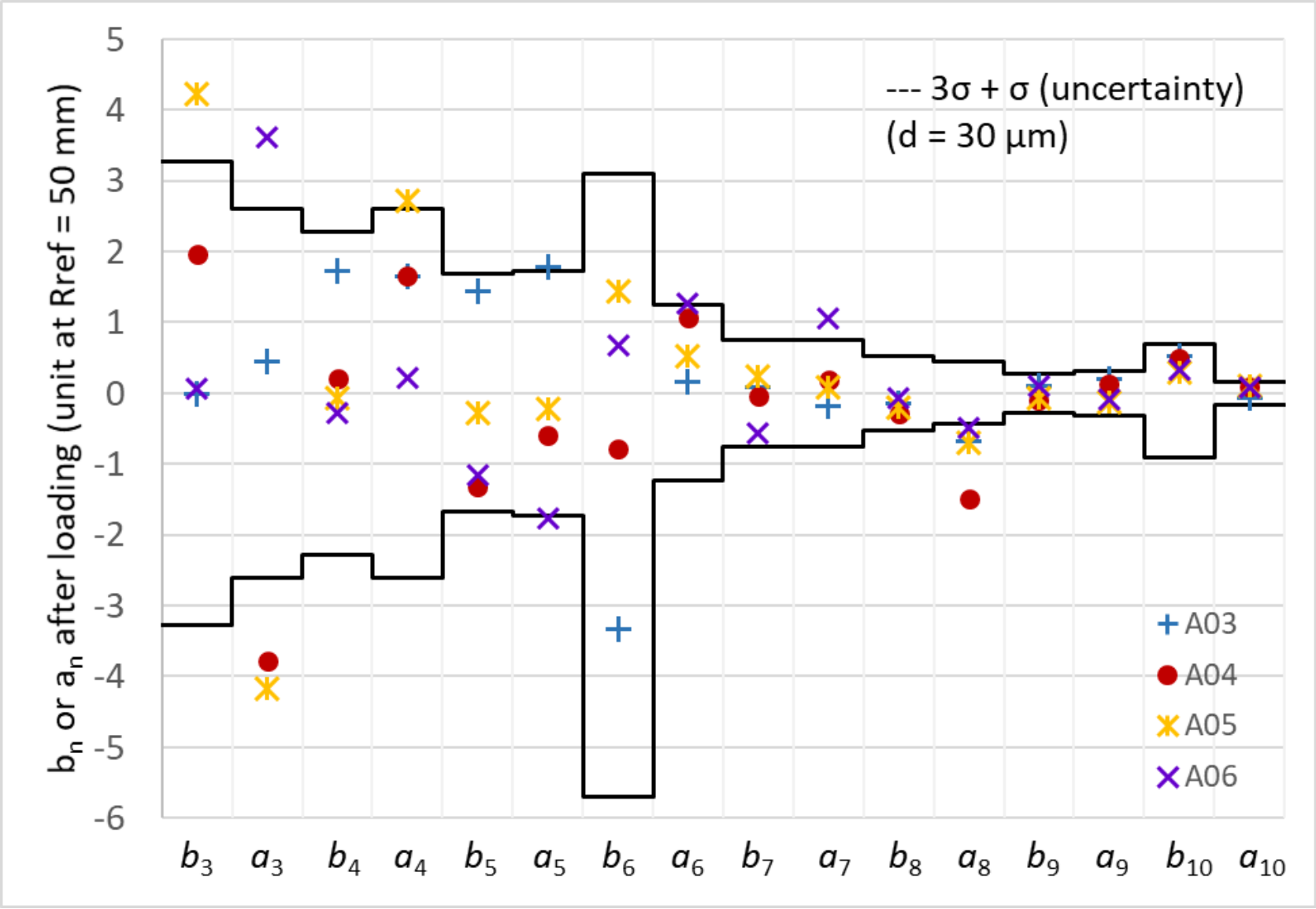

Fig.6.9: Field errors after loading. Data of A03 to A06 were measured using a 110 mm long rotating coil.

The field errors of the assembled magnets MQXFA03 to A06, measured at room temperature using a 110 mm long rotating coil, are shown in Fig. 6.9.

### 6.8.3. Transfer function

<u>For the transfer function, the following specifications are set:</u>

- ***The absolute value of the main field transfer function averaged over the straight part of magnets measured at room temperature after loading shall be within +8.810 T/(m kA) and +8.910 T/(m kA).***

The absolute value of the main field transfer function in the straight part of the magnets MQXFA03 to A06 is plotted in Fig. 6.10.

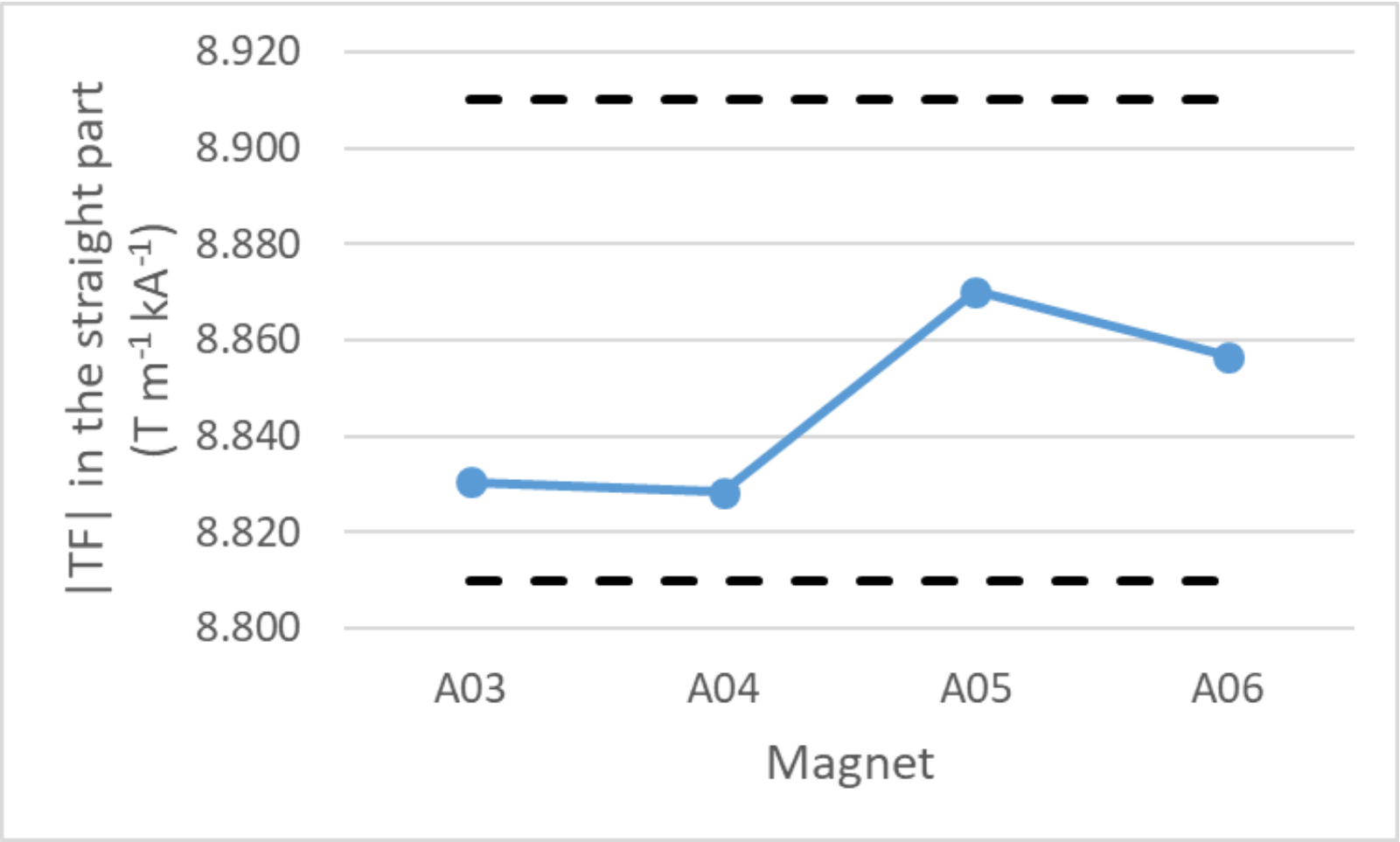

Fig.6.10: Absolute value of the main field transfer function in the straight part of the magnets after loading. Data of A03 to A06 were measured using a 110 mm long rotating coil. The lower bound is 8.810 T/m/kA and the upper bound is 8.910 T/m/kA.

## 6.9. Total length

<u>For the total length, the following specifications are set:</u>

- ***The total magnet length after loading at room temperature shall be 4906.7 ±3.5 mm***
    - *179.7 ±0.5 + 4563 ±1 + 164 ±2 mm = 4906.7 ±3.5 mm (see Fig. 6.11 and [28])*

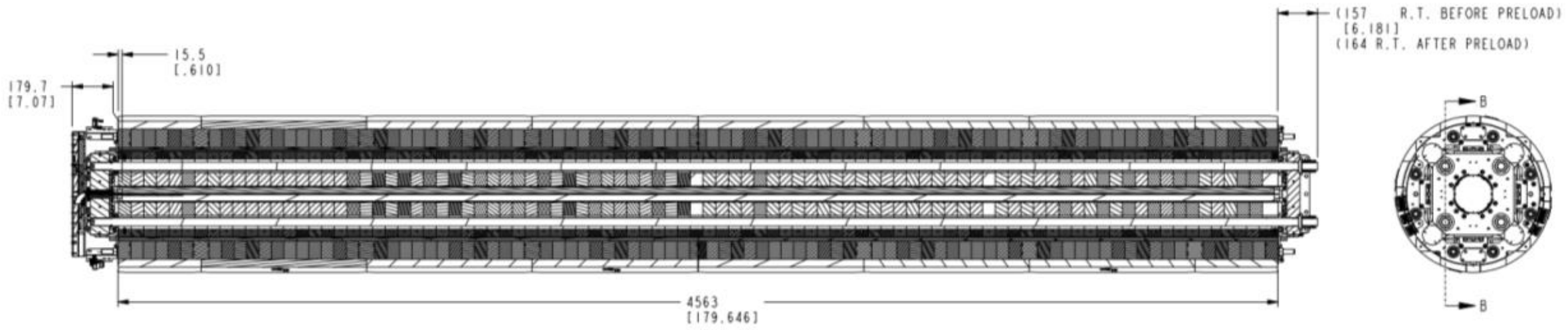

Fig. 6.11: Magnet length, excerpted from drawing SU-1011-0518.

The ±2 mm tolerance on the 164 mm of the return end after pre-load has been chosen to account for the rotational orientation of the rods.

# 7. QA/QC

The magnet QA plan will adhere to the US HL-LHC AUP at LBNL Quality Assurance Plan [29]. Fabrication QC tests will adhere to the QC Plan for MQXFA Structures Parts [4]. Assembly measurements (electrical, dimensional, magnetic, and strain gauge measurements, as well as by inspections made throughout the fabrication process) will follow the QC processes described in the Work Instructions mentioned in the respective sections above, and listed in full in Table 8.1.

After fabrication is complete magnet summary data must be uploaded per the LBNL MQXFA Magnet Fabrication Manufacturing & Inspection Plan (MIP) [30] together with all Discrepancy Reports, including Non-Conformity Reports (NCR), and travelers for L3 approval. After reviewing data, the MQXFA Magnets L2 must approve the magnet for shipment to BNL. If there are critical NCRs [31] the MQXFA Magnets L2 must be informed as soon as possible.

Table 8.1: List of work instruction

| # | Title | ID |
|---|---|---|
| 1 | Coil Acceptance WI | SU-1011-5294 |
| 2 | Coil CMM WI | SU-1012-3958 |
| 3 | Yoke Pre-Stack WI | SU-1008-8072 |
| 4 | Yoke Half Stack WI | SU-1009-7829 |
| 5 | Shell Instrumentation WI | SU-1009-3745 |
| 6 | Shell-Yoke Assy WI | SU-1008-2169 |
| 7 | Loadpad Pre-Stack WI | SU-1008-8075 |
| 8 | Dressed Coil WI | SU-1008-8073 |
| 9 | Pad-Collar Assy WI | SU-1010-1610 |
| 10 | Coil Pack Subassy WI | SU-1008-8074 |
| 11 | Magnet Electrical QA at LBNL | SU-1010-1903 |
| 12 | Magnetic Measurements | SU-1010-2018 |
| 13 | Integration and Loading WI | SU-1011-5637 |
| 14 | Axial Rods Instrumented WI | SU-1008-8069 |
| 15 | Connectors and Finishing WI | SU-1012-5517 |
| 16 | Splice Box WI | SU-1008-8067 |
| 17 | Magnet Shipping WI | SU-1011-3284 |

# 8. Magnet handling and shipment requirement

Specification for MQXFA magnet handling and shipment are in “MQXFA Magnet Handling and Shipping Requirements” [32].

# 9. Summary of magnet specifications

## 9.1. Structure component specifications

For the shell ID and OD, the following specifications are set:

- ***The shell ID at 5 evenly spaced rings shall have an average diameter of 556 mm +0.1/-0.0***
- ***The shell OD at 5 evenly spaced rings shall have an average diameter of 614 mm +0.1/-0.0***

For the yoke outer radius, the following specifications are set:

- ***The yoke radial profile shall maintain a 0.03 [.001”] tolerance.***

For the yoke master key interface, the following specifications are set:

- ***The yoke master key interface profile shall maintain a 0.03 [.001”] tolerance.***

For the yoke pre-stacks, the following specifications are set:

- ***588 mm Pre-stacks width shall be 588 mm +/- 0.13 with a Parallelism of 0.03 [.001"]***
- ***245 mm Pre-stacks width shall be 245 mm +/- 0.13 with a Parallelism of 0.03 [.001"]***

For the yoke half-stacks, the following specifications are set:

- ***Yoke half-stacks width shall be 2281.5 mm +/- 0.25 with a Parallelism of 0.05***

For the master key outer profile, the following specifications are set:

- ***The master key outer profile scan shall maintain a .05 [.002"] tolerance.***

For the load pad master key interface profile, the following specifications are set:

- ***The Load Pad master key interface profile shall maintain a 0.03 [.001"] tolerance.***

For the load pad collar key interface profile, the following specifications are set:

- ***The Load Pad collar interface profile shall maintain a 0.03 [.001"] tolerance.***

For the stainless-steel section of the pad pre-stack, the following specifications are set:

- ***On the LE pre-stack, the LE pads shall have a stainless-steel section of 300 ± 0.6 mm and an overall pre-stack length of 1050 ±0.13 mm***
- ***On the RE pre-stack, the RE pads shall have a stainless-steel section of 200 ± 0.4 mm and an overall pre-stack length of 1088 ±0.13 mm***
- ***The 1100 pre-stack shall have an overall length of 1100 ±0.13 mm***

For the collar to load pad interface profile, the following specifications are set:

- ***The entire collar to load pad interface profile shall maintain a 0.03 [.001"] tolerance all around.***

For the collar to coil interface profile, the following specifications are set:

- ***The collar to coil interface profile shall maintain a 0.03 [.001"] tolerance all around.***

## 9.2. Sub-assembly specifications

For the shell azimuthal and axial strain after shell-yoke sub-assembly, the following specifications are set:

- ***The shell strain in the azimuthal direction after shell-yoke sub-assembly shall be within +50/+250 micro-strain***
- ***The shell strain in the axial direction after shell-yoke sub-assembly shall be within -150/+50 micro-strain***

For the yoke cavity size after shell-yoke sub-assembly, the following specifications are set:

- ***The variation of yoke cavity dimension with respect to nominal shall be within +0.100/+0.400 mm***

For the pre-tension of the full-length tie rods of the yoke, the following specifications are set:

- ***The piston (RSM 200 cylinder with effective area of 4.43 in$^2$) shall be pressurized to 4100 ± 200 PSI and the nuts shall by turned to contact.***

For the straightness of the shell-yoke sub-assembly, the following specifications are set:

- ***The maximum spread (max – min) of the vertical and horizontal position of the fiducials in the top shell cut-outs along 7 longitudinal locations shall be ≤ 0.250 mm.***
- ***The maximum spread (max – min) of the vertical and horizontal position of the fiducials in one side of the mid-plane shell cut-outs along 7 longitudinal locations shall be ≤ 0.250 mm.***

For the yoke protrusion with respect to the shell, the following specifications are set:

- ***The yoke protrusion with respect to the shell shall be within +0.600/+1.200 mm***

For the radial gap between collars and insulated coils, the following specifications are set:

- ***The variation along the longitudinal direction of the radial gap between collars and insulated coils, considering for each z location the average among the four coils, shall be -0.125 mm -0.100 / +0.050 mm.***

For the coil-pack uniformity and squareness the following specifications are set:

- ***The uniformity of the vertical and horizontal dimensions along the z axis shall be within ±0.200 mm***
- ***The squareness of the vertical and horizontal dimensions along the z axis shall be within +0.900 mm***

For the pole key gap, the following specifications are set:

- ***The average pole key gap (per side) along the magnet length shall be +0.400 ±0.050 mm in each quadrant.***
- ***The minimum average pole key gap (per side) in any quadrant and in any longitudinal location shall be > +0.300 mm.***

For the main field transfer function, the following specifications are set:

- ***The absolute value of the main field transfer function averaged over the straight part of the coil pack shall be within +8.760 T/(m kA) and +8.860 T/(m kA).***

For the field errors, the following targets are set:

- ***The field errors in units averaged in the straight part of the coil pack with a reference radius of 50 mm should be within the upper and lower bounds determined according to the Triplet Field Quality Table in Table 1 of [8]***

For the electrical tests of the coil pack, the following specifications are set:

- ***1. Resistance check with hand-held multimeter***
  - ***a. Coil to structure***
  - ***b. Coil to coil***
  - ***c. Coil to PH***
- ***2. Hipot.***
  - ***b. Individual coil to structure, 3680 V.***
- ***3. Impulse test up to 2500 V, Direct- and Reversed-polarity.***
  - ***a. 1000 V, 1500 V, 2000 V, 2500 V***

## 9.3. Magnet assembly specifications

For the pre-load operation sequence, the following specifications are set:

- ***The loading operation shall follow the sequence described below:***
  - ***Pre-load axial rods with average to a measured of strain ~50 με***
  - ***Apply 50% azimuthal pre-load***
  - ***Apply 50% axial pre-load***
  - ***Apply 100% azimuthal pre-load***
  - ***Apply 100% axial pre-load***

For the pre-load targets at the end of the pre-load operations, the following specifications are set:

- ***The target average measured stress on coils, shells and rods at the end of the loading (after at least 24 h) shall be***
  - ***Shell average azimuthal stress: +58 ± 6 MPa***
  - ***Coil (winding pole) average azimuthal stress: -80 ± 8 MPa***
  - ***Rod average strain: +950 με ± 95 με***

For the maximum stress reached during the pre-load operations, the following specifications are set:

- ***During the entire pre-load operation, the maximum compression measured on each coil shall never exceed -110 MPa***

For the maximum coil size (arc length) variation along the coil length, the following specifications are set:

- ***Average coil size (arc length) after shimming shall range within ±100 µm with respect to average coil size measured in strain gauge location.***

## 9.4. Assembled magnet specifications

For the magnet aperture, the following specifications are set (requirement R-T-01 in [8])

- ***The MQXFA coil aperture at room temperature without preload is 149.5 mm. Coil bumpers (pions) of 1.2 mm thickness shall be installed on the four coil poles. The minimum free coil aperture at room temperature after coil bumpers installation, magnet assembly and preload shall be 146.7 mm. This guarantees an annulus free for HeII of minimum thickness 1.2 mm, average thickness larger or equal to 1.5 mm.***

For the magnet outer aperture, the following specifications are set (requirement R-T-02 in [8])

- ***The MQXFA nominal outer diameter without preload is 614 mm. The MQXFA outer diameter best-fit after assembly and preload shall not exceed 615 mm.***

For the straightness of the magnet, the following specifications are set:

- ***The maximum spread (max – min) of the vertical and horizontal position of the fiducials in the top shell cut-outs along 7 longitudinal locations shall be ≤ 0.250 mm.***
- ***The maximum spread (max – min) of the vertical and horizontal position of the fiducials in both sides of the mid-plane shell cut-outs along 7 longitudinal locations shall be ≤ 0.250 mm.***

For the magnet cooling, the following specifications are set (requirement R-T-06 and R-T-08 in [8])

- ***The MQXFA cooling channels shall be capable of accommodating two (2) heat exchanger tubes running along the length of the magnet in the yoke cooling channels. The minimum diameter of the MQXFA yoke cooling channels that will provide an adequate gap around the heat exchanger tubes is 77 mm.***
- ***The MQXFA shall have provisions for the following cooling passages: (1) Free passage through the coil pole and subsequent G-11 alignment key equivalent of 8 mm diameter holes repeated every 50 mm; (2) free helium paths interconnecting the four yoke cooling channels holes; and (3) a free cross-sectional area of at least 150 $cm^2$***

For the splice and instrumentation, the following specifications are set (requirement R-T-14 and R-T-15 in [8])

- ***Splices are to be soldered with CERN approved materials.***
- ***Voltage Taps: the MQXFA magnet shall be delivered with three redundant (3x2) quench detection voltage taps located on each magnet lead and at the electrical midpoint of the magnet circuit; and two (2) voltage taps for each internal MQXFA Nb3Sn-NbTi splice. Each voltage tap used for critical quench detection shall have a redundant voltage tap.***

For the coils and quench heaters, the following specifications are set (requirement R-T-16 in [8])

- ***The MQXFA magnet coils and quench protection heaters shall pass the hi-pot test specified in Table 3 [8] before cold test.***

For the radiation resistance, the following specifications are set (requirement R-T-20 in [8])

- ***All MQXFA components must withstand a radiation dose of 35 MGy, or shall be approved by CERN for use in a specific location as shown in MQXFA Materials List [EDMS 1786261].***
- ***The MQXFA magnets shall meet the interface specifications with the following systems: (1) other LMQXFA Cold Mass components; (2) the CERN supplied power system; (3) the CERN supplied quench protection system, and (4) the CERN supplied instrumentation system. These interfaces are specified in Interface Control Document [25].***

For the Positions of the local magnetic center and magnetic field angle, the following specifications are set (requirement MQXFA-R-O-01in [8])

- ***The positions of the local magnetic center and magnetic field angle are measured along the magnet axis with values obtained by averaging sections ≤ 500 mm. In each section, the local magnetic center is to be within ±0.5 mm from the magnet magnetic axis both in horizontal and in vertical direction. The local magnetic field angle in each section is to be ±2 mrad from the average magnetic field angle of the whole magnet, with the exception of the first measurement in the connection side where field angle is affected by magnet leads.***

For the field errors, the following targets are set (requirement MQXFA-R-O-02 in [8]):

- ***The field errors in units averaged in the straight part of the assembled magnet at a reference radius of 50 mm should be within the upper and lower bounds determined according to the Triplet Field Quality Table in Table 1 of [8]***
    - *A systematic component (0.9 units) associated with assembly and cooldown for b6 are considered.*

For the transfer function, the following specifications are set:

- ***The absolute value of the main field transfer function averaged over the straight part of magnets measured at room temperature after loading shall be within +8.810 T/(m kA) and +8.910 T/(m kA).***

For the total length, the following specifications are set:

- ***The total magnet length after loading at room temperature shall be 4906.7 ±3.5 mm***